\documentclass[epj]{svjour}
\usepackage{graphics}
\usepackage{tikz}
\usepackage{graphicx}
\usepackage{amsmath}

\usepackage[colorlinks=true, allcolors=blue]{hyperref}
\usepackage{scalerel,stackengine}
\usepackage[british]{babel}
\usepackage{epsfig}
\usepackage{appendix}
\usepackage[switch,columnwise]{lineno}

\usepackage[space]{cite}
\usepackage{amsmath}
\usepackage{multirow}

\stackMath
\newcommand\reallywidehat[1]{%
\savestack{\tmpbox}{\stretchto{%
  \scaleto{%
    \scalerel*[\widthof{\ensuremath{#1}}]{\kern.1pt\mathchar"0362\kern.1pt}%
    {\rule{0ex}{\textheight}}
  }{\textheight}%
}{2.4ex}}%
\stackon[-6.9pt]{#1}{\tmpbox}%
}
\usepackage{braket}
\usepackage[utf8]{inputenc}

\begin{document}\sloppy
\raggedbottom
\hugehead
\title{The potential of $\Lambda$ and $\Xi^-$ studies with PANDA at FAIR}

%
%
\author{
G.~Barucca\inst{1} \and 
F.~Davì\inst{1} \and 
G.~Lancioni\inst{1} \and 
P.~Mengucci\inst{1} \and 
L.~Montalto\inst{1} \and 
P. P.~Natali\inst{1} \and 
N.~Paone\inst{1} \and 
D.~Rinaldi\inst{1} \and 
L.~Scalise\inst{1} \and 
W.~Erni\inst{2}
% ...
}
\institute{
Università Politecnica delle Marche-Ancona,{ \bf Ancona}, Italy \and 
Universität Basel,{ \bf Basel}, Switzerland
% ...
}

\date{August 2020}
\abstract{The antiproton experiment PANDA at FAIR is designed to bring hadron physics to a new level in terms of scope, precision and accuracy. In this work, its unique capability for studies of hyperons is outlined. We discuss ground-state hyperons as diagnostic tools to study non-perturbative aspects of the strong interaction, and fundamental symmetries. New simulation studies have been carried out for two benchmark hyperon-antihyperon production channels: $\bar{p}p \to \bar{\Lambda}\Lambda$ and $\bar{p}p \to \bar{\Xi}^+\Xi^-$. The results, presented in detail in this paper, show that hyperon-antihyperon pairs from these reactions can be exclusively  reconstructed with high efficiency and very low background contamination. In addition, the polarisation and spin correlations have been studied, exploiting the weak, self-analysing decay of hyperons and antihyperons. Two independent approaches to the finite efficiency have been applied and evaluated: one standard multidimensional efficiency correction approach, and one efficiency independent approach. The applicability of the latter was thoroughly evaluated for all channels, beam momenta and observables. The standard method yields good results in all cases, and shows that spin observables can be studied with high precision and accuracy already in the first phase of data taking with PANDA.
}
\maketitle

\section{Introduction}
\label{sec:intro}
The Standard Model of particle physics has proven successful in describing the elementary particles and their interactions \cite{pdg}. However, it still falls short in explaining many of the basic features of the nucleon, features that to this day remain the objects of intensive research: spin \cite{pspin,pspin2}, size \cite{pradius1,pradius2,pradius3}, intrinsic structure \cite{pstructure,pstructure2} and abundance, \textit{i.e.} the excess of nucleons compared to antinucleons in the universe \cite{pasym}. 

One of the nucleons, the proton, is the most stable composite system we know. In order to study its properties, we therefore need to distort or break it by scattering something on it, for example an electron, or by adding some energy and thereby excite it. A third option is to replace one or several of the building blocks \cite{granados}. The latter is the main concept of \textit{hyperon physics}: one or several light $u$ or $d$ quarks in the nucleon is replaced by strange ones.\footnote{In principle, one can also replace it with a charm or a bottom quark, but the scope of this paper is strange hyperons.} The mass of the strange quark is $\approx$ 95 MeV, which is $\geq$ 20 times larger than the light $d$ and $u$ quark masses. The strange quark in a hyperon is therefore expected to behave a bit differently than the light quarks, for example it will be less relativistic. Furthermore, a larger part of the mass of a hyperon comes from the quarks compared to the nucleon. However, the mass of the strange quark is much smaller than the mass of the hyperon itself, in contrast to the more than ten times heavier charm quark. Hence, strange hyperons are sufficiently similar to nucleons for comparisons to be valid, for example assuming approximate SU(3) flavour symmetry.

By being unstable, hyperons reveal more of their features than protons. In particular, the weak, parity violating and thereby \textit{self-analysing} decay of many ground-state hyperons make their spin properties experimentally accessible. This makes hyperons a powerful diagnostic tool that can shed light on various physics problems, \textit{e.g.} non-perturbative production dynamics, internal structure and fundamental symmetries.

In this paper, we outline the assets of hyperon physics to be exploited by the future PANDA (antiProton ANnihilation at DArmstadt) experiment with an antiproton beam at FAIR (Facility for Antiproton and Ion Research) in Darmstadt, Germany. We describe in detail a comprehensive simulation study that demonstrates the feasibility of the planned hyperon physics programme, and discuss the impact and long-term perspectives.

\section{The PANDA experiment}
\label{sec:panda}

\begin{figure*}[t]
\centering
\begin{tikzpicture}
\node[anchor=south west, inner sep=0] (image) {\includegraphics[width=\textwidth]{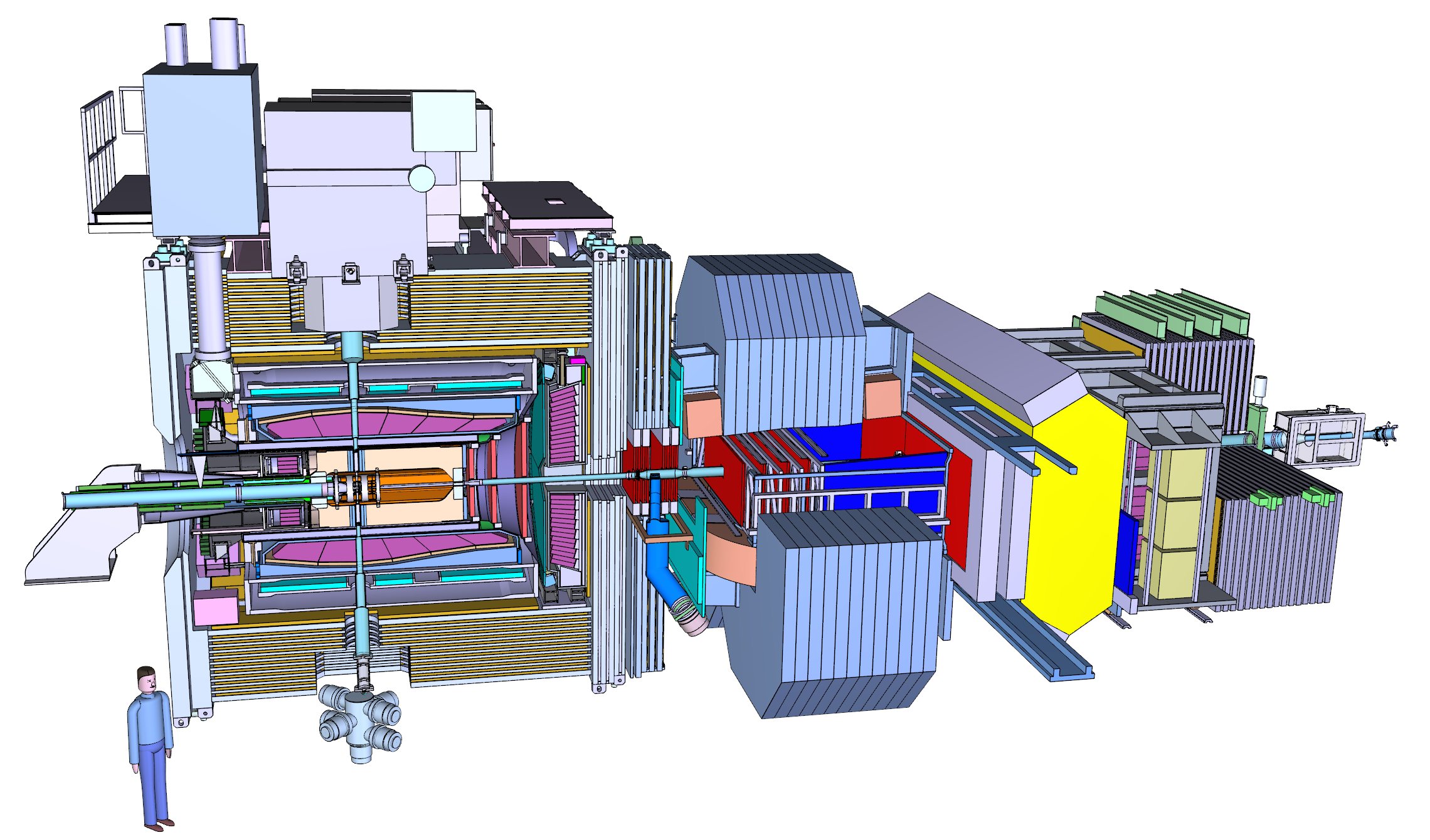}};
	\begin{scope}[x={(image.south east)},y={(image.north west)}]
        
        \begin{scope}
            \node[align=center, red] at (0.25,0.95) {TS};
            \node[align=center, red] at (0.65,0.7) {FS};
        \end{scope}
	\end{scope}
\end{tikzpicture}
\caption{(Colour online) Overview of the full PANDA setup. The antiproton beam will go from left to right, whereas the target jets/pellets from top to bottom. The left part of the detector surrounds the interaction point and is the Target Spectrometer (TS), whereas the right part is the Forward Spectrometer (FS).}
\label{fig:panda}
\end{figure*} 

The PANDA experiment, which is currently under construction at FAIR \cite{fair} in Darmstadt, Germany, offers a broad programme for studies of the strong interaction and fundamental symmetries \cite{panda}. The unique combination of an antiproton beam in the intermediate energy range and a nearly 4$\pi$ detector with vertex- and tracking devices, multiple particle identification (PID) detectors and calorimeters, give excellent conditions for a new generation of hyperon physics experiments.

The High Energy Storage Ring (HESR) will deliver an antiproton beam with momenta ranging from 1.5 GeV/$c$ up to 15 GeV/$c$ \cite{hesr}. In the start-up phases, referred to as \textit{Phase One} and \textit{Phase Two}, the HESR will be able to accumulate up to $10^{10}$ antiprotons in 1000~s. In the final \textit{Phase Three}, the luminosity will be ramped up by the Recuperated Experimental Storage Ring (RESR), allowing up to $10^{11}$ antiprotons to be injected and stored in the HESR. The HESR will offer stochastic cooling resulting in a beam momentum spread of better than $5\cdot 10^{-5}$. The antiproton beam will impinge on a hydrogen cluster jet or pellet target, which during Phase One will result in an average luminosity of $\approx 10^{31}$ cm$^2$s$^{-1}$~\cite{target}. At low energies, the luminosity will be about a factor of two lower. During Phase Three, the design luminosity of $\approx 2\cdot10^{32}$ cm$^2$s$^{-1}$ will be achieved. 

The PANDA detector, shown in Fig. \ref{fig:panda} and described in detail in Ref.~\cite{pandadet}, is divided into two parts: the target spectrometer (TS) and the forward spectrometer (FS). The TS covers polar angles of $> 10^{\mathrm{o}}$ in the horizontal direction and $> 5^{\mathrm{o}}$ in the vertical direction, whereas the FS covers polar angles $<10^{\mathrm{o}}$. The TS provides timing and vertexing by the silicon micro vertex detector (MVD). The MVD is also used for tracking together with the gas-filled straw tube trackers (STT). The polar angle range of the latter is $22^{\mathrm{o}} < \theta < 140^{\mathrm{o}}$. In order to bridge the acceptance between the STT and the FS, the gas electron multiplier detectors (GEM) is designed to track particles emitted below 22$^\circ$. Time-of-flight detectors (TOF), made of scintillating tiles, offer excellent time resolution. By providing the reaction time $t_0$, it improves the resolution of the track parameters, and increases the particle identification capabilities. Detection of internally reflected Cherenkov light (DIRC) offer independent PID and an electromagnetic calorimeter (EMC) with lead-tungstate (PbWO$_4$) crystals will measure energies between 10 MeV and 7 GeV. The laminated yoke of the solenoid magnet, outside the barrel EMC, is interleaved with sensitive layers to act as a range system for the detection and identification of muons. Measurement of the charge and momenta are possible thanks to the bending of particle trajectories by a solenoid magnet providing a field of up to 2.0 Tesla. 

The FS will consist of six straw tube stations for tracking, a dipole magnet, a ring imaging Cherenkov counter (RICH) detector for PID as well as a TOF for timing and PID. The energies of the forward going, electromagnetically interacting particles, will be measured by a Shashlyk electromagnetic calorimeter. A muon range system, using sensors interleaved with absorber layers, is placed at the end of the FS.

The luminosity will be determined by using elastic antiproton-proton scattering as the reference channel. The differential cross section of this process can be calculated with extremely high precision at small angles, where the Coulomb component dominates~\cite{luminosity}. At polar angles within 3-8~mrad, the scattered antiproton will be measured by a luminosity detector consisting of four layers of thin monolithic active pixels sensors made of silicon~\cite{luminosity}. 

PANDA will feature, as one of the first experiments, a time-based data acquisition system (DAQ) without hard-ware triggers. Data will instead be read out as a continuous stream using an entirely software-based selection scheme. This change of paradigm is driven by the large foreseen reaction rates, resulting in huge amounts of data to be stored. 


The feasibility studies presented in this work are performed within the common simulation and analysis framework PandaROOT \cite{PANDAROOT}. It comprises the complete simulation chain, including Monte Carlo event generation, particle propagation and detector response, hardware digitization, reconstruction and calibration, and data analysis. PandaROOT is derived from the FairROOT framework \cite{FAIRROOT} which in turn is based on ROOT \cite{ROOT}. 

\section{Hyperon production with antiproton probes}
\label{sec:hyperonphys}
The focus of this paper is $\Lambda$ and $\Xi^-$ hyperon production in the $\bar{p}p \to \bar{Y}Y$ reaction, where $Y$ refers to the octet hyperons $\Lambda$, $\Xi^-$. Understanding the production and decay of these hyperons is crucial in order to correctly interpret experimental analyses of heavier hyperons. The study of excited multi-strange hyperons constitutes an important part of the PANDA physics programme and is described in more detail in Ref. \cite{jennypthesis}. However, octet hyperons are interesting in their own right. The $\Lambda$ and $\Xi^-$ hyperons considered in this work, predominantly decay into charged final state particles which makes them straight-forward to measure experimentally.
In the following, we will discuss how the self-analysing decays can shed light on various aspects of fundamental physics and the advantages of antiproton probes in hyperon studies.

\subsection{Weak two-body decays}
\label{sec:features}
\begin{figure}
\begin{center}
\includegraphics[width=0.35\textwidth]{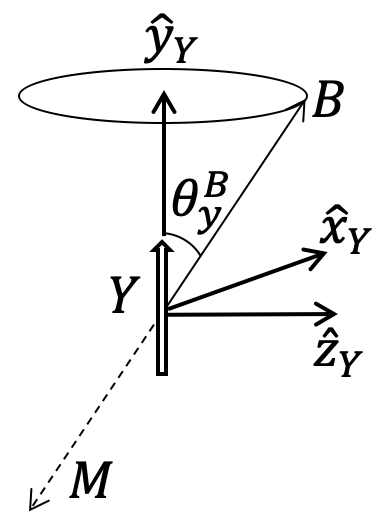}
\end{center}
\caption{The $Y \to BM$ decay, with the spin direction of $Y$ along the $y$-axis.}
\label{fig:hypdecay}
\end{figure}

All ground-state hyperons except the $\Sigma^0$ decay weakly through a process that has a parity violating component. This means that the direction of the decay products depends on the spin direction of the mother hyperon. In Fig. \ref{fig:hypdecay}, the two-body decay of a spin 1/2 hyperon $Y$ into a spin 1/2 baryon $B$ and a pseudoscalar meson $M$, is illustrated. The angular distribution of $B$ in the rest system of $Y$ is given by \cite{pdg,bigibook}
\begin{equation}
W(\cos\theta^B)=\frac{1}{4\pi}(1+\alpha P_y^{Y}(\cos\theta_{Y}) \cos\theta^B),
\label{eq:decay}
\end{equation}

\noindent where $P^Y_y(\cos\theta_{Y})$ is the polarisation with respect to some reference axis $\hat y$. $P^Y_y$ carries information about the production process and therefore depends on the collision energy and the scattering angle. The decay asymmetry parameter $\alpha$ is the real part of the product between the parity violating and the parity conserving decay amplitudes, $T_s$ and $T_p$ \cite{leeyang}. 
Eq. (\ref{eq:decay}) demonstrates how the experimentally measurable decay angular distribution is related to quantities with physical meaning, \textit{i.e.} $P^Y_y$ and $\alpha$. This feature makes hyperons a powerful diagnostic tool. 

\subsection{Scientific case}
\label{sec:scientific}

Antihyperon-hyperon pair production in antiproton-proton annihilation, $\bar{p}p \to \bar{Y}Y$, provides excellent conditions for hyperon studies, since
\begin{itemize}
    \item Antihyperons and hyperons, also with double- and triple strangeness, can be produced in two-body processes at low energies, where the number of partial waves is small. This makes the production process parameterizable in a close to model-independent way.
    \item Antihyperons and hyperons can be studied simultaneously, under symmetric conditions.
    \item The production cross sections for single- and double-strange hyperons are known to be large \cite{tord} which results in large count rates also for modest luminosities.
    \end{itemize}

The scale of strangeness production is governed by the mass of the strange quark, $m_s \approx 95$ MeV/$c^2$. This is far below the scale where perturbative QCD breaks down ($\approx$ 1 GeV) but close to the QCD cut-off scale ($\Lambda_{QCD}$). Therefore, the relevant degrees of freedom in processes involving strange quarks are unclear: quarks and gluons, or hadrons?  

Single-strange hyperon production in $\bar{p}p \to \bar{Y}Y$ has been modeled using quark-gluon degrees of freedom \cite{quarkgluon}, meson exchange \cite{kaonexchange} and a combination of the two \cite{quarkgluonhadron}. The production of double-strange hyperons requires interactions at shorter distances, since it either implies annihilation of two quark-antiquark pairs \cite{XiQG,kaidalov}, or exchange of two kaons \cite{XiMEX}. Spin observables, accessible for hyperons through their self-analysing decays, are particularly powerful in differentiating between models, since they are sensitive to the production mechanism. Spin observables can also give information about possible polarized strangeness content in nucleons \cite{alberg} and final state interactions \cite{haidenbauerFSI}. It is important to have a solid understanding of the latter also when interpreting results from antihyperon-hyperon pair production with other probes. One example is the $e^+e^- \to Y\bar{Y}$ process, from which the time-like electromagnetic form factors are determined. In Ref. \cite{haidenbauerFSI}, the $\Lambda$ complex form factors are predicted based on potential models fitted to PS185 data \cite{PS185} on spin observables in the $\bar{p}p \to \bar{\Lambda}\Lambda$ reaction \cite{haidenbauerPot}. It was found that the form factors are sensitive to the $\bar{\Lambda}\Lambda$ final state interaction and that spin observables are necessary in order to discriminate between them \cite{haidenbauerFSI}. This has been done in a recent measurement of $\Lambda$ form factors by the BESIII collaboration \cite{bes3prl}.

Hyperon decays can provide one piece to the puzzle of nucleon abundance, more commonly referred to as the \textit{matter-antimatter asymmetry} puzzle. According to the present paradigm, equal amounts of matter and antimatter should have been produced in the Big Bang. Unless the initial matter-antimatter imbalance was fine-tuned, a dynamical enrichment of matter with respect to antimatter must have occurred, \textit{i.e.} Baryogenesis. However, this is only possible if i) processes exist that violate baryon number conservation ii) processes exist that violate C and CP symmetry, and iii) the aforementioned processes occured outside thermal equilibrium \cite{pasym}. With hyperons, criterion ii) can be tested. CP symmetry means that hyperons and antihyperons have the same decay patterns, but with reversed spatial coordinates. For two-body hyperon decays, it means that the decay asymmetry parameters, \textit{e.g.} $\alpha$ in Eq. (\ref{eq:decay}), have exactly the same value but with opposite sign compared to the corresponding antihyperon parameter, \textit{i.e.} $\alpha = -\bar{\alpha}$. The large production rates and the symmetric hyperon and antihyperon conditions make the $\bar{p}p \to \bar{Y}Y$ a suitable reaction for searching for CP violation. Hyperon-antiproton studies have been carried out recently with BESIII in $e^+e^- \to Y\bar{Y}$, a reaction that is similar to $\bar{p}p \to \bar{Y}Y$ in the sense that it is a two-body reaction that is symmetric in particle-antiparticle observables. These studies show that the precision can be greatly improved by several orders of magnitudes if the production process can be pinned down \cite{goran,goranand,bes3nature}. Hence, a proper understanding of the $\bar{p}p \to \bar{Y}Y$ reaction mechanism constitutes a crucial milestone in future large-scale CP studies with PANDA at FAIR.

\subsection{State of the art}
\label{sec:stateoftheart}

\subsubsection{Hyperon production in $\bar{p}p$ annihilations}

The large amount of high-quality data on single-strange hyperons \cite{tord,PS185,PS185164} produced in antiproton-proton annihilation, partly with a polarised target, led to important insights. For instance, it was found that the $\bar{\Lambda}\Lambda$ pair is produced almost exclusively in a spin triplet state. From this, conclusions about the $\Lambda$ quark structure can be drawn: the spin of the $\Lambda$ is carried by the strange quark, while the light $u$ and $d$ quarks form a spin-0 \textit{di-quark}. Theoretical investigations based on the aforementioned quark-gluon approach \cite{quarkgluon}, kaon exchange \cite{kaonexchange} and a combined approach \cite{quarkgluonhadron}, reproduced this finding. However, no model so far describes the complete spin structure of the reaction. The models extensions into the double-strange sector \cite{XiQG,XiMEX} have not been tested due to the lack of data -- only a few bubble-chamber events exist for $\Xi^-$ and $\Xi^0$ from $\bar{p}p$ annihilations \cite{Musgrave1965}. In Ref. \cite{XiMEX}, $\bar{\Xi}^+$ emitted in the forward-direction in the center of mass frame are predicted to be in a triplet state, while backward-going $\bar{\Xi}^+$ are in a singlet state, in contrast to the $\Lambda$ case that is in a spin-triplet state irrespective of the angle \cite{PS185}. With future data from PANDA, this prediction can be tested. The hope is also that new spin structure data of $\bar{p}p \rightarrow \bar{Y}Y$ reactions will trigger the activity of the theory community and lead to a deeper understanding of strange reaction dynamics.  

\subsubsection{CP symmetry in hyperon decays}

The existence of CP violation for spinless mesons is experimentally well-established in the strange and bottom sector \cite{pdg} and recently also in the charm sector \cite{LHCbcharm}. It is also incorporated in the Standard Model, through the Cabibbo-Kobayashi-Maskawa mechanism \cite{cabibbo,kobayashi}. However, Standard Model deviation from CP symmetry would result in a matter-antimatter asymmetry of eight orders of magnitude smaller than the observed one \cite{werner}. Hence, this problem is intimately connected to the search for physics beyond the Standard Model. The spin-carrying baryons could give new insights into CP violation, since spin behaves differently from momentum under parity flip. However, the only indication of CP violation in a baryon decay, observed very recently by the LHCb collaboration \cite{LHCb}, was not confirmed in a later study with larger precision by the same experiment \cite{LHCb2}. Two-body decays of strange hyperons provide a cleaner search-ground, but require large data samples. The most precise CP test in the strange sector so far is provided by the HyperCP collaboration. The proton angular distributions from the $\Xi^- \to \Lambda \pi^-, \Lambda \to p\pi^-$ chain was studied, along with the corresponding antiproton distributions from the $\bar{\Xi}^+$ decay chain. The result was found to be consistent with CP symmetry with a precision of $10^{-4}$ \cite{hyperCP}. The most precise test for the $\Lambda$ hyperon was obtained recently by the BESIII collaboration \cite{bes3nature}. They analysed $\bar{\Lambda}\Lambda$ pair production from $J/\Psi$ using a multi-dimensional method. The good precision for a relatively modest sample size ($\approx$ 420 000 $\bar{\Lambda}\Lambda$ events) demonstrates the merits of exclusive measurements of polarised and entangled hyperon-antihyperon pairs, produced in two-body reactions. The most remarkable finding was however that the decay asymmetry parameter $\alpha_{\Lambda}$ was found to be $0.750 \pm 0.009 \pm 0.004$, \textit{i.e.} 17\% larger than the PDG world average of $0.642$ at the time \cite{pdg}. This average was calculated from measurements made in the 1960s and 1970s, based on the proton polarimeter technique \cite{cronin}. In the 2019 update of the PDG, the old measurements are discarded and instead, the BESIII value is established as the recommended one. In a re-analysis of CLAS data, the $\alpha_{\Lambda}$ was calculated to be $0.721 \pm 0.006 \pm 0.005$. This is between the old average and the new BESIII value, though much closer the to the latter \cite{claslambda}. More high-precision measurements from independent experiments will be valuable not only to establish the correct decay asymmetry, but also to understand the difference between old and new measurements.

\section{Formalism}
\label{sec:formalism}

\begin{figure}[htbp]
\begin{center}

\includegraphics[width=0.95\linewidth]{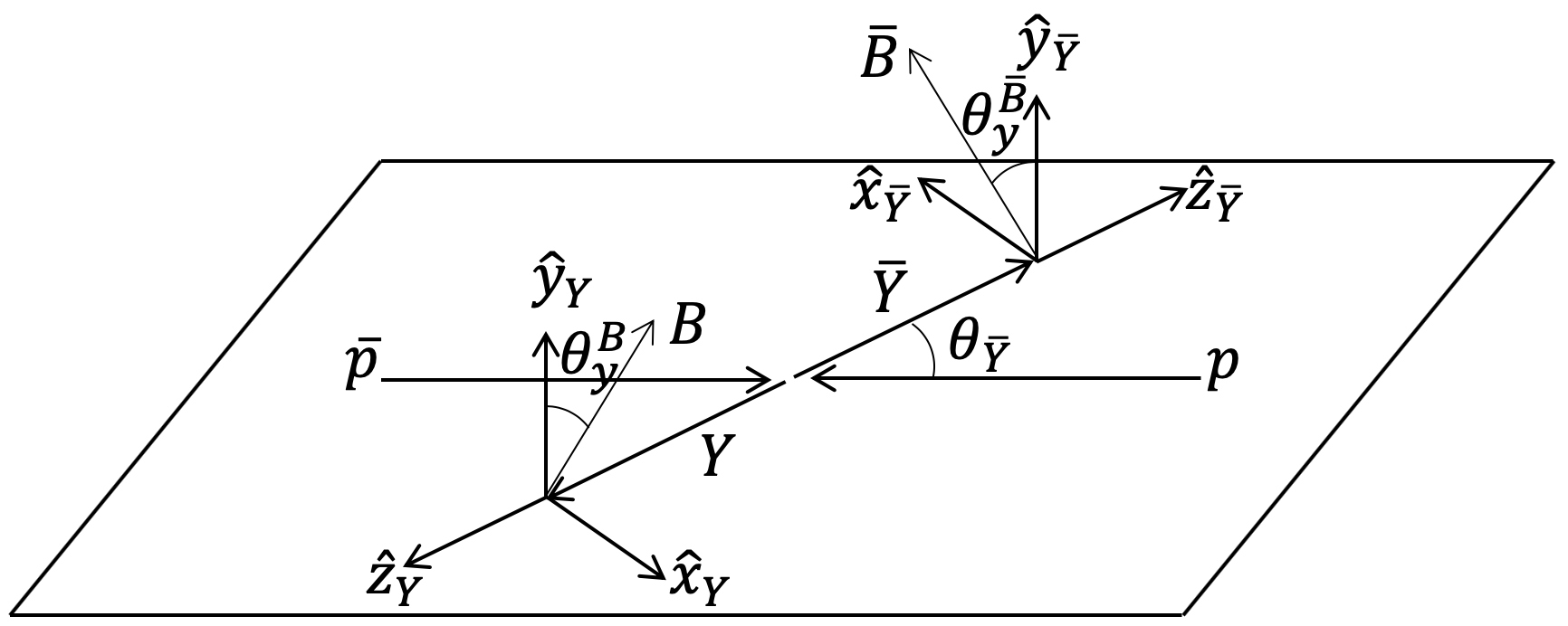}

\caption{The reference system of the $\bar{p}p \to \bar{Y}Y$ reaction.}
\label{fig:refsys}
\end{center}
\vspace{-11pt}
\end{figure}
\noindent Consider an antiproton beam impinging on a hydrogen target, producing a $\bar{Y}Y$ pair. Then the rest systems of the  outgoing hyperons can be defined as in Fig. \ref{fig:refsys}: the $\hat{y}_Y$ and $\hat{y}_{\bar{Y}}$ axes as the normal of the production plane, spanned by the incoming antiproton beam and the outgoing antihyperon in the centre of mass system of the reaction. The $\hat{z}_Y$ ($\hat{z}_{\bar{Y}}$) is defined along the direction of the outgoing hyperon (antihyperon) and the $\hat{x}_Y$ ($\hat{x}_{\bar{Y}}$) is obtained by the cross product of the $y$ and $z$ direction:  
\begin{equation}
	\hat{z}_Y = \frac{\vec{p}_{Y}}{|\vec{p}_{Y}|}, \hat{y}_Y = \frac{\vec{p}_{beam} \times \vec{p}_{Y}}{|\vec{p}_{beam} \times \vec{p}_{Y}|}, \hat{x}_Y = \hat{y}_Y \times \hat{z}_Y,
\end{equation}
where $\vec{p}_Y$ is the momentum vector of the outgoing hyperon and $\vec{p}_{beam}$ is the momentum of the initial beam.

Interference between complex production amplitudes has a polarising effect on the outgoing hyperon and antihyperon, even if the initial state is unpolarised. In PANDA, the beam and target will be unpolarised. Since the $\bar{p}p \to \bar{Y}Y$ reaction is a strong, parity-conserving process, the polarisation of the outgoing hyperon and antihyperon can only be non-zero in the direction along the normal of the production plane, \textit{i.e.} $\hat{y}_Y$($\hat{y}_{\bar{Y}})$ in Fig. \ref{fig:refsys}. In the case of a spin 1/2 hyperon $Y$ (antihyperon $\bar{Y}$) decaying into a spin 1/2 baryon $B$ (antibaryon $\bar{B}$) and a meson $M$ (antimeson $\bar{M}$), the angular distribution of the decay baryon and antibaryon can be parameterised as:
\begin{align}
\label{eq:ppbar}
    &I(\theta_Y,\theta^B,\theta^{\bar{B}}) = N[1 + \alpha\sum_i P^Y_{i}(\theta_Y)\cos\theta_{i}^{B} \nonumber\\ &+ \bar{\alpha}\sum_j P^{\bar{Y}}_{j}(\theta_Y)\cos\theta_{j}^{\bar{B}} + \alpha\bar{\alpha} \sum_{ij} C^{Y\bar{Y}}_{ij}(\theta_Y)\cos\theta_{i}^{B}\cos\theta_{j}^{\bar{B}}]
 \end{align}
where $i,j = x,y,z$ and the opening angle $\cos\theta_{i}^B$ $(\cos\theta_{j}^{\bar{B}})$ is taken between the direction of the final state baryon $B$ (antibaryon $\bar{B}$) and the axis $i$ $(j)$ in the rest system of the hyperon (antihyperon). The $P^Y_i(\theta_Y)$ denote the vector polarisation and the $C^{\bar{Y}Y}_{ij}(\theta_Y)$ the spin correlation of the antihyperon and hyperon with respect to the axes $i,j = x,y,z$. The $\theta_Y$ angle is defined in the reaction CMS system. With the unpolarised beam and target foreseen with PANDA and the reference system defined in Fig. \ref{fig:refsys}, most spin variables must be zero due to parity conservation. The only non-zero spin variables are $P^Y_y$, $P^{\bar{Y}}_y$, $C^{Y\bar{Y}}_{xz}$, $C^{Y\bar{Y}}_{zx}$, $C^{Y\bar{Y}}_{xx}$, $C^{Y\bar{Y}}_{yy}$ and $C^{Y\bar{Y}}_{zz}$ \cite{erikthesis,paschke}. Of these, only five are independent since $P^Y_y = P^{\bar{Y}}_y$ and $C^{Y\bar{Y}}_{xz} = C^{Y\bar{Y}}_{zx}$.

The angular distribution can also be expressed with a matrix formulation. Then, one first defines the 4D vectors
\begin{align}
	&k_{\bar{B}}=(1,\cos\theta_{x}^{\bar{B}},\cos\theta_{y}^{\bar{B}},\cos\theta_{z}^{\bar{B}}) \\
	&k_B=(1,\cos\theta_{x}^{B},\cos\theta_{y}^{B},\cos\theta_{z}^{B}).
\end{align}

\noindent In addition, a matrix with spin observables and decay parameters can be defined in the following way

\begin{equation}
	D_{\mu\nu}=\begin{pmatrix}
		1 & \alpha P^Y_{x} & \alpha P^Y_{y} & \alpha P^Y_{z} \\
		\bar{\alpha} P^{\bar{Y}}_{x} & \bar{\alpha}\alpha C^{\bar{Y}Y}_{xx} & \bar{\alpha}\alpha C^{\bar{Y}Y}_{xy} & \bar{\alpha}\alpha C^{\bar{Y}Y}_{xz} \\
		\bar{\alpha} P^{\bar{Y}}_{y} & \bar{\alpha}\alpha C^{\bar{Y}Y}_{yx} & \bar{\alpha}\alpha C^{\bar{Y}Y}_{yy} & \bar{\alpha}\alpha C^{\bar{Y}Y}_{yz} \\
		\bar{\alpha} P^{\bar{Y}}_{z} & \bar{\alpha}\alpha C^{\bar{Y}Y}_{zx} & \bar{\alpha}\alpha C^{\bar{Y}Y}_{zy} & \bar{\alpha}\alpha C^{\bar{Y}Y}_{zz}
	\end{pmatrix},
\end{equation}

\noindent where $\mu= 0,1,2,3$ or $0, x, y, z$ for the antihyperon and $\nu= 0,1,2,3$ or $0, x, y, z$ for the hyperon. Since parity is conserved in strong interactions, the spin observables matrix in the $\bar{p}p \to \bar{Y}Y$ reduces to 

\begin{align}
	&D_{\mu\nu}=\begin{pmatrix}
		1 & 0 & D_{02} & 0 \\
		0 & D_{11} & 0 & D_{13} \\
		D_{20} & 0 & D_{22} & 0 \\
		0 & D_{31} & 0 & D_{33}
	\end{pmatrix}\nonumber\\
	&=\begin{pmatrix}
		1 & 0 & \alpha P^Y_{y} & 0 \\
		0 & \bar{\alpha}\alpha C^{\bar{Y}Y}_{xx} & 0 & \bar{\alpha}\alpha C^{\bar{Y}Y}_{xz} \\
		\bar{\alpha} P^{\bar{Y}}_{y} & 0 & \bar{\alpha}\alpha C^{\bar{Y}Y}_{yy} & 0 \\
		0 & \bar{\alpha}\alpha C^{\bar{Y}Y}_{zx} & 0 & \bar{\alpha}\alpha C^{\bar{Y}Y}_{zz}
	\end{pmatrix}.
\end{align}

\noindent Then the angular distribution, expressed in matrix form, becomes

\begin{equation}
	I(\theta_{\bar{B}},\phi_{\bar{B}},\theta_B,\phi_B)=\frac{1}{16\pi^2}k_{\bar{B}}D_{\mu\nu}k_B^T.\label{eq:pdfspincorrmatrix}
\end{equation}

From the spin correlations, one can calculate the \textit{singlet fraction}:

\begin{equation}
	F_S = \frac{1}{4}(1+C^{\bar{Y}Y}_{xx}-C^{\bar{Y}Y}_{yy}+C^{\bar{Y}Y}_{zz}).
	\label{eq:spinfrac}
\end{equation}

\noindent In its original form, derived in Ref.\cite{durand}, it equals the expectation value of the product of the Pauli matrices \textbf{$\sigma_{\bar{Y}}$}$\cdot$\textbf{$\sigma_{Y}$} which is a number between -3 and 1. In Eq. \ref{eq:spinfrac}, it has been rewritten to stay between 0 and 1. If $F_S = 0$, all $\bar{Y}Y$ states are produced in a spin triplet state whereas $F_S = 1$ means they are all in a singlet state. If the spins are completely uncorrelated, the singlet fraction equals 0.25.

\section{Simulations of hyperon production in PANDA}
\label{sec:simulations}

In order to estimate the expected hyperon reconstruction efficiency with PANDA, and to quantify its sensitivity to spin observables, a comprehensive simulation study of two key channels has been performed. We have simulated the reactions

\begin{itemize}
    \item $\bar{p}p \rightarrow \bar{\Lambda}\Lambda, \bar{\Lambda} \to \bar{p}\pi^+, \Lambda \to p \pi^-$ at $p_{beam} = 1.64$ GeV/$c$;
    \item $\bar{p}p \rightarrow \bar{\Xi}^+\Xi^-, \bar{\Xi}^+ \to \bar{\Lambda}\pi^+, \bar{\Lambda} \to \bar{p}\pi^+, \Xi^- \to \Lambda \pi^-, \\\Lambda \to p \pi^-$ at $p_{beam} = 4.6$ GeV/$c$ and $p_{beam} = 7.0$ GeV/$c$.
\end{itemize}

\noindent The channels have been chosen since the most prominent decay channel in each case leaves only charged particles in the final state. Even though PANDA will be capable of measuring both neutral and charged final states, charged final states are more straight-forward and can be reconstructed with better resolution. Hence, channels with charged final state particles serve as a first benchmark in the overall PANDA hyperon performance check-list. The beam momentum $p_{beam} = 1.64$ GeV/$c$ for the $\bar{p}p \rightarrow \bar{\Lambda}\Lambda$ was chosen since it coincides with a large data set collected by the PS185 experiment \cite{PS185164}. The PS185 measurement of the cross section, the angular distribution and the spin observables can be compared with new data from one of the first foreseen data taking periods with PANDA. This allows for a systematic comparison between PANDA and a completely independent previous experiment, hence providing important guidance for all future hyperon studies with PANDA.

Neither differential cross sections nor spin observables of the $\bar{p}p \rightarrow \bar{\Xi}^+\Xi^-$ reaction have been studied before, and the goal of PANDA is therefore to contribute with completely new insights. For the double-strange $\Xi^-$, the chosen beam momenta coincide with the hyperon spectroscopy campaign (4.6 GeV/$c$, see Ref. \cite{jennynstar}) and the $X(3872)$ line-shape campaign (7 GeV/$c$, see Ref. \cite{xscan}). 

Since hyperons have relatively long life-time ($10^{-10}$ s), they travel a measurable distance before decaying. This makes the track reconstruction a challenging task \cite{walter,michael,jenny} since most standard algorithms assume that all tracks originate in the beam-target interaction point. The simulation study presented here is focused on Phase One of PANDA. A realistic PandaROOT implementation of the Phase One conditions was used \cite{phaseone}. Some simplifications were made due to limitations in the current version of the simulation software:
\begin{itemize}
    \item The general track reconstruction algorithms that can handle tracks originating far from the interaction point, are still under development and have not yet been deployed as a part of the standard PandaROOT package. Therefore, an ideal pattern recognition algorithm has been used, combined with some additional criteria on the number of hits per track in order to mimic realistic conditions.
    
    \item The particle identification method is not yet stabilised and therefore, ideal PID matching was used. It was shown in Ref. \cite{erikthesis} that the event selection of non-strange final state particles (\textit{i.e.} decay products of $\Lambda$ and $\Xi^-$) can be performed without PID thanks to the distinct topology of hyperon events. Ideal PID however considerably reduces the run-time due to combinatorics, and was therefore used in the reconstruction. 
    
\end{itemize}  
\noindent In order to mimic the conditions of real pattern recognition, each track in the target spectrometer was required to contain either 4 hits in the MVD, or in total 6 hits in the MVD + STT + GEM. Tracks in the forward spectrometer are required to contain at least 6 hits in the FTS. \footnote{The minimum number of hits to fit a circle is three, but additional hits are needed in order to verify that the hits come from a real track and to resolve ambiguities.}
\subsection{Signal sample}
In total, $10^6$ events were generated for $\bar{\Lambda}\Lambda$ and $\bar{\Xi}^+\Xi^-$ \cite{walter} using the EvtGen generator \cite{evtgen}. 
The $\bar{\Lambda}\Lambda$ sample was weighted using a parameterization of data from PS185, that revealed a strongly forward-peaking $\bar{\Lambda}$ distribution in the CMS system of the reaction \cite{PS185,PS185164}. The $\bar{\Xi}^+\Xi^-$ final state has never been studied and was therefore generated both with an isotropic angular distribution and with a forward-peaking distribution, using a parameterisation from $\bar{p}p \to \bar{\Sigma}^0\Lambda$ production in Ref. \cite{sigma6}. In this way, we can estimate the sensitivity of the reconstruction efficiency to the underlying angular distribution. This is particularly important in a fixed-target, two-spectrometer experiment like PANDA.

In hyperon-antihyperon pair production in $\bar{p}p$ annihilations, the $\theta_Y$ dependence of the spin observables is not straight-forward to parameterize in contrast to the $e^+e^-$ case \cite{goran}, since more than two production amplitudes can contribute. However, the spin observables must satify some constraints: i) they need to stay within the interval $[-1,1]$ and ii) they need to go to zero at extreme angles, \textit{i.e.} $\theta_Y = 0^{\mathrm{o}}$ and $\theta_Y = 180^{\mathrm{o}}$. The latter is because at these angles, the incoming beam is either parallel or anti-parallel to the outgoing antihyperon. Their cross product, giving the direction of the normal of the production plane, is thus not defined.

The data in this study were weighted according to 

\begin{equation}
P^Y_y(\theta_Y) = \sin2\theta_Y  
\end{equation}

and 

\begin{equation}
C^{\bar{Y}Y}_{ij}(\theta_Y) = \sin\theta_Y. 
\end{equation}

\noindent since they satisfy the constraints and since this gives a polarisation with a shape that resembles real data \cite{PS185,PS185164}. 

\subsection{Background samples}

\subsubsection{Background to $\bar{p}p \to \bar{\Lambda}\Lambda$}

Generic hadronic background is denoted $\bar{p}p\to X$, where $X$ refers to \textit{any} allowed final state. Such processes are simulated using the Dual Parton Model (DPM) generator \cite{pbarx}, based on a phenomenological model that incorporates Regge theory, topological expansions of QCD, and concepts from the parton model. From this, energy dependencies are obtained of hadron-hadron cross sections with large number of particles with small transverse momenta with respect to the collision axis. Since the strong coupling is large for such processes, perturbation theory is not applicable. Instead, a topological expansion is employed, where the number of colours $N_c$ or flavours $N_f$ is the expansion parameter.

The total cross section of all $\bar{p}p\to X$ processes is around three orders of magnitude larger than that of the $\bar{p}p\to\bar{\Lambda}\Lambda$. The expected ratio of produced generic background and signal events can be estimated from simulations using:
\begin{equation}
	\frac{N_{X}}{N_{\mathrm{signal}}} = \frac{\sigma(\bar{p}p \to X)}{\sigma (\bar{p}p \to \bar{\Lambda}\Lambda)\mathrm{BR}(\Lambda \to p \pi)^2},
\end{equation}
where $\sigma (\bar{p}p \to \bar{\Lambda}\Lambda) = 64.1 \pm 0.4 \pm 1.6$ $\mu$b is the production cross section \cite{PS185164_1}, $\mathrm{BR}(\Lambda \to p \pi) = 63.9 \pm 0.5 \%$ is the branching ratio \cite{pdg} and $\sigma(\bar{p}p \to X) = 96 \pm 3$ mb \cite{CERNHERA}.

 In order to estimate the expected background contamination, one should ideally produce a realistic amount of background events with respect to the signal. This would however require $3.6\times 10^3$ DPM events per signal event which in turn implies more than $10^9$ DPM events. Since this would take an unreasonably long time to simulate, a smaller background sample has been generated and then weighted to give the expected signal-to-background ratio.

Among the numerous channels included in the generic background, the non-resonant  $\bar{p}p \to \bar{p}p\pi^+\pi^-$ process is particularly important. This is because it has the same final state particles as the process of interest $i.e.$ $\bar{p}p \to \bar{\Lambda}\Lambda\to \bar{p}p\pi^+\pi^-$ and a cross section that is of the same order of magnitude as the signal process \cite{CERNHERA,EASTMAN197329,LYS1973610}. Though included in the DPM generator, its cross section has not been tuned to real data. Therefore, this reaction has been simulated separately. The number of simulated events, the cross sections and the weights when calculating signal-to-background ratios, are given in Table \ref{tab:samplesize1}

\begin{table}[ht]
		\centering
		\begin{tabular}{c|c|c|c}
		Channel & $\bar{\Lambda}\Lambda$ & $\bar{p}p\pi^+\pi^-$ & DPM \\ \hline
		Sample & $9.75\cdot10^5$ & $9.74\cdot10^5$ & $9.07\cdot10^6$\\
		Cross section [$\mu$b] & 64.1 \cite{PS185164_1} & 15.4 & 96 000\\
		Weight & 1.00 & 0.590 & 395 \\	
		\hline
		\end{tabular}
		\caption{Sample sizes, cross sections and weights for the simulation study at 1.64 GeV/$c$. The non-resonant cross section has been calculated from the average of Refs. \cite{CERNHERA,EASTMAN197329,LYS1973610}}
		\label{tab:samplesize1}
	\end{table}

\subsubsection{Background to $\bar{p}p \to \bar{\Xi}^+\Xi^-$}

Also in this case, generic $\bar{p}p\to X$ processes are studied with the DPM generator to understand the background. 

The expected production ratio of generic background and signal is given by

\begin{equation}
	\frac{N_{X}}{N_{\mathrm{signal}}} = \frac{\sigma(\bar{p}p \to X)}{\sigma (\bar{p}p \to \bar{\Xi}^+\Xi^-)\mathrm{BR}(\Xi\to\Lambda\pi)^2 \mathrm{BR}(\Lambda\to p\pi)^2}. \label{xsecratio}
\end{equation}

The cross sections $\sigma(\bar{p}p \to X)$ at the beam momenta 7.3 GeV/$c$ (the tabulated value closest to 7.0 GeV/$c$) and 4.6 GeV/$c$ are $58.3 \pm 1.3$ mb at $p_{\mathrm{beam}} = 7.3$ GeV/c \cite{dpm7} and $68.8 \pm 0.8$ mb \cite{dpm46}, respectively. From Eq. (\ref{xsecratio}), we see that for each simulated signal event, at least $4.76\cdot10^5$ DPM events must be simulated to obtain the correct signal-to-background ratio. This would be even more computationally demanding than in the case of $\bar{p}p \to \bar{\Lambda}\Lambda$. The weighting method presented in the previous section can be applied, but the weights need to be about two orders of magnitude larger. This means that if very few DPM pass the selection criteria, then the signal-to-background ratio becomes very sensitive to fluctuations. Therefore, the most important background channels are considered separately. These are identified based on their final state particles, vertex topology and invariant masses of particle combinations, and found to be $\bar{p}p\to\bar{\Sigma}^*(1385)^+ \Sigma^*(1385)^-$, $\bar{p}p \to\bar{\Lambda}\Lambda\pi^+\pi^-$ and $\bar{p}p \to \bar{p}p 2 \pi^+ 2 \pi^-$. Events from these channels are removed from the DPM sample at the analysis stage, to avoid double-counting of background. Out of the $9.80\cdot10^7$ simulated DPM events at $p_{\mathrm{beam}} = 7.0$ GeV/c, $\sim 7\cdot 10^4$ events were removed. For the DPM sample at $p_{\mathrm{beam}} = 4.6$ GeV/c, $\sim10
^4$ were removed from the $9.8\cdot10^7$ generated events. The simulated samples, cross sections and weights are summarised in Table \ref{tab:samplesize2}.

\begin{table*}[ht]
		\centering
		\begin{tabular}{c|c|c|c|c|c}
		Channel at 7 GeV/$c$ & $\bar{\Xi}^+\Xi^-$ & $\bar{\Sigma}^*(1385)^+ \Sigma^*(1385)^-$ & $\bar{\Lambda}\Lambda \pi^+\pi^-$ & $\bar{p}p 2 \pi^+ 2 \pi^-$ & DPM \\ \hline
		Sample & $8.54\cdot10^5$ & $9.87\cdot10^6$ & $9.85\cdot10^6$ & $9.78\cdot10^6$ & $9.73\cdot10^7$\\
		$\sigma_{\mathrm{eff}}$ [$\mu$b] & 0.123 & 1.39 & 24.1 & 390 & $5.83\cdot10^4$\\	
		Weight factor & 1.00 & 0.98 & 17.1 & 278 & $4.18\cdot10^3$ \\	
		\hline
		Channel at 4.6 GeV/$c$ & $\bar{\Xi}^+\Xi^-$ & $\bar{\Sigma}^*(1385)^+ \Sigma^*(1385)^-$ & $\bar{\Lambda}\Lambda \pi^+\pi^-$ & $\bar{p}p 2 \pi^+ 2 \pi^-$ & DPM \\ \hline
		Sample & $8.80\cdot10^5$ & $9.86\cdot10^6$ & $9.88\cdot10^6$ & $9.80\cdot10^6$ & $9.82\cdot10^7$\\
		$\sigma_{\mathrm{eff}}$ [$\mu$b] & 0.41 & 1.39 & 14.7 & 143 & $6.88\cdot10^4$\\	
		Weight factor & 1.00 & 0.304 & 3.21 & 31.4 & $1.51\cdot10^3$ \\ \hline
		\end{tabular}
		\caption{Sample sizes, cross sections and weights for the simulation study at $p_{\mathrm{beam}} = 7$ GeV/c and $p_{\mathrm{beam}} = 4.6$ GeV/c. The $\bar{\Sigma}^*(1385)\Sigma^*(1385)$ cross section is obtained from Ref. \cite{llbarpippim46}, and the $\bar{\Lambda}\Lambda\pi^+\pi^-$ cross sections from Ref. \cite{llbarpippim7} and \cite{llbarpippim46} at 7 GeV/$c$ and 4.6 GeV/$c$, respectively. The non-resonant $\bar{p}p 2\pi^+2\pi^-$ cross section is obtained from Ref. \cite{nonres7} at 7 GeV/$c$ and the average of Refs. \cite{nonres46} and \cite{nonres462} at 4.6 GeV/$c$.}
		\label{tab:samplesize2}
	\end{table*}

\subsection{Event selection} 

\begin{figure*}[ht]
\begin{minipage}{0.4\textwidth}
	\centering
	\includegraphics[width = \textwidth]{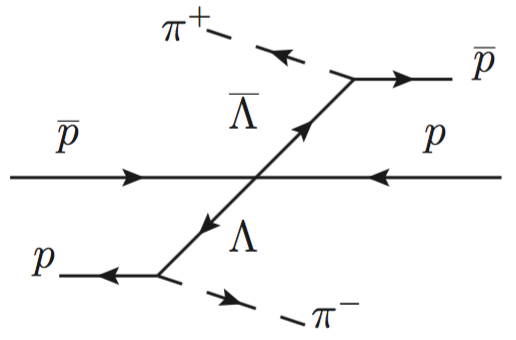}
	\end{minipage}
	\begin{minipage}{0.6\textwidth}
	\includegraphics[width =\textwidth]{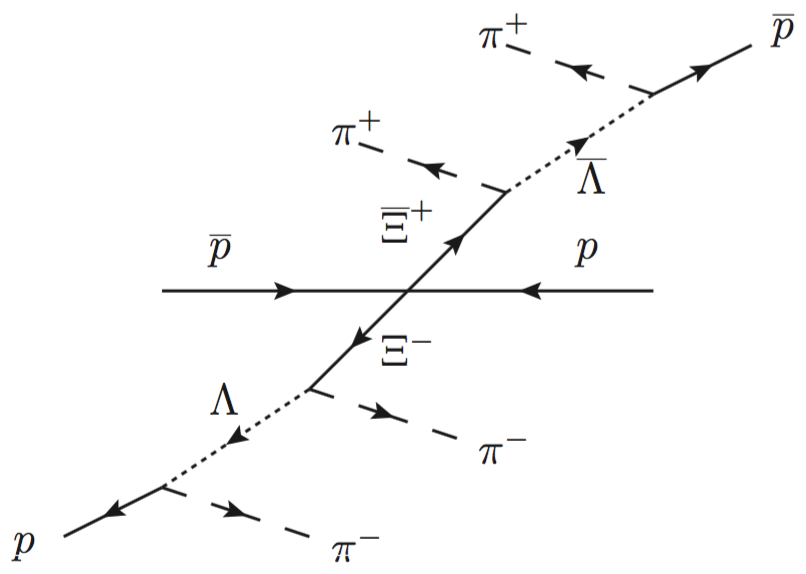}
		\end{minipage}
	\caption{Signal event topology of $\bar{p}p \to \bar{\Lambda}\Lambda, \bar{\Lambda} \to \bar{p}\pi^+, \Lambda \to p\pi^-$ (left) and $\bar{p}p \to \bar{\Xi}^+\Xi^-, \bar{\Xi}^+ \to \bar{\Lambda}, \bar{\Lambda} \to \bar{p}\pi^+, \Xi^- \to \Lambda\pi^-, \Lambda \to p\pi^-$ (right).}
	\label{fig:topology}
	\end{figure*}

Reactions involving hyperons have a very distinct topology, since the long-lived hyperons decay a measurable distance from the point of production. The topology of each reaction and subsequent decay chain studied in this work, are shown in Fig. \ref{fig:topology}. This can be exploited in the event selection procedure, as outlined in this chapter.

The event selection is performed in two stages: a pre-selection and a fine selection. The pre-selection comprises a set of basic topological criteria, that reduces the total simulated sample and hence the analysis run-time. The fine selection involves kinematic fits and fine-tuned mass windows.

\subsubsection{The $\bar{p}p \to \bar{\Lambda}\Lambda$ reaction}
\label{sec:lambda}

\noindent The pre-selection criteria for this reaction are:
\begin{itemize}
    \item Each event must contain at least one each of the following: $p$, $\bar{p}$, $\pi^+$ and $\pi^-$.
    \item Each event contains at least one $p\pi^-$ and one $\bar{p}\pi^+$ combination that can be successfully fitted to one common vertex, with a probability of $> 0.01$. If more than one such $\Lambda$ or $\bar{\Lambda}$ candidate exist in one event (occurs in 6\% of the cases for $\Lambda$ and 2\% of the cases for $\bar{\Lambda}$), then the one with the smallest $\chi^2$ is kept for further analysis.
    \item Each event must contain at least one $p\pi^-$ and one $\bar{p}\pi^+$ combination with an invariant mass that satisfies $|m_{\Lambda}-m(p\pi)| < 0.3$ GeV/$c^2$. This mass window is very wide and is further tightened in the final selection.
    \item The four-vectors of the $\Lambda$ and the $\bar{\Lambda}$ candidate can be fitted successfully to the initial beam momentum, with a \textit{four-constraints} ($4C$) fit.
\end{itemize}

\noindent The event filtering is further improved by the fine selection. The criteria of the fine selection were tuned and optimised using as a figure of merit the significance, \textit{i.e.} $S/\sqrt{S+B}$, where $S$ refers to the number of signal events and $B$ the number of generic hadronic events generated by DPM. The criteria are the following:

\begin{itemize}
    \item The $\chi^2$ of the $4C$ fit is required to be $< 100$. 
    \item The total distance $z_{tot}$ from the interaction point in the beam direction of the $\Lambda$ and $\bar{\Lambda}$ candidate must fulfill $z_{tot} = |z_{\Lambda} + z_{\bar{\Lambda}}| > 2$ cm.
    \item The invariant mass of the $p\pi^-$ and $\bar{p}\pi^+$ system must not differ from the PDG $\Lambda$ mass by more than 5$\sigma$, where $\sigma$ is the width of a Gaussian fitted to the invariant mass peak. 
\end{itemize}
\noindent The mass resolution differs between $\Lambda$ ($\sigma = 2.864\cdot10^{-3}$ GeV/$c^2$) and $\bar{\Lambda}$ ($\sigma = 2.980\cdot10^{-3}$ GeV/$c^2$). This is because the decay products from $\Lambda$ are primarily emitted in the acceptance of the MVD and STT, while the decay products of $\bar{\Lambda}$ to a larger extent hit the FTS. The $\bar{p}\pi^-$ invariant mass for signal and background are shown in the left panel of Figure \ref{fig:invmass}.

The reconstruction efficiency of the signal reaction and the most important background sources for the different selection criteria are given in Table \ref{tab:effllbar}. In addition, the number of expected background events for a given number of signal events has been calculated taking the cross sections into account. It is clear that background can be very successfully suppressed. A signal-to-background ratio of $S/B \approx 106$ is obtained. We conclude that the PANDA detector will be capable of collecting very clean $\bar{\Lambda}\Lambda$ samples, which is essential when extracting spin observables. 
\begin{table}[ht]
	\centering
	\begin{tabular}{c|c|c|c}
	 Channel & $\bar{\Lambda}\Lambda$ & $\bar{p}p\pi^+\pi^-$ & DPM \\ \hline
	 Generated & $9.75\cdot10^5$ & $9.74\cdot10^5$ & $9.07\cdot10^6$ \\
	Preselection & $2.129\cdot10^5$ & 292700 & 651 \\
	$\chi^2 < 100$ & $1.879\cdot10^5$ & $249190$ & $136$ \\
	$\Delta m < 5\sigma$ & $1.685\cdot10^5$ & $29180$ & $3$ \\
	$z_{\bar{\Lambda}} + z_{\Lambda} > 2$ cm & $1.572\cdot10^5$ & 470 & 2 \\ 
	Eff. (\%) & $16.0\pm0.4$ & $0.05$ & $2.2\cdot10^{-7}$\\
	\hline
		$N_{exp}$ & $1.572\cdot10^5$ & 277 & 790 \\
		\hline
		\end{tabular}
	\caption{Reconstruction efficiency after the final selection for signal events as well as non-resonant and generic hadronic background. In the bottom row, the proportion of expected events are shown. These numbers were calculated by applying the weights in Table \ref{tab:samplesize1}. \label{tab:effllbar}}
\end{table}

\subsubsection{The $\bar{p}p \to \bar{\Xi}^+\Xi^-$ reaction}
\label{sec:xi}

The $\bar{p}p \to \bar{\Xi}^+\Xi^-$ reaction is more complicated than the $\bar{p}p \to \bar{\Lambda}\Lambda$ reaction since i) there are more particles in the final state ii) there are several identical particles in the final state and iii) each event contains four displaced decay vertices instead of two. In addition, the cross section is smaller and at the larger beam momenta necessary for $\Xi$ studies, the cross section of background channels are larger. Hence, the selection procedure is by necessity a bit more involved. In the following, we summarise the pre-selection criteria. For simplicity, the charge conjugated mode is implied unless otherwise stated.

\noindent\textbf{Final State Reconstruction and Combinatorics}

\noindent The first step is to combine the final state particles into $\Lambda$ and $\Xi$ candidates:
\begin{itemize}
	\item All possible $p \pi^-$ combinations are combined to form $\Lambda$ candidates.
	\item All combinations fulfilling $|m_{\Lambda} - M(p\pi^-) | < 0.05$ GeV/c$^2$ are accepted and stored for further analysis.
	\item All possible $\Lambda \pi^-$ combinations are combined to form $\Xi^-$ candidates.
	\item All combinations fulfilling $|m_{\Xi} - M(p\pi^-\pi^-) | < 0.05$ MeV/c$^2$ are accepted and stored for further analysis. This mass window is very wide and is tightened in the final selection.
\end{itemize}
\textbf{Fit of the $\mathbf{\Xi^-\to \Lambda \pi^-, \Lambda \to p \pi^-}$ decay chain}

\noindent The second step is to exploit the distinct topology of the $\bar{p}p \to \bar{\Xi}^+\Xi^-, \bar{\Xi}^+ \to \bar{\Lambda}\pi^+, \bar{\Lambda} \to \bar{p}\pi^+, \Xi^-\to \Lambda \pi^-, \Lambda \to p \pi^-$ process, that imposes many constraints. Therefore, all $\Xi^-$ candidates from the previous step are fitted under the $\Xi^-\to \Lambda \pi^-, \Lambda \to p \pi^-$ hypothesis where the $\Lambda$ mass is constrained to its PDG value. This is achieved using the Decay Chain Fitting package \cite{HULSBERGEN2005566}, designed to perform kinematic fitting on a sequence of decays with at least one over-constraint. Taking all constraints and unknown parameters in the fit into account, results in three effective degrees of freedom. The advantage of this approach compared to multiple sequential fits, is that all constraints in a reaction are taken into account simultaneously, on an equal basis. This feature is not available in conventional fitters. In our case, the procedure is the following:

\begin{itemize}
	\item The decay chain $\Xi^- \to \Lambda \pi^-, \Lambda \to p \pi^-$ is fitted. The constraints are provided by momentum conservation, the two vertex positions and the $\Lambda$ mass. All momentum components of all particles are modified in the fit.
	\item Candidates with a fit probability $< 0.01$ are rejected.
\end{itemize}
\textbf{Reconstructing the $\bar{p}p$ system} \\

	\begin{figure*}[ht]
		        \includegraphics[width=0.33\linewidth]{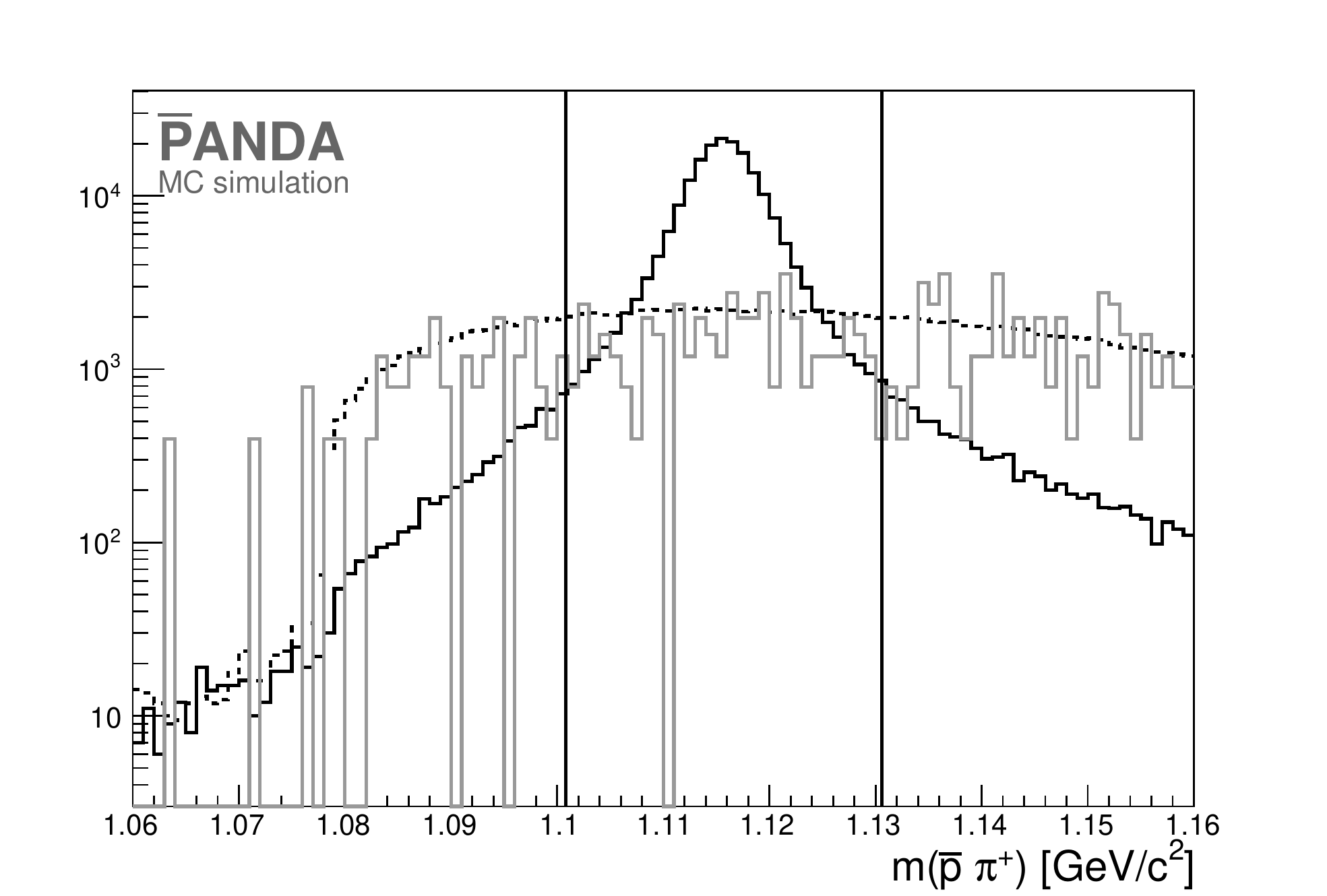}
		        \includegraphics[width=0.33\linewidth]{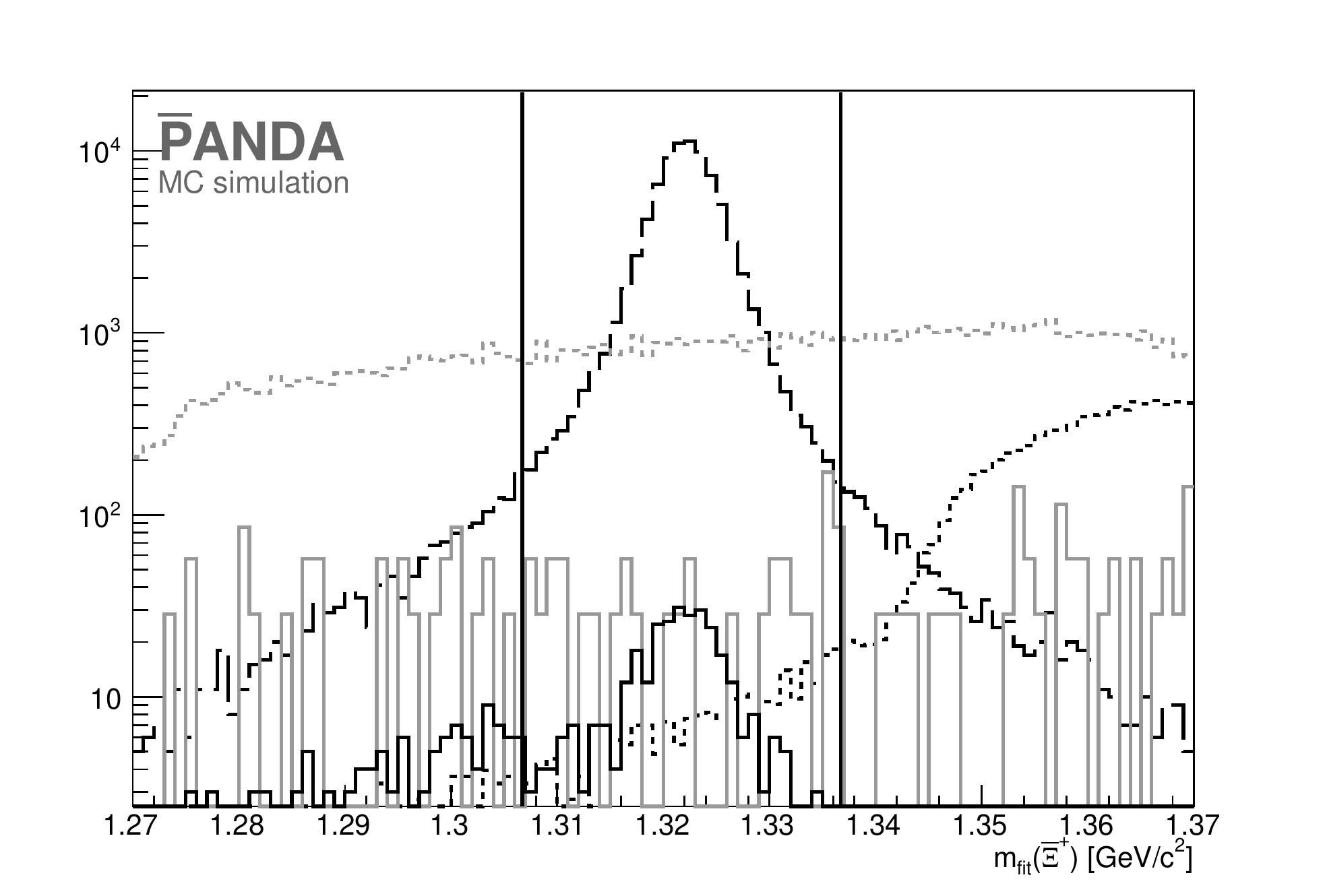}
		        \includegraphics[width=0.33\linewidth]{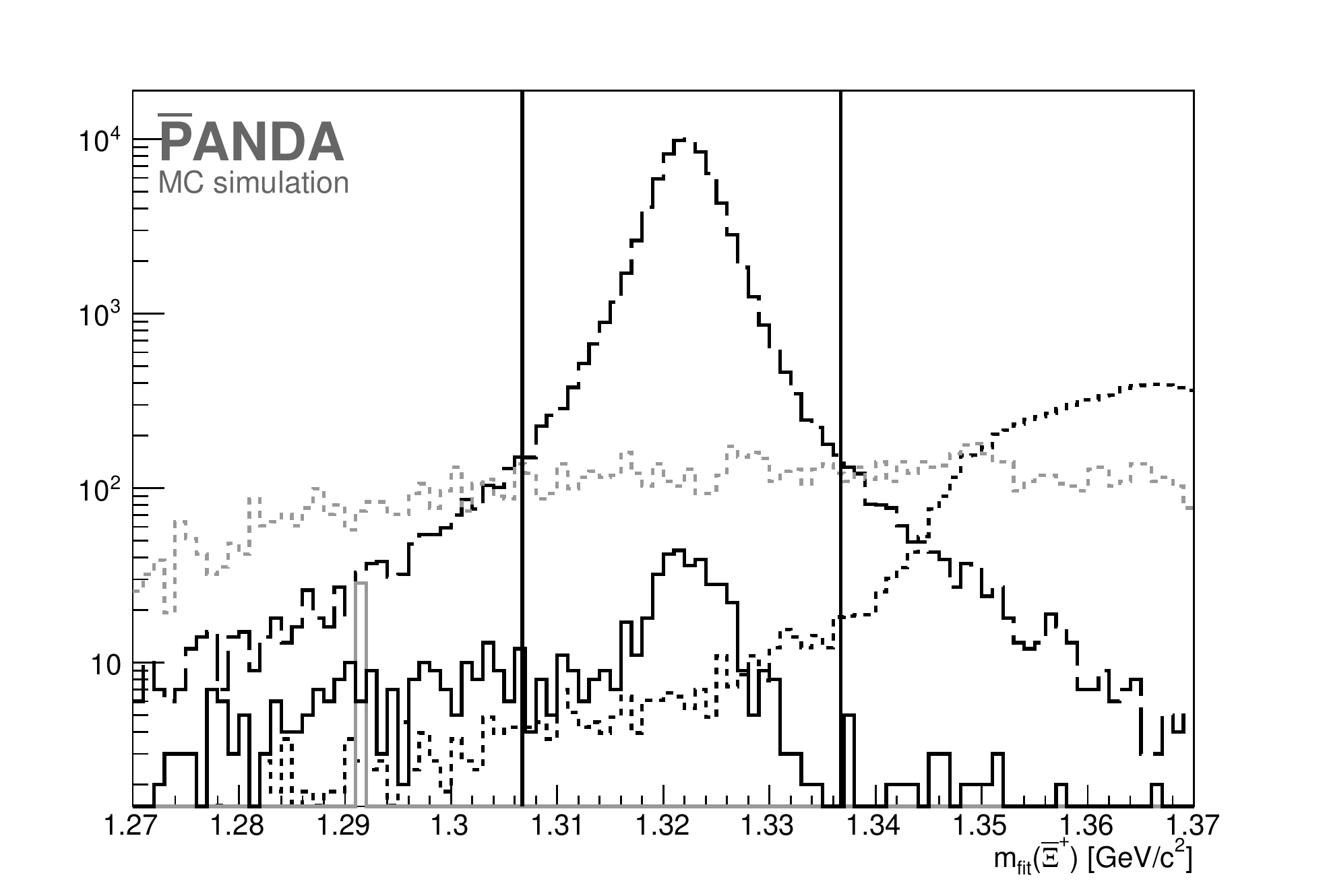}
		        
			\caption{Invariant mass distributions of signal and background samples in the final selection state. Left: The $\bar{p}\pi^+$ invariant mass at $p_{\mathrm{beam}}$ = 1.64 GeV/c for the $\bar{p}p \to \bar{\Lambda}\Lambda$ reaction (black), non-resonant $\bar{p}p \to \bar{p}p\pi^+\pi^-$ (dotted) and DPM (grey). Middle: The $\bar{p}\pi^+\pi^+$ invariant mass at $p_{\mathrm{beam}}$ = 4.6 GeV/c for the $\bar{p}p \to \bar{\Xi}^+\Xi^-$ reaction (black dashed), the $\bar{p}p \to \bar{\Sigma(1385)}^+\Sigma(1385)^-$ (black dotted), $\bar{p}p \to \bar{\Lambda}\Lambda\pi^+\pi^-$ (grey dotted), $\bar{p}p \to \bar{p}p2\pi^+2\pi^-$  (grey solid) and combinatorial (black solid). Right: Same as in the middle panel but at $p_{\mathrm{beam}}$ = 7.0 GeV/c. The vertical lines mark the final selection mass window. All distributions are normalised to previously measured cross sections.}
			\label{fig:invmass}
		\end{figure*}

\noindent The decay chain fitter results in a list of $\Xi^-$ and $\bar{\Xi}^+$ candidates in each event. The next step is to combine these candidates and test the hypothesis that they come from a common production vertex, and fulfill the kinematics of the initial $\bar{p}p$ system.
\begin{itemize}
	\item All possible $\bar{\Xi}^+\Xi^-$ combinations form a hypothetical $\bar{p}p$ system.
	\item A vertex fit of $\bar{\Xi}^+\Xi^-$ pairs is performed to reconstruct the interaction point.
	\item Candidates with a fit probability $ < 0.01$ are rejected.
	\item Candidates where the opening angle of the $\bar{\Xi}^+\Xi^-$ pair is $> 3$ rad in the CMS system are selected for further analysis.This is because in the two-body reaction of interest, the $\bar{\Xi}^+$ and the $\Xi^-$ are emitted back to back.
	\item Events where $\Lambda$ and $\Xi^-$ candidates satisfy $\Delta z = z(\Lambda) - z(\Xi) > 0$ cm are selected, where $z(Y)$ is the $z$-position of the hyperon decay vertex.
	\item A kinematic fit of $\bar{\Xi}^+\Xi^-$ pairs is performed, where energy and momentum are constrained to the initial system. 
	\item In case there is more than one $\bar{\Xi}^+\Xi^-$ combination in an event, fulfilling all previous criteria, the candidate with the smallest $\chi^2$ value from the kinematic fit is chosen for further analysis.
\end{itemize}

\noindent In the fine selection, additional criteria are applied after careful studies of the significance $S/\sqrt{S+B}$:
\begin{itemize}
	\item Combinations of $\bar{p}\pi^+\pi^+$ must fulfill \\ $|m_{fit}(\bar{p}\pi^+\pi^+) - m_{PDG}(\Xi^-)| < 5\cdot0.003 $ GeV/c$^2$, where 0.003 GeV/c$^2$ is the $\sigma$ width of the broader Gaussian component of the curve fitted to the data in the peak region.
	\item Combinations of $p\pi^-\pi^-$ must fulfill \\$| m_{fit}(p\pi^-\pi^-) - m_{PDG}(\Xi^-)| < 5\cdot0.003 $ GeV/c$^2$, where 0.003 GeV/c$^2$ is the $\sigma$ width of the broader Gaussian component of the curve fitted to the data in the peak region.
	\item The total distance in the beam direction from the reconstructed interaction point (IP) in an event must satisfy $(z_{fit}(\bar{\Xi}^+) - z_{fit}(IP)) + (z_{fit}(\Xi^-) - z_{fit}(IP)) > 3$ cm.
\end{itemize}

\noindent Invariant mass plots of the $\bar{p}\pi^+\pi^-$ system for signal and various background channels are shown in the middle (at $p_{\mathrm{beam}}$ = 4.6 GeV/c) and right ($p_{\mathrm{beam}}$ = 7.0 GeV/c) panel of Figure \ref{fig:invmass}. The resulting reconstruction efficiency for each criterion, or set of criteria, are shown in Table \ref{tab:effxixibar}. The proportion of expected events, calculated from the cross sections, are also given. No non-resonant nor any generic background events satisfy the selection criteria. Therefore, the Poisson upper limit of 2.3 events has been used to estimate the number of background events at a confidence level of 90\%.

\begin{table*}[ht]
		\centering
		\resizebox{1.\textwidth}{!}{%
		\begin{tabular}{c|c|c|c|c|c}
			$p_{beam}$ = 7.0 GeV/$c$ & $\bar{\Xi}^+\Xi^-$ & $\bar{\Sigma}(1385)^+ \Sigma(1385)^-$ & $\bar{\Lambda}\Lambda\pi^+\pi^-$ & $\bar{p}p 2 \pi^+ 2 \pi^-$ & DPM \\ \hline
Generated & $8.54\cdot10^5$ & $9.87\cdot10^6$ & $9.85\cdot10^6$ & $9.78\cdot10^6$ & $9.73\cdot10^7$\\ 
			Pre-selection & $7.83\cdot10^4$ & $3.45\cdot10^4$ & $3.51\cdot10^3$ & $1$ & $100$  \\
			Mass cut & $7.27\cdot10^4$ & $23$ & $379$ & $<2.3$ & $7.0$  \\
			$\Delta d>3$ & $6.76\cdot10^4$ & $3.0$ & $14$ & $<2.3$ & $<2.3$  \\ 
			Efficiency \% & $7.95\pm0.03$ & $(3.0\pm0.2)\cdot10^{-5}$ & $(1.4\pm0.4)\cdot10^{-4}$ & $<2.3\cdot 10^{-5}$ & $<2.3\cdot 10^{-6}$\\ \hline
			$N_{exp}$ weighted & $6.76\cdot10^4$ & $2.9$ & $239$ & $<640$ & $<9.61\cdot10^3$ \\
			\hline 
			$p_{beam}$ = 4.6 GeV/$c$ &  &  &  &  &  \\ \hline
			Generated & $8.80\cdot10^5$ & $9.86\cdot10^6$ & $9.88\cdot10^6$ & $9.80\cdot10^6$ & $9.82\cdot10^7$\\ 
			Pre-selection & $8.65\cdot10^4$ & $3.29\cdot10^4$ & $2.61\cdot10^4$ & $105$ & $44$ \\ 
			Mass cut & $8.06\cdot10^4$ & $21$ & $2.49\cdot10^3$ & $13$ & $6.0$ \\ 
			$\Delta d >3$ & $7.23\cdot10^4$ & $1.0$ & $39$ & $<2.3$ & $<2.3$  \\ 
			Efficiency (\%) & $8.22\pm0.03$ & $(1.0\pm1.0)\cdot10^{-5}$ & $(4.0\pm0.6)\cdot10^{-4}$ & $<2.3\cdot10^{-5}$ & $<2.3\cdot10^{-6}$\\ \hline
			$N_{exp}$ weighted & $7.23\cdot10^4$ & $0.30$ & $125$ & $<72$ & $<3.47\cdot10^3$	\\
			\hline
		\end{tabular}}%
		\caption{Reconstruction efficiency after the final selection for signal events as well as non-resonant and generic hadronic background. $N_{exp}$ is the expected proportion of events, applying weights in Table \ref{tab:samplesize2}. The Poisson upper limits are given at a 90\% confidence level. \label{tab:effxixibar}}
		
	\end{table*}

\section{Parameter estimation}
\label{sec:paramest}

To estimate the physics parameters $\alpha$, $\bar{\alpha}$, $P^Y_y$, $P^{\bar{Y}}_y$, $C^{Y\bar{Y}}_{xz}$, $C^{Y\bar{Y}}_{xx}$, $C^{Y\bar{Y}}_{yy}$ and $C^{Y\bar{Y}}_{zz}$ from the measured quantities, \textit{i.e.} the hyperon scattering angle and the baryon and antibaryon decay angles, methods like Maximum Log Likelihood or the Method of Moments can be used. In the very first phase of data taking with PANDA, the samples will be relatively modest and the measurements will be focused on the production related parameters, \textit{i.e.} the polarisation and the spin correlations. These can be obtained for any given beam momentum and scattering angle by fixing $\alpha$ and $\bar{\alpha}$ to the already measured value of $\alpha$ \cite{pdg}, assuming CP symmetry \textit{i.e.} $\alpha$ = $-\bar{\alpha}$. 

In this study, the Method of Moments has been chosen as parameter estimation method, due to its computational simplicity.
At a given antihyperon scattering angle $\theta_{\bar{Y}}$, it can be shown \cite{erikthesis} that the first moment of $\cos\theta_y^B$ is proportional to the polarisation at this angle:
\begin{align}
    <\cos\theta_y^B>_{\theta_{\bar{Y}}}=&\frac{\int I(\theta_y^B,\theta_y^{\bar{B}})_{\theta_{\bar{Y}}}\cos\theta_y^B d\Omega_B d\Omega_{\bar{B}}}{\int I(\theta_y^B,\theta_y^{\bar{B}})_{\theta_{\bar{Y}}} d\Omega_B d\Omega_{\bar{B}}}\\
    & = \frac{\alpha P^Y_{y,\theta_{\bar{Y}}}}{3}
\end{align}
\noindent Hence, the polarisation can be calculated from the moment
\begin{equation}
    P_{y,\theta_{\bar{Y}}}^{Y/\bar{Y}}=\frac{3<\cos\theta_y^{B/\bar{B}}>_{\theta_{\bar{Y}}}}{\alpha}
    \label{eq:pol}
\end{equation}
\noindent where the estimator of the moment is the arithmetic mean of $\cos\theta_y^{B/\bar{B}}$ obtained from a sample of $N$ events:

\begin{equation}
	<\reallywidehat{\cos\theta_y^{B/\bar{B}}}>_{\theta_{\bar{Y}}} = \frac{1}{N} \sum_{i=1}^N \cos\theta_{y,i}^{B/\bar{B}} \Bigg \rfloor _{\theta_{\bar{Y}}}.
	\label{eq:polest}
\end{equation}

\noindent In the following, we always refer to moments and spin observables at a given $\theta_Y$, unless explicitly stated otherwise. That means $P^{\bar{Y}}_{y,\theta_Y} = P^{\bar{Y}}_y$ and so on.

The variance of the first moment is given by the difference between the second moment and the square of the first moment. In our case, we have

\begin{equation}
    V(<\cos_y^B>) = \frac{1}{N(N-1)}[<\cos^2\theta_y^B>-<\cos\theta_y^B>^2]
\end{equation}

\noindent that after error propagation and some algebra becomes 

\begin{equation}
    V(P_y) = \frac{3-(\alpha P_{y})^2}{\alpha(N-1)}.
    \label{eq:polvar}
\end{equation}

In a similar way, the spin correlations at a given hyperon scattering angle $\theta_Y$ can be obtained from the moments of the product of the cosines with respect to the different reference axes $i,j = x,y,z$  \cite{erikthesis}:

\begin{equation}
    C^{\bar{Y}Y}_{i,j} = \frac{9<\cos\theta_{i}^B\cos\theta_{j}^{\bar{B}}>}{\alpha\bar{\alpha}}
    \label{eq:spincorr}
\end{equation}

\noindent where the estimator of the moment is given by the arithmetic mean of the cosine product from the data sample at a given scattering angle:

\begin{equation}
	<\reallywidehat{\cos\theta_{i}^B\cos\theta_{j}^{\bar{B}}}> = \frac{1}{N} \sum_{k=1}^N \cos\theta_{i,k}^B \cos\theta_{j,k}^{\bar{B}}.
	\label{eq:spinest}
\end{equation}

\noindent The variance of the spin correlations can be calculated in the same way as that of the polarisation and is found to be

\begin{equation}
    V(C^{\bar{Y}Y}_{i,j}) = \frac{9-(\alpha\bar{\alpha}C^{\bar{Y}Y}_{i,j})^2}{\alpha\bar{\alpha}(N-1)},
    \label{eq:spinvar}
\end{equation}

\noindent for $i,j=x,y,z$.

\subsection{Efficiency corrections}
In reality, detectors and reconstruction algorithms have finite efficiencies. This needs to be taken into account in the parameter estimation. However, the efficiency is a complicated function of all measured variables. In the case of exclusive $\bar{p}p \to \bar{\Lambda}\Lambda, \bar{\Lambda} \to \bar{p}\pi^+, \Lambda \to p\pi^-$ measurements, there are five independent measured variables: the $\bar{\Lambda}$ scattering angle, the proton decay angles $\theta_p$ and $\phi_p$ and the antiproton decay angles $\theta_{\bar{p}}$ and $\phi_{\bar{p}}$. In principle, this means that parameter estimation methods, which rely on integration, such as the Method of Moments, should employ efficiency corrections in all five independent variables. In the case of $\bar{p}p \to \bar{\Xi}^+\Xi^-, \bar{\Xi}^+ \to \bar{\Lambda}\pi^+, \bar{\Lambda} \to \bar{p}\pi^+, \Xi^- \to \Lambda\pi^-, \Lambda \to p\pi^-$, the efficiency depends on nine independent variables. This is however difficult to achieve in practice, since the number of Monte Carlo simulated events required for a five or nine-dimensional correction matrix is very large and thus unfeasible.

Instead, different approximations have to be made, based on reasonable and testable assumptions. In this work, we have treated the efficiency with two independent methods, the \textit{efficiency dependent} and \textit{efficiency independent method}, as outlined in the following.

\subsubsection{Efficiency dependent method}

With this method, the efficiency is corrected for on an event-by-event basis. The efficiency corrected estimator of the moment $<\cos\theta_y^B>$ is given by

\begin{equation}
	<\reallywidehat{\cos\theta_y^B}> = \frac{1}{N} \sum_{i=1}^N \cos\theta_{y,i}^B w_i(\theta_y,\Omega_B,\Omega_{\bar{B}})
	\label{eq:polesteff}
\end{equation}

where 
\begin{equation}
   w_i(\theta_Y,\Omega_B,\Omega_{\bar{B}}) = \frac{1}{\epsilon_i(\theta_y,\Omega_B,\Omega_{\bar{B}})} 
   \label{eq:weight}
\end{equation}.

In the polarisation extraction, we assume for computational simplicity that the efficiency of the $\Lambda$ as a function of the $\Lambda$ angles is independent of the $\bar{\Lambda}$ angles, and vice versa. Then we can reduce $\epsilon(\theta_Y,\Omega_B,\Omega_{\bar{B}})$ to $\epsilon(\theta_Y,\Omega_B)$. Furthermore, our simulations show that the efficiency is symmetric with respect to the azimuthal angle $\phi_y$, which means that we can integrate over $\phi_y$ without introducing a bias. This means that our efficiency is simplified to $\epsilon(\theta_Y,\cos\theta^B_{y})$. Hence, we can represent the efficiency by two-dimensional matrices: the $\bar{Y}$ scattering angle $\theta_{\bar{Y}}$ in the CMS system of the reaction \textit{versus} the decay proton angle $\theta^B_{y}$ with respect the $y$ axis in Fig. \ref{fig:refsys}, in the rest frame of the decaying hyperon. 

For the spin correlation $C^{\bar{Y}Y}_{i,j}$, we need to take into account the decay angles from the hyperon and the antihyperon. We then assume a 3D efficiency \\ $\epsilon(\theta_Y,\cos\theta^B_{i}, \cos\theta^{\bar{B}}_{j})$ and hence, we use 3D matrices. Here, $i,j = x,y,z$ in Fig. \ref{fig:refsys}. These three-dimensional correction matrices were also used in a cross check analysis of the polarisation estimation, with consistent results.

In the $\bar{p}p \to \bar{\Xi}^+\Xi^-$ case, we have assumed that the efficiency is symmetric with respect to the $\Lambda \to p\pi^-$ and $\bar{\Lambda} \to \bar{p}\pi^+$ decay angles which were integrated out. 

The estimator for the polarisation is given by

\begin{equation}
	\reallywidehat{P_y}^{Y/\bar{Y}} = \frac{3}{\alpha}\frac{\sum^N_{i=1}\cos\theta_{y,i}^{B/\bar{B}}\cdot{w_i(\cos\theta_{y,i}^{B/\bar{B}},\cos\theta_{\bar{Y}})}}{\sum^N_{i=1}w_i(\cos\theta_{y,i}^{B/\bar{B}},\cos\theta_{\bar{Y}})}.
	\label{eq:recpol}
\end{equation}

\noindent where $w(\cos\theta_{y,i},\cos\theta_{\bar{Y}})$ is the weight (Eq. (\ref{eq:weight})) at the given $\cos\theta_y$ and $\cos\theta_{\bar{Y}}$. $N$ is the number of events in the sample. For the spin correlations, the estimators are given by 

\begin{equation}
	\reallywidehat{C^{\bar{Y}Y}_{\mu\nu}} = \frac{9}{\alpha\bar{\alpha}}\frac{\sum^N_{i=1} \cos\theta_{\mu,i}^{\bar{B}}\cos\theta_{\nu,i}^B\cdot w_i(\cos\theta_{\mu,i}^{\bar{B}},\cos\theta_{\nu,i}^B,\cos\theta_{Y})}{\sum^N_{i=1} w_i(\cos\theta_{\mu,i}^{\bar{B}},\cos\theta_{\nu,i}^B,\cos\theta_{Y})}.
	\label{eq:recspincorr}
\end{equation}

\subsubsection{Efficiency independent method}
\label{sec:effind}

For special cases, alternative estimators can be defined which do not require efficiency corrections. These have been treated thoroughly in Refs. \cite{erikthesis,tayloethesis} and will be briefly summarised here. 

Here, it is most convenient to use the matrix formulation, see Sect. \ref{sec:formalism}. The first order moments of the angles and their products can be gathered in a $4\times4$ matrix as follows:

\begin{equation*}
	E=\begin{pmatrix}
	\braket{1} & \braket{\cos\theta_{x}^{B}} & \braket{\cos\theta_{y}^{B}} & \braket{\cos\theta_{z}^{B}} \\
	\braket{\cos\theta_{x}^{\bar{B}}} & \braket{\cos\theta_{x}^{\bar{B}}\cos\theta_{x}^{B}} & \braket{\cos\theta_{x}^{\bar{B}}\cos\theta_{y}^{B}} & \braket{\cos\theta_{x}^{\bar{B}}\cos\theta_{z}^{B}} \\
	\braket{\cos\theta_{y}^{\bar{B}}} & \braket{\cos\theta_{y}^{\bar{B}}\cos\theta_{x}^{B}} & \braket{\cos\theta_{y}^{\bar{B}}\cos\theta_{y}^{B}} & \braket{\cos\theta_{y}^{\bar{B}}\cos\theta_{z}^{B}} \\
	\braket{\cos\theta_{z}^{\bar{B}}} & \braket{\cos\theta_{z}^{\bar{B}}\cos\theta_{x}^{B}} & \braket{\cos\theta_{z}^{\bar{B}}\cos\theta_{y}^{B}} & \braket{\cos\theta_{z}^{\bar{B}}\cos\theta_{z}^{B}} 
	\end{pmatrix}
\end{equation*}

and some additional moments in vector form

\begin{align}
	& F = \begin{pmatrix}
		\braket{1} & \braket{\cos^2\theta_{x}^{B}} & \braket{\cos^2\theta_{y}^{B}} & \braket{\cos^2\theta_{z}^{B}}
	\end{pmatrix}\\
	& \bar{F} = \begin{pmatrix}
		\braket{1} & \braket{\cos^2\theta_{x}^{\bar{B}}} & \braket{\cos^2\theta_{y}^{\bar{B}}} & \braket{\cos^2\theta_{z}^{\bar{B}}}
	\end{pmatrix}.
\end{align}

\noindent We assume that the efficiency of the antihyperon and its decay is independent of that of the hyperon, \textit{i.e.} 

\begin{equation}
    \epsilon(\Omega_{\bar{B}},\Omega_B) = \epsilon(\Omega_{\bar{B}})\cdot\epsilon(\Omega_B).
\end{equation}
We then define the following matrices of efficiency weighted moments

\begin{align}
	& \bar{\mathcal{A}}_{\mu,\nu} \equiv \int \cos\theta_{\mu}^{\bar{B}} \cos\theta_{\nu}^{\bar{B}} \epsilon(\Omega_{\bar{B}}) d\Omega_{\bar{B}} \\
	& \mathcal{A}_{\mu,\nu} \equiv \int \cos\theta_{\mu}^{B} \cos\theta_{\nu}^{B} \epsilon(\Omega_{B}) d\Omega_{B} \\
	& \bar{\mathcal{B}}_{\mu,\nu} \equiv \int \cos^2\theta_{\mu}^{\bar{B}} \cos\theta_{\nu}^{\bar{B}} \epsilon(\Omega_{\bar{B}}) d\Omega_{\bar{B}} \\
	& \mathcal{B}_{\mu,\nu} \equiv \int \cos\theta_{\mu}^{B} \cos^2\theta_{\nu}^{B} \epsilon(\Omega_{B}) d\Omega_{B} \\
	& \bar{\mathcal{C}}_{\mu} \equiv \bar{\mathcal{A}}_{\mu,0} \\
	& \mathcal{C}_{\nu} \equiv \mathcal{A}_{0,\nu}.
\end{align}

\noindent By definition, the $\mathcal{A}$ matrices are symmetric in $\mu$ and $\nu$. Furthermore, some elements of $\mathcal{B}$ are identical to those of $\mathcal{A}$ $e.g.$ $\mathcal{B}_{10}=\mathcal{A}_{11}$, $\mathcal{\bar{B}}_{01}=\mathcal{\bar{A}}_{11}$. With these definitions, the moments can be related to the spin observables in the following way:
\begin{equation}
	E = \frac{1}{16\pi^2} \bar{\mathcal{A}} D \mathcal{A}
	\label{eq:Esolve}
\end{equation}
\begin{equation}
	\bar{F} = \frac{1}{16\pi^2} \bar{\mathcal{B}} D \mathcal{C}
	\label{eq:Fbarsolve}
\end{equation}
\begin{equation}
	F = \frac{1}{16\pi^2} \bar{\mathcal{C}} D \mathcal{B}
	\label{eq:Fsolve}
\end{equation}

If the efficiency is symmetric with respect to $\cos\theta_y$ for both the antibaryon and the baryon, \textit{i.e.}
\begin{align}
	&\epsilon(\cos\theta_{x}^{\bar{B}},\cos\theta_{y}^{\bar{B}},\cos\theta_{z}^{\bar{B}})=\epsilon(\cos\theta_{x}^{\bar{B}},-\cos\theta_{y}^{\bar{B}},\cos\theta_{z}^{\bar{B}})\\
	&\epsilon(\cos\theta_{x}^{B},\cos\theta_{y}^{B},\cos\theta_{z}^{B})=\epsilon(\cos\theta_{x}^{B},-\cos\theta_{y}^{B},\cos\theta_{z}^{B}).
\end{align}
\noindent then all matrix elements in $\mathcal{A}$ and $\mathcal{B}$ with odd powers of $\cos\theta_y$ are zero. The matrices then reduce to
\begin{equation}
	\mathcal{A} = \begin{pmatrix}
		\mathcal{A}_{00} & \mathcal{A}_{01} & 0 & \mathcal{A}_{03} \\
		\mathcal{A}_{01} & \mathcal{A}_{11} & 0 & \mathcal{A}_{13} \\
		0 & 0 & \mathcal{A}_{22} & 0 \\
		\mathcal{A}_{03} & \mathcal{A}_{13} & 0 & \mathcal{A}_{33} \\
	\end{pmatrix}
\end{equation}
\begin{equation}
	\mathcal{\bar{B}} = \begin{pmatrix}
		\mathcal{\bar{B}}_{00} & \mathcal{\bar{B}}_{01} & 0 & \mathcal{\bar{B}}_{03} \\
		\mathcal{\bar{B}}_{10} & \mathcal{\bar{B}}_{11} & 0 & \mathcal{\bar{B}}_{13} \\
		\mathcal{\bar{B}}_{20} & \mathcal{\bar{B}}_{21} & 0 & \mathcal{\bar{B}}_{23} \\
		\mathcal{\bar{B}}_{30} & \mathcal{\bar{B}}_{31} & 0 & \mathcal{\bar{B}}_{33} \\
	\end{pmatrix},\;\;\; B = \begin{pmatrix}
		\mathcal{B}_{00} & \mathcal{B}_{01} & \mathcal{B}_{02} & \mathcal{B}_{03} \\
		\mathcal{B}_{10} & \mathcal{B}_{11} & \mathcal{B}_{12} & \mathcal{B}_{13} \\
		0 & 0 & 0 & 0 \\
		\mathcal{B}_{30} & \mathcal{B}_{31} & \mathcal{B}_{32} & \mathcal{B}_{33} \\
	\end{pmatrix}
\end{equation}

With these simplifications, the right hand side of Eqs. \ref{eq:Esolve}, \ref{eq:Fbarsolve} and \ref{eq:Fsolve} can be solved, resulting in terms that consist of products of $\mathcal{A}_{\mu,\nu}$, $\mathcal{B}_{\mu,\nu}$ and $D_{\mu,\nu}$. We find that some of these terms are small in magnitude. If these terms can be neglected, then the non-zero spin observables are shown in Ref. \cite{tayloethesis} to be

\begin{align}
	&D_{20} = \frac{E_{20}}{\bar{F}_2},\;\;\; D_{02} = \frac{E_{02}}{F_2} \\
	&D_{22} = \frac{E_{22}}{\bar{F}_2 F_2}\\
	&D_{11} = \frac{E_{11}-E_{10}E_{01}}{\bar{F}_1 F_1},\;\;\;D_{13} = \frac{E_{13}-E_{10}E_{03}}{\bar{F}_1 F_3}\\
	&D_{31} = \frac{E_{31}-E_{30}E_{01}}{\bar{F}_3 F_1},\;\;\;D_{33} = \frac{E_{33}-E_{30}E_{03}}{\bar{F}_3 F_3},
\end{align}
which translates to
\begin{align}
	& P^{\bar{Y}}_{y} = \frac{1}{\bar{\alpha}}\frac{\braket{\cos\theta_{y,\bar{B}}}}{\braket{\cos^2\theta_{y,\bar{B}}}}\label{eq:pybarfinal} \\
	& P^Y_{y} = \frac{1}{\alpha}\frac{\braket{\cos\theta_{y,B}}}{\braket{\cos^2\theta_{y,B}}}\label{eq:pyfinal} \\
	& C^{\bar{Y}Y}_{yy} = \frac{1}{\bar{\alpha}\alpha} \frac{\braket{\cos\theta_{y,\bar{B}}\cos\theta_{y,B}}}{\braket{\cos^2\theta_{y,\bar{B}}}\braket{\cos^2\theta_{y,B}}}\label{eq:cyyfinal} \\
	& C^{\bar{Y}Y}_{xx} = \frac{1}{\bar{\alpha}\alpha} \frac{\braket{\cos\theta_{x,\bar{B}}\cos\theta_{x,B}} - \braket{\cos\theta_{x,\bar{B}}} \braket{\cos\theta_{x,B}} }{\braket{\cos^2\theta_{x,\bar{B}}}\braket{\cos^2\theta_{x,B}}}\label{eq:cxxfinal} \\
	& C^{\bar{Y}Y}_{xz} = \frac{1}{\bar{\alpha}\alpha} \frac{\braket{\cos\theta_{x,\bar{B}}\cos\theta_{z,B}} - \braket{\cos\theta_{x,\bar{B}}} \braket{\cos\theta_{z,B}} }{\braket{\cos^2\theta_{x,\bar{B}}}\braket{\cos^2\theta_{z,B}}}\label{eq:cxzfinal} \\
	& C^{\bar{Y}Y}_{zx} = \frac{1}{\bar{\alpha}\alpha} \frac{\braket{\cos\theta_{z,\bar{B}}\cos\theta_{x,B}} - \braket{\cos\theta_{z,\bar{B}}} \braket{\cos\theta_{x,B}} }{\braket{\cos^2\theta_{z,\bar{B}}}\braket{\cos^2\theta_{x,B}}}\label{eq:czxfinal} \\
	& C^{\bar{Y}Y}_{zz} = \frac{1}{\bar{\alpha}\alpha} \frac{\braket{\cos\theta_{z,\bar{B}}\cos\theta_{z,B}} - \braket{\cos\theta_{z,\bar{B}}} \braket{\cos\theta_{z,B}} }{\braket{\cos^2\theta_{z,\bar{B}}}\braket{\cos^2\theta_{z,B}}}\label{eq:czzfinal}.
\end{align}

To summarize, the efficiency independent method is viable if the following three conditions are met:
\begin{enumerate}
	\item The detection efficiency of the antibaryon is independent of that of the baryon.
	\item The efficiency is symmetric in $\cos\theta_{y}^{B}$ and $\cos\theta_{y}^{\bar{B}}$ 
	\item Higher order terms emerging from Eqs.~\eqref{eq:Esolve}, \eqref{eq:Fbarsolve} and \eqref{eq:Fsolve} can be neglected.
\end{enumerate}

Simulations show that the first criterion is fulfilled for both channels at both momenta, whereas the second and third criteria are channel- and momentum dependent. For $\bar{p}p \to \bar{\Lambda}\Lambda$ at 1.642 GeV/$c$, the second criterion is fulfilled. Furthermore, the higher order terms appearing in the expressions for $P^Y_y$, $P^{\bar{Y}}_y$ and $C^{\bar{Y}Y}_{yy}$ can be neglected whereas they are large for $C^{\bar{Y}Y}_{xx}$, $C^{\bar{Y}Y}_{zz}$ $C^{\bar{Y}Y}_{xz}$ and $C^{\bar{Y}Y}_{zx}$. This means we expect the efficiency independent method to work for $P
^Y_y$, $P^{\bar{Y}}_y$ and $C^{\bar{Y}Y}_{yy}$ but not for the other observables.

For the $\bar{p}p \to \bar{\Xi}^+\Xi^-$ channel at 4.6 GeV/$c$, all three criteria are fulfilled for all spin observables. Thus the efficiency independent method can be used without restrictions in this case. At 7.0 GeV/$c$, the second criterion is not fulfilled which means that the efficiency independent method cannot be used to extract the polarisation of neither the hyperon nor the antihyperon. However, it can be applied to estimate all spin correlations.

\section{Results}
\label{sec:results}

\subsection{Reconstruction rates}
\label{sec:rates}
With the reconstruction efficiencies obtained from the simulations, the measured $\bar{p}p \to \bar{\Lambda}\Lambda$ cross section from Ref. \cite{PS185164} and the predicted cross sections from Ref. \cite{kaidalov}, we can calculate the expected rate at which hyperons can be reconstructed exclusively in PANDA. We have performed the calculations for two different scenarios: with the Phase One luminosity, which will be around $10^{31}$cm$^{-2}$s$^{-1}$, and with the 20 times larger design luminosity. The results are presented in Table \ref{hypprod}. 
However, during the very first period of data taking, the luminosity at low beam momenta will be smaller by about a factor of two. This means that in the first $\bar{p}p \to \bar{\Lambda}\Lambda$ benchmark study, the actual luminosity will be about $5\cdot10^{30}$cm$^{-2}$s$^{-1}$, giving a two times smaller reconstruction rate than at the nominal Phase One luminosity. The $S/B$ ratios are calculated using all remaining signal events $S$ and background events from all sources, weighted using their corresponding weight factors given in Tables \ref{tab:samplesize1} and \ref{tab:samplesize2}.
\begin{table*}
\centering
\begin{tabular}{llllllll}
\hline
$p_{\bar{p}}$ (GeV/$c$) & Reaction & $\sigma$ ($\mu$b) & Eff (\%) & Decay & S/B & Rate  ($s^{-1}$) & Rate ($s^{-1}$)\\&&&&&& at $10^{31}$cm$^{-2}$s$^{-1}$ & at 2$\cdot10^{32}$cm$^{-2}$s$^{-1}$\\\hline
1.64 & $\bar{p}p \rightarrow \bar{\Lambda}\Lambda$ & 64.1 $\pm$ 1.6~\cite{PS185164_1} & 16.04 $\pm$ 0.04 & $\Lambda \rightarrow p \pi^-$ & 114 & 44 & 880  \\\hline

4.6 & $\bar{p}p \rightarrow \bar{\Xi}^+\Xi^-$ & $\approx$1~~\cite{kaidalov} & 8.22 $\pm$ 0.03 & $\Xi^- \rightarrow \Lambda \pi^-$ & 270 & 0.3  & 6 \\\hline

7.0 & $\bar{p}p \rightarrow \bar{\Xi}^+\Xi^-$ & $\approx$0.3~~\cite{kaidalov} & 7.95 $\pm$ 0.03 & $\Xi^- \rightarrow \Lambda \pi^-$ & 170 & 0.1 & 2 \\\hline
\end{tabular}
\caption{Results from simulation studies of the various production reactions of ground state hyperons. The efficiencies are for exclusive reconstruction, and are presented with statistical uncertainties. The $S/B$ denotes the signal-to-background ratio.}
\label{hypprod}       
\end{table*}

\subsubsection{Effects from the $\bar{\Xi}^+$ angular distribution}

The distribution of the $\bar{\Xi}^+$ scattering angle is not known, since so far, only a few bubble-chamber events exist from the $\bar{p}p \to \bar{\Xi}^+\Xi^-$ reaction \cite{Musgrave1965}. The nominal simulations in this work were therefore performed for isotropically distributed $\bar{\Xi}^+$ antihyperons. However, in reality, the angular distribution in the CMS system of the reaction may be forward peaking in a similar way as for $\bar{p}p \to \bar{\Lambda}\Lambda$ \cite{PS185} and $\bar{p}p \to \bar{\Sigma}^0\Lambda + c.c.$ \cite{sigma6}. Since the $\bar{\Xi}^+$ share one less quark with the initial $\bar{p}$ compared to $\bar{\Lambda}$ and $\bar{\Sigma}^0$, the forward peak is expected to be less pronounced for $\bar{\Xi}^+$. Investigations with meson exchange models have resulted in a fairly strong anisotropy for $\bar{\Xi}^0$ while almost flat for the $\bar{\Xi}^+$ \cite{XiMEX}. This can have an impact on the total reconstruction efficiency, partly because decay products of the $\bar{\Xi}^+$ may escape detection by being emitted along the beam pipe, and partly because a backward-going $\Xi^-$ in the CMS system is almost at rest in the lab system. Its decay products may then have too low energy to reach the detectors. 

\begin{figure*}[ht]
		        \includegraphics[width=0.5\linewidth]{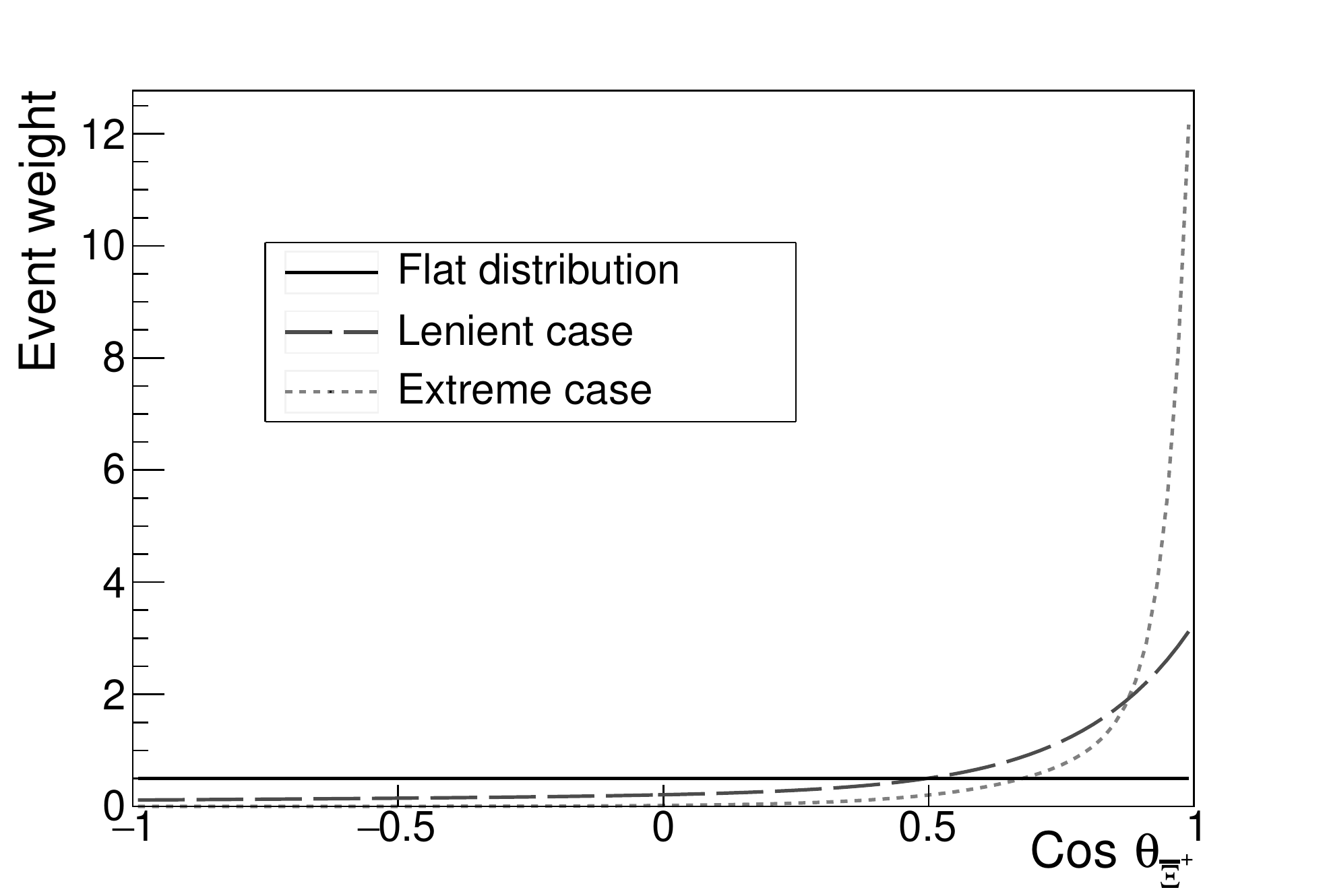}
		        \includegraphics[width=0.5\linewidth]{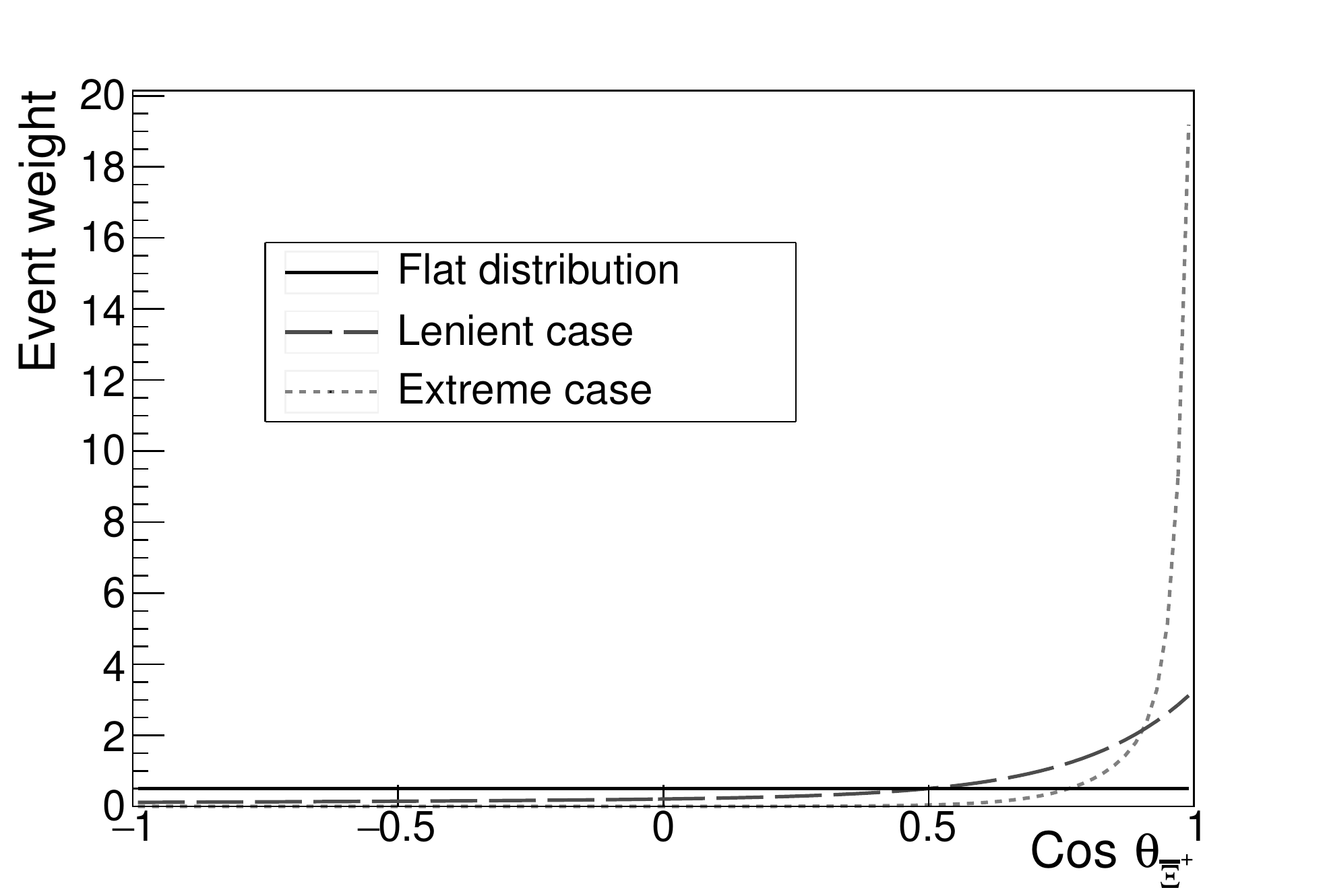}
			\caption{Simulated angular distributions of the $\bar{p}p\to\bar{\Xi}^+\Xi^-$ reaction using a flat distribution (black), lenient case (dashed), and the extreme case (dotted) at $p_{\mathrm{beam}}=4.6$ GeV/$c$ (left) and $p_{\mathrm{beam}}=7.0$ GeV/$c$ (right). Note the different scales on the $y$-axes.}
			\label{fig:costhtfwp}
		\end{figure*}

In order to investigate the sensitivity of the total reconstruction efficiency to the $\bar{\Xi}^+$ angular distribution, additional simulations were carried out for two other scenarios with more forward-going antihyperons. The \textit{extreme case} employs angular distribution parameters from the most forward-peaking distributions that have been observed so far, namely in $\bar{p}p \to \bar{\Lambda}\Sigma^0 + c.c.$ \cite{sigma6}. The \textit{lenient case} represents an intermediate scenario with parameters between those of a flat distribution and those of an extreme one. The distributions are shown in Fig. \ref{fig:costhtfwp} and the results from the simulations are presented in Table \ref{tab:fwpeff}. Indeed, the reconstruction efficiency decreases for a strongly forward peaking $\bar{\Xi}^+$ distribution. However, the most extreme case results in a reduction of 25-35\% and the total efficiency -- 5-6\% -- is still feasible for $\bar{p}p \to \bar{\Xi}^+\Xi^-$ studies.

			\begin{table}[ht]
			\centering
			\begin{tabular}{c|c|c|c}
				\textbf{$p_{\bar{p}}$ (GeV/c)} & $\epsilon_{Isotropic}$ (\%) & $\epsilon_{Lenient}$ (\%) & $\epsilon_{Extreme}$ (\%) \\ \hline
				4.6 & 8.22 $\pm$ 0.03 & 7.7 $\pm$ 0.03 & 6.1 $\pm$ 0.03 \\
				7.0 & 7.95 $\pm$ 0.03 & 7.5 $\pm$ 0.03 & 5.0 $\pm$ 0.03 \\
				\hline
			\end{tabular}
			\caption{Reconstruction efficiency of the $\bar{p}p\to\bar{\Xi}^+\Xi^-$ reaction with an isotropic angular distribution, a lenient one and an extremely forward peaking distribution. \label{tab:fwpeff}}
		\end{table}

\subsection{Spin observables}
\label{sec:spin}

The spin observables defined in Sect. \ref{sec:formalism} have been reconstructed with two independent methods to handle the efficiency, described in Sect. \ref{sec:paramest}. In both cases, we have used data samples that are realistic during the first year of data taking with PANDA, given the reconstruction rates estimated in Sect. \ref{sec:rates}. Since the background can be suppressed to a very low level, background effects are neglected in these spin studies.

\subsubsection{The $\bar{p}p \to \bar{\Lambda}\Lambda$ reaction}

In this study, $1.5\cdot10^6$ reconstructed $\bar{p}p \to \bar{\Lambda}\Lambda$ events were used. This amount can be collected in 24 hours during the first phase of data taking with PANDA, where the luminosity at the lowest beam momenta will be about half of that of intermediate and high momenta, \textit{i.e.} 5$\cdot10^{30}$cm$^{-2}$s$^{-1}$. 

\begin{figure*}[ht]
    \centering
    \begin{minipage}{.5\textwidth}
        \centering
        \includegraphics[width=1.\linewidth]{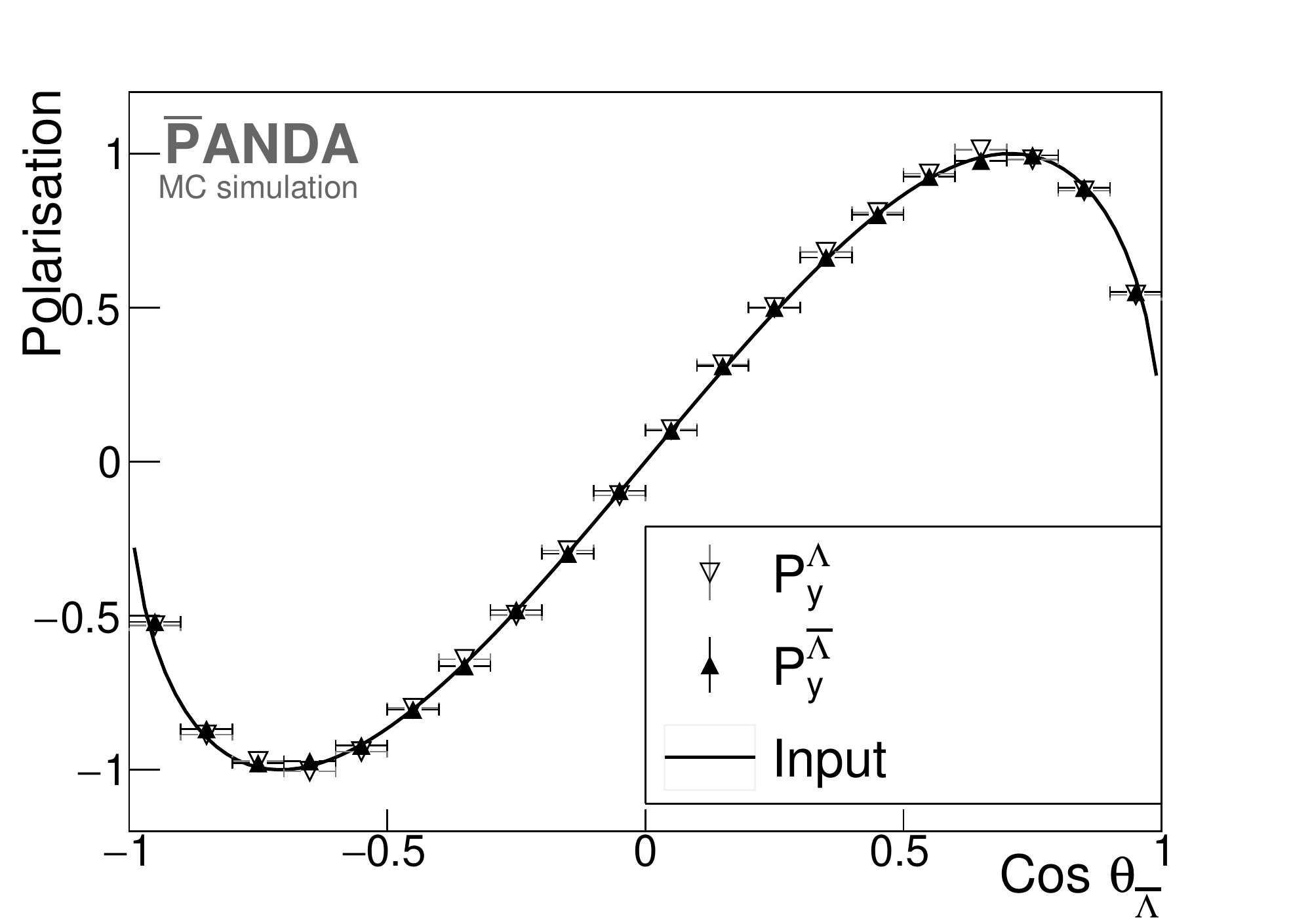}\\
        \includegraphics[width=1.\linewidth]{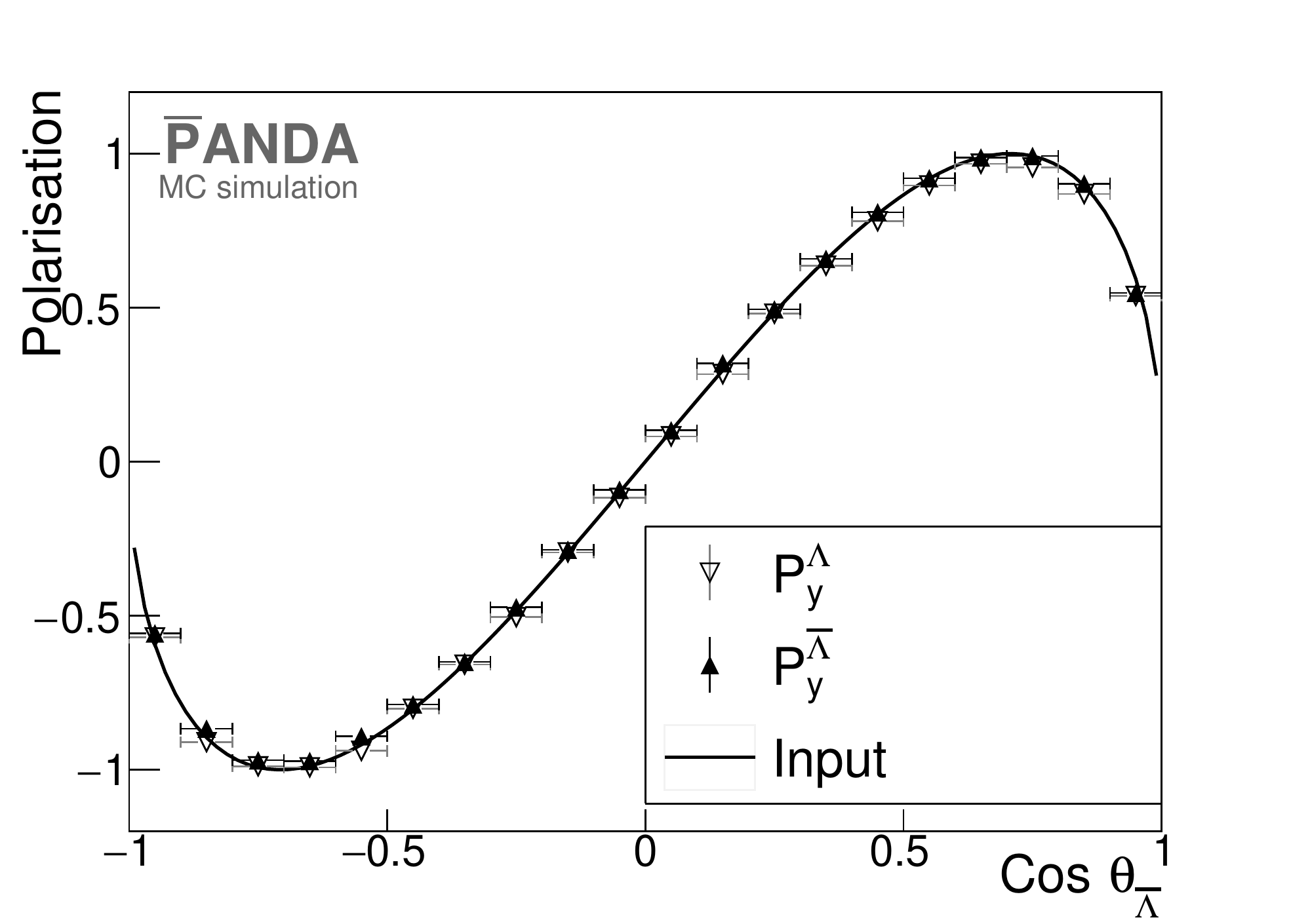}
    \end{minipage}%
    \begin{minipage}{0.5\textwidth}
        \centering
        \includegraphics[width=1.\linewidth]{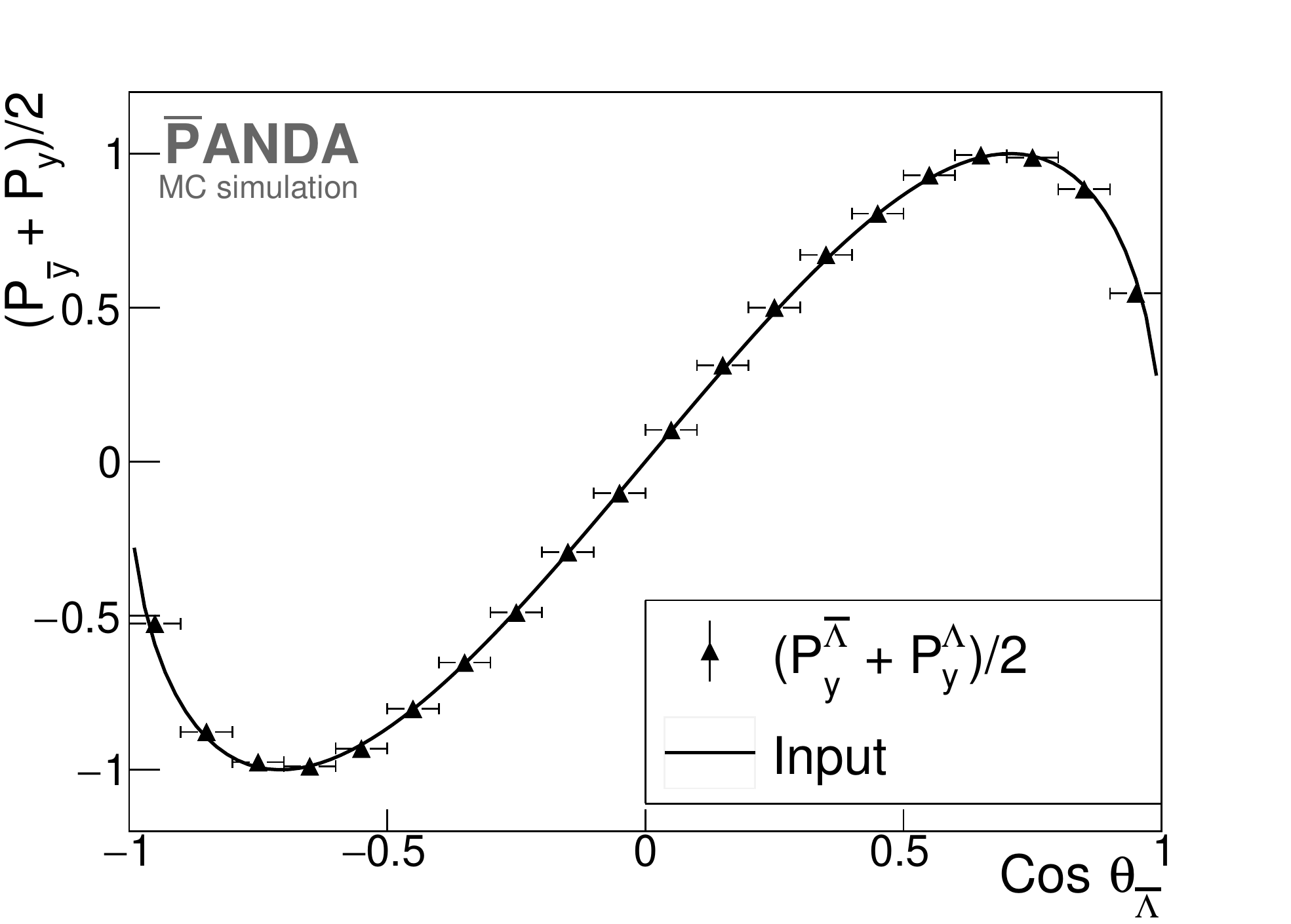}\\
        \includegraphics[width=1.\linewidth]{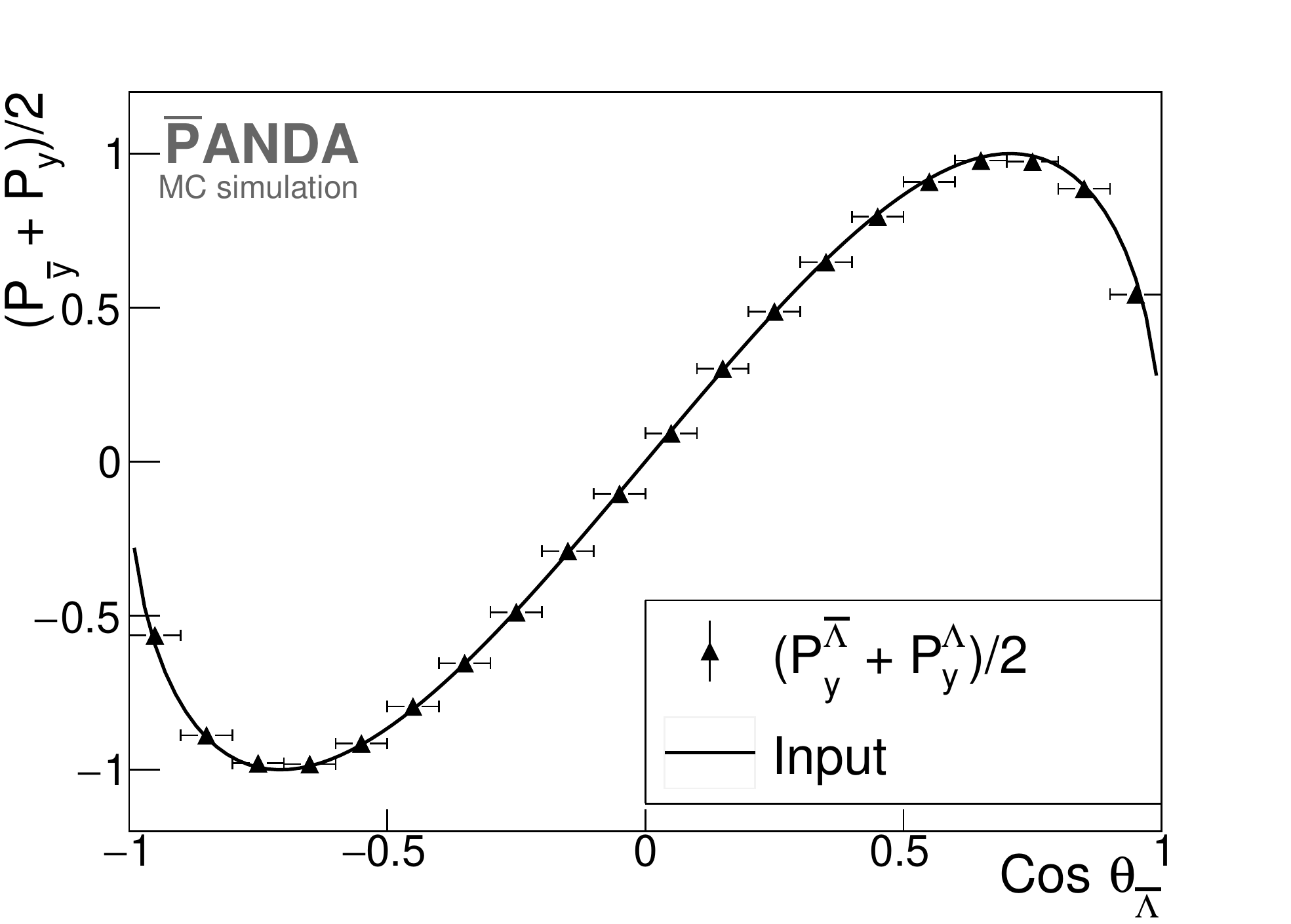}
    \end{minipage}
	\caption{Top left: Polarization of the $\bar{\Lambda}$ (black) and the $\Lambda$ (open) at $p_{\mathrm{beam}}=1.642$ GeV/c, reconstructed using the efficiency dependent method with 2D efficiency matrices. Top-right: Average values of the two reconstructed polarisations. Bottom-left: Polarisations reconstructed using the efficiency independent method. Bottom-right: Average of the polarisations reconstructed with the efficiency independent method. The vertical error bars represent statistical uncertainties, the horizontal bars the bin widths and the solid curves the input model.}
	\label{fig:llbarpol}
\end{figure*}

The polarisation of $\bar{\Lambda}$ and $\Lambda$ as a function of the $\bar{\Lambda}$ scattering angle in the CMS system are shown in Fig. \ref{fig:llbarpol}. The $\bar{\Lambda}$ and $\Lambda$ polarisation are shown to the left in the same plot. Since charge conjugation invariance requires $P^{Y} = P^{\bar{Y}}$, deviations from this equality could indicate artificial bias from the detector or the reconstruction procedure. However, the agreement is excellent. In the right panels, the average of the $\bar{\Lambda}$ and $\Lambda$ polarisation is shown. The top panels show the polarisations extracted with efficiency corrections, estimated by Eq. (\ref{eq:pol}). The bottom panels are extracted using the efficiency independent method, applying Eqs. (\ref{eq:pybarfinal}) and (\ref{eq:pyfinal}). The polarisations reconstructed with the two techniques agree very well with the input distributions, shown as solid curves. The statistical uncertainties are found to be very small.

\begin{figure*}[ht]
    \centering
    \begin{minipage}{.5\textwidth}
        \centering
        \includegraphics[width=1.\linewidth]{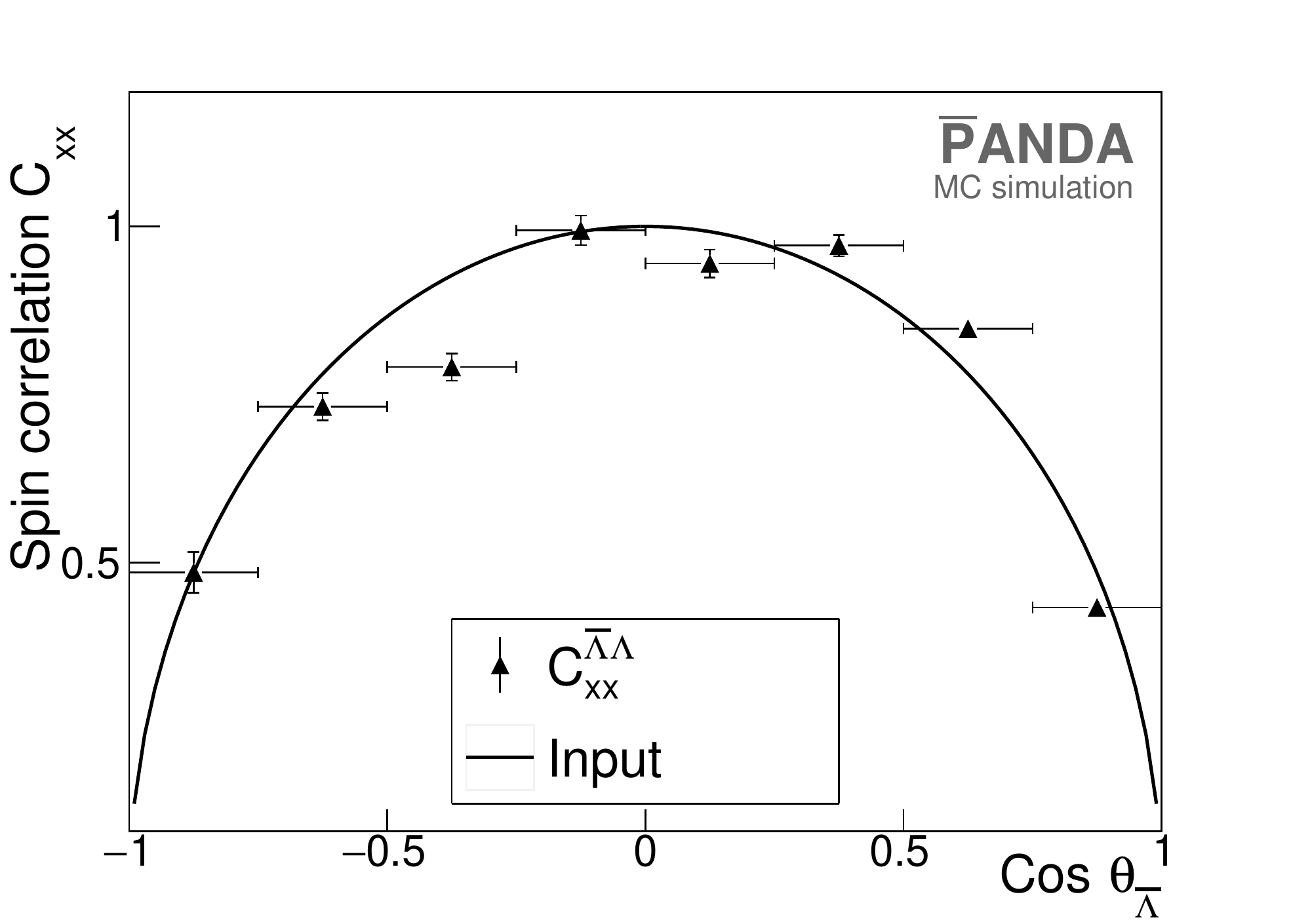}\\
        \includegraphics[width=1.\linewidth]{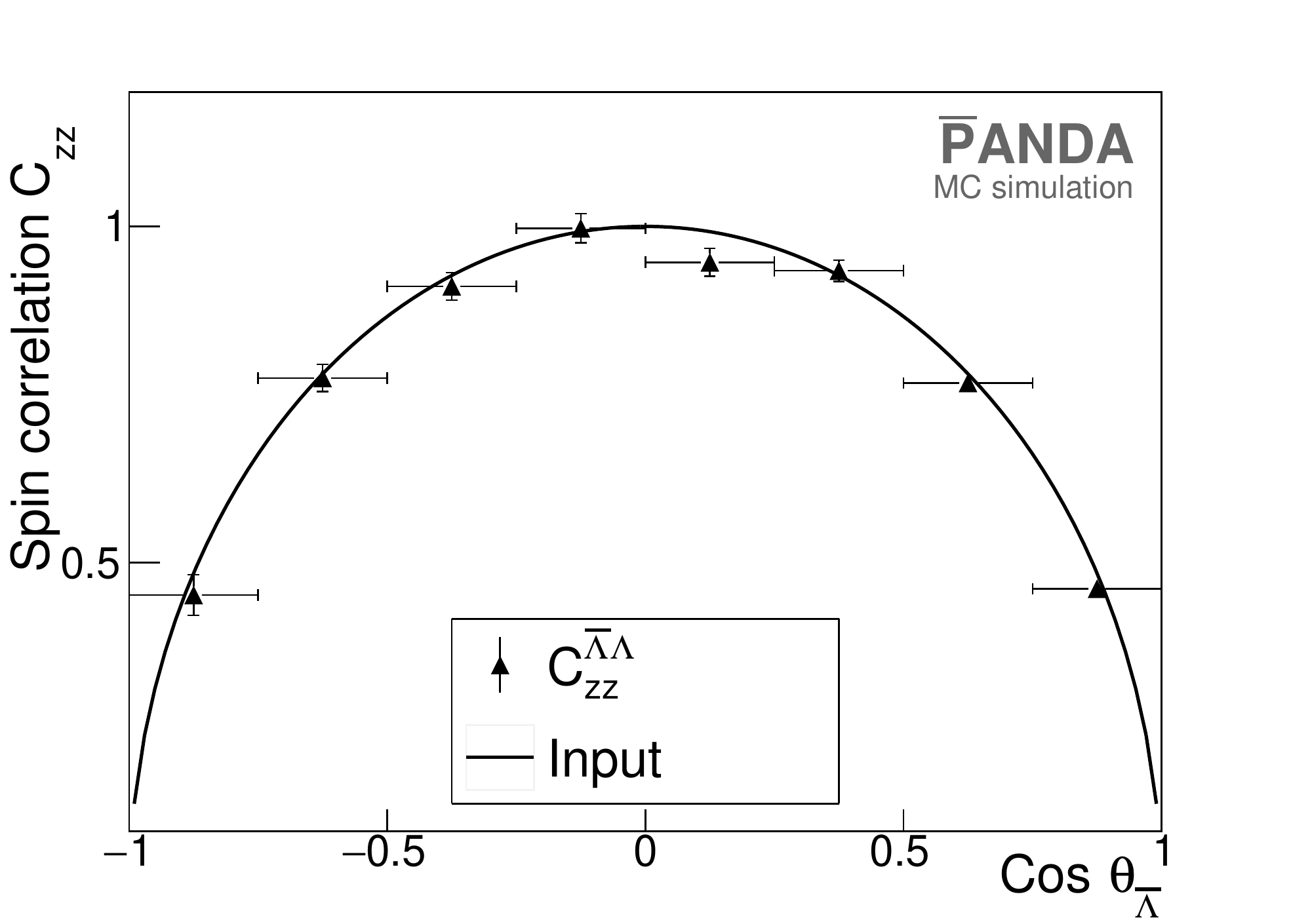}
    \end{minipage}%
    \begin{minipage}{0.5\textwidth}
        \centering
        \includegraphics[width=1.\linewidth]{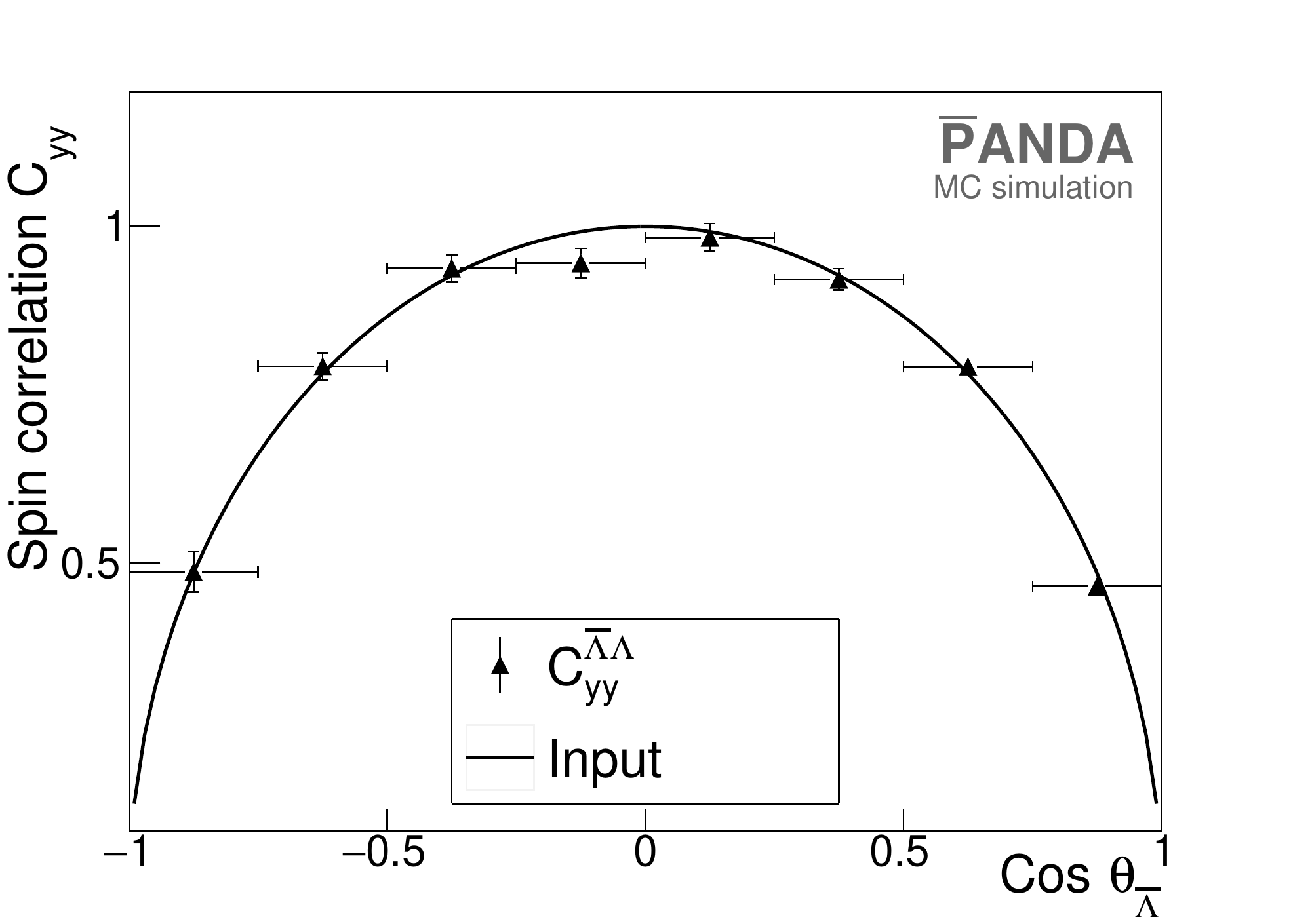}\\
        \includegraphics[width=1.\linewidth]{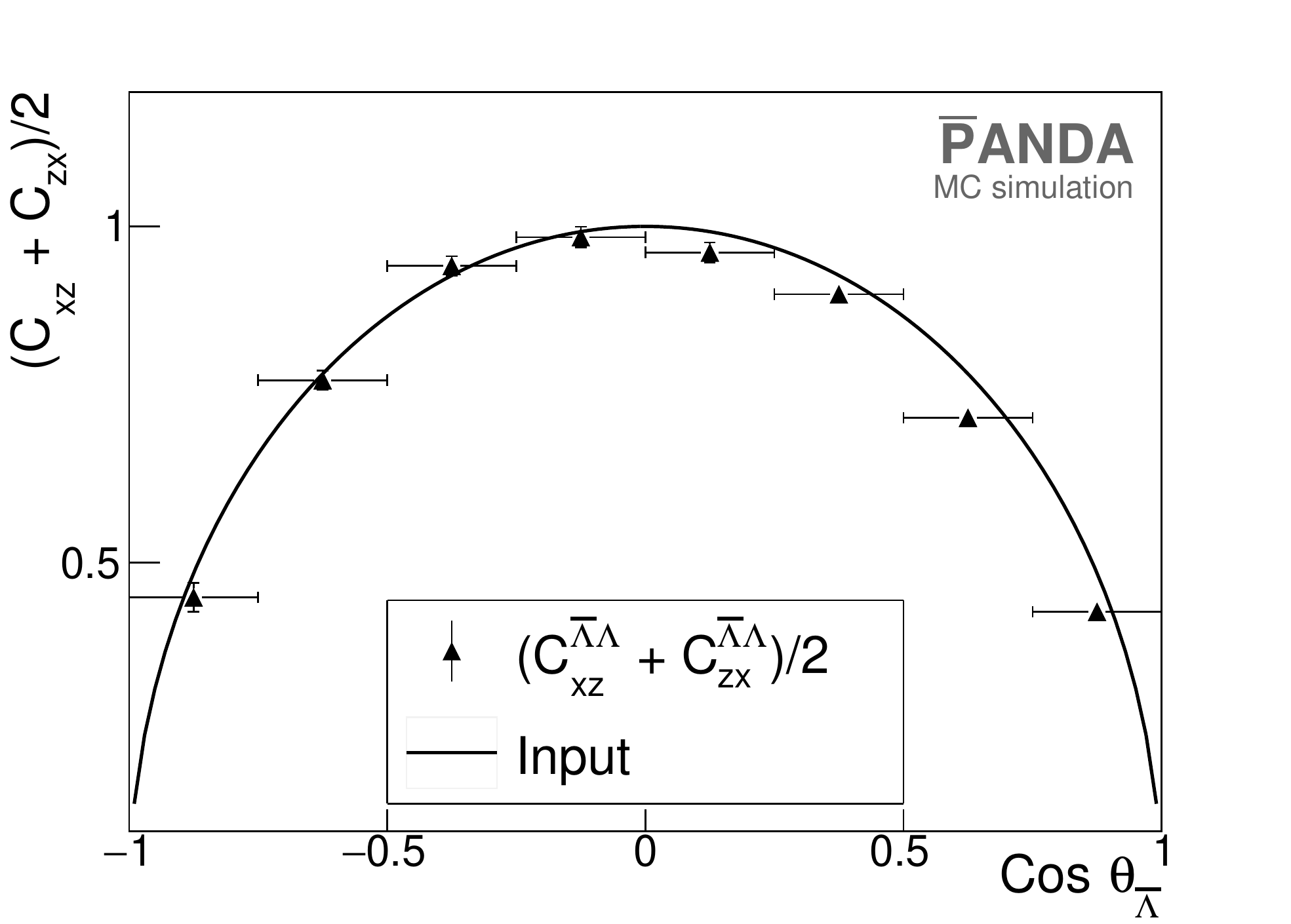}
    \end{minipage}
	\caption{Spin correlations of the $\bar{\Lambda}\Lambda$ pair produced at $p_{\mathrm{beam}}=1.642$ GeV/c. These observables were estimated with the efficiency dependent method, using 3D efficiency matrices. Top-left: $C^{\bar{Y}Y}_{xx}$, top-right: $C^{\bar{Y}Y}_{yy}$ and bottom-left: $C^{\bar{Y}Y}_{zz}$ of the $\bar{\Lambda}\Lambda$ pair. Bottom-right: The average $(C^{\bar{Y}Y}_{xz} + C^{\bar{Y}Y}_{zx})/2$. The vertical error bars represent statistical uncertainties, the horizontal bars the bin widths and the solid curve the input distributions.}
	\label{fig:llbarspincorrii}
\end{figure*}

\begin{figure*}[ht]
    \centering
    \begin{minipage}{.5\textwidth}
        \centering
        \includegraphics[width=1.\linewidth]{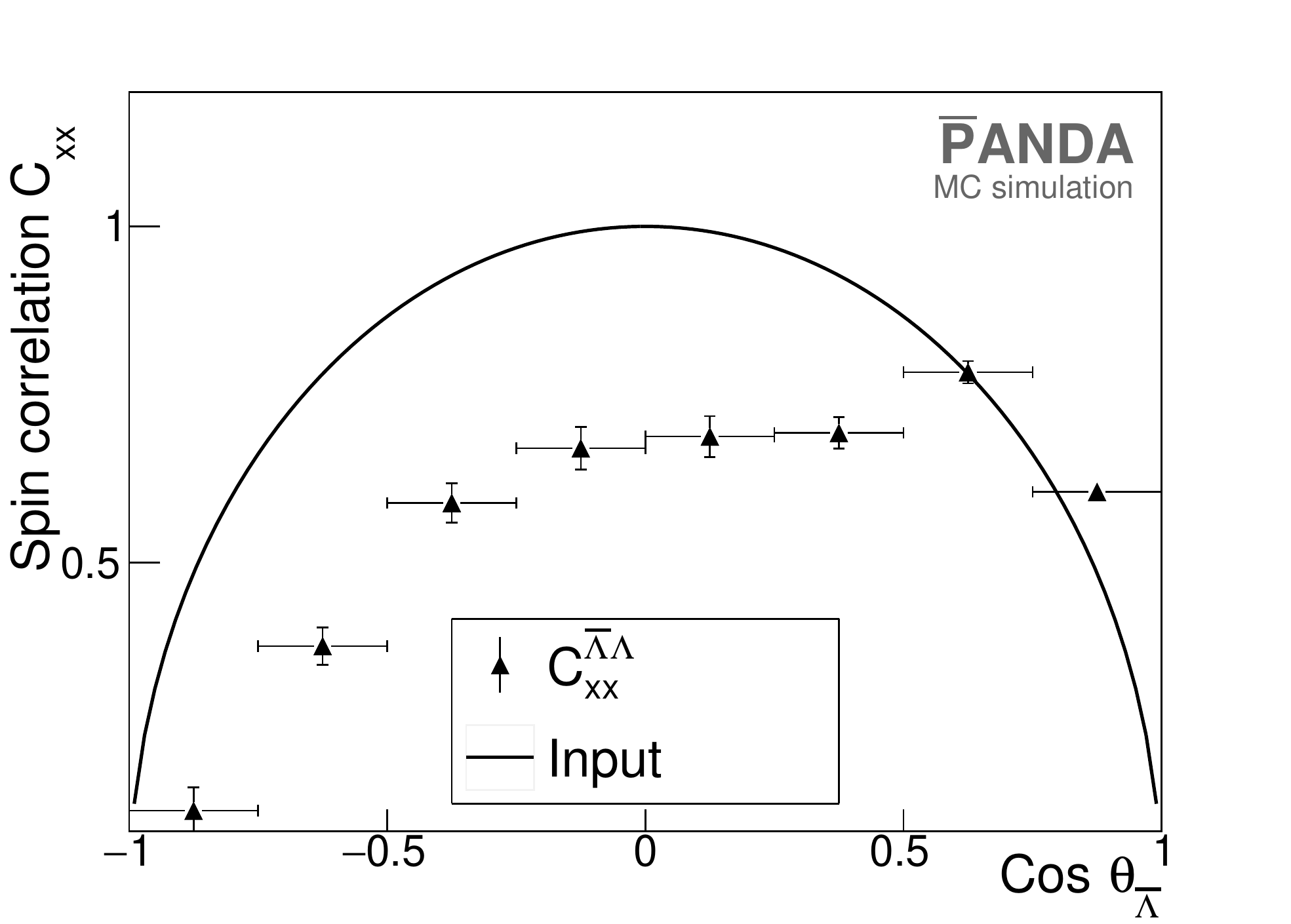}\\
        \includegraphics[width=1.\linewidth]{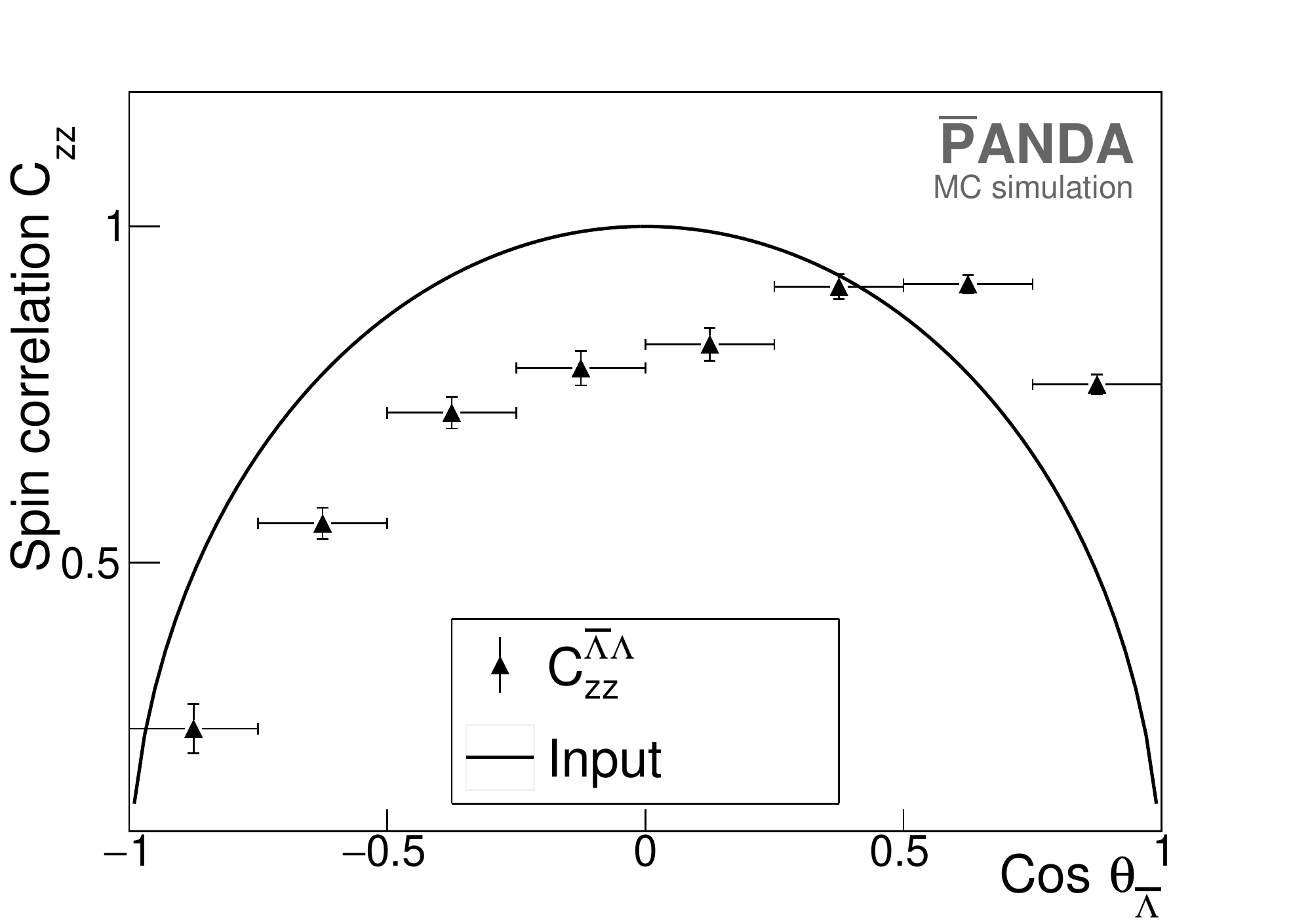}\\
    \end{minipage}%
    \begin{minipage}{0.5\textwidth}
        \centering
        \includegraphics[width=1.\linewidth]{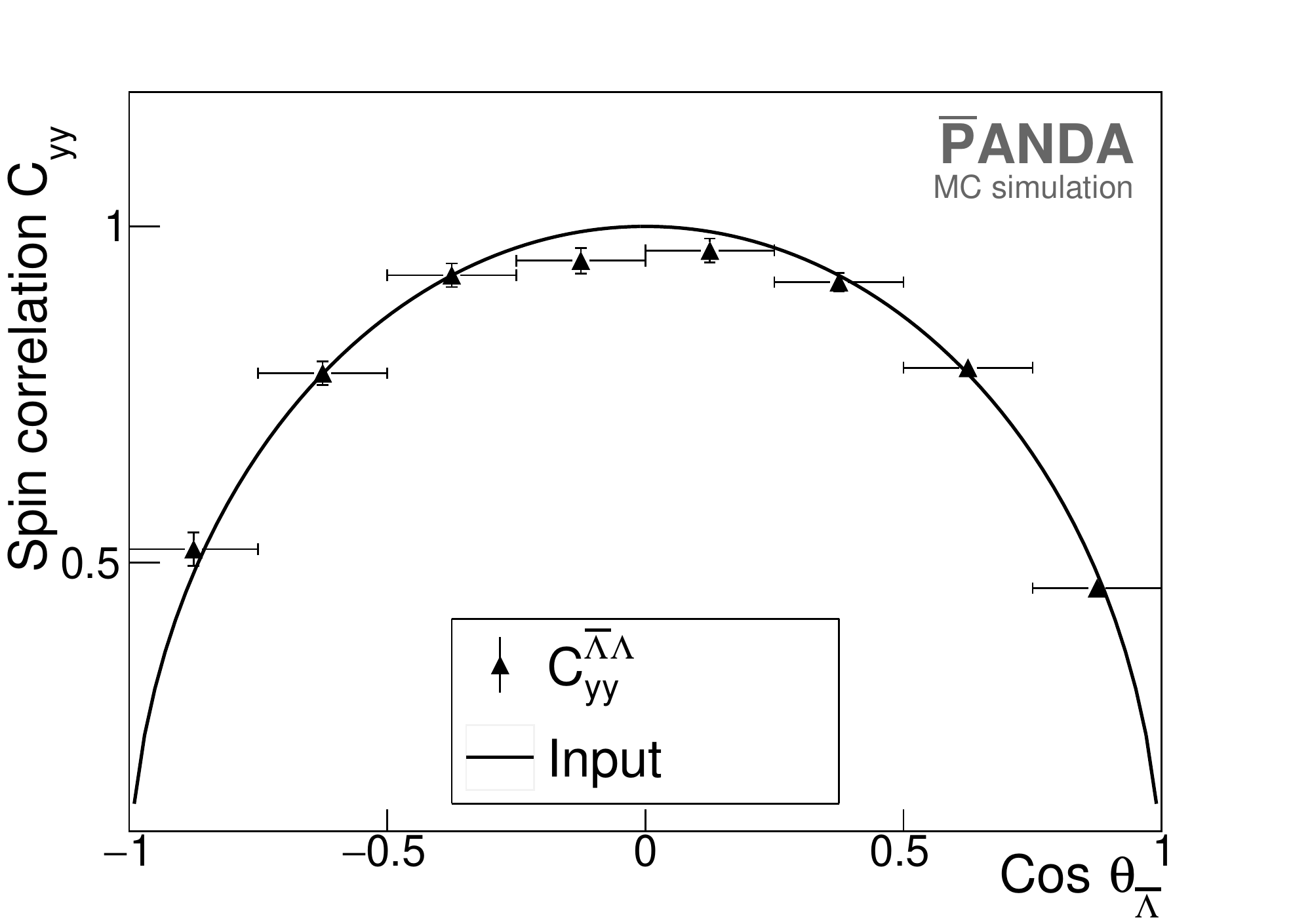}\\
        \includegraphics[width=1.\linewidth]{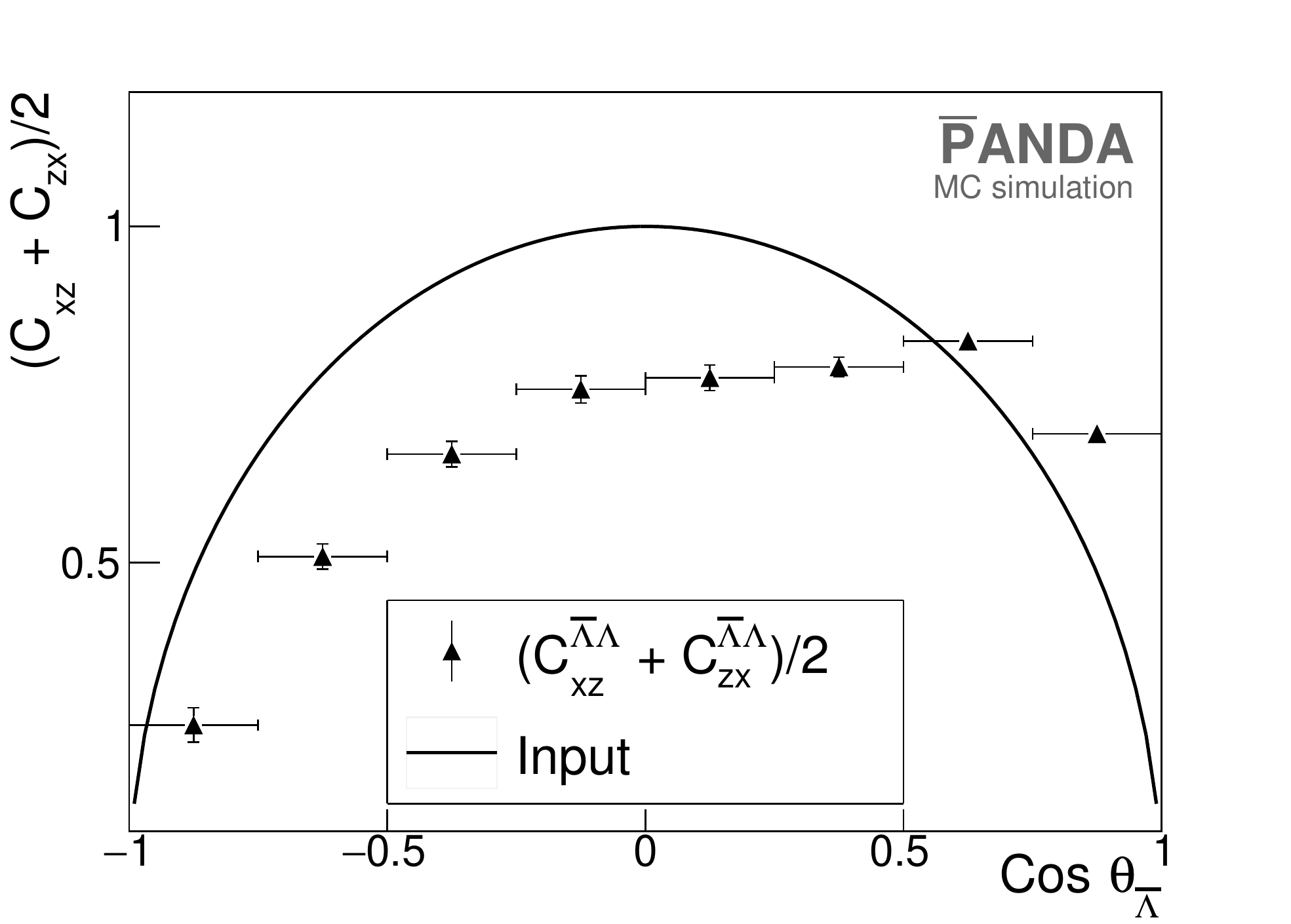}\\
	
    \end{minipage}
	\caption{Spin correlations of the $\bar{\Lambda}\Lambda$ pair produced at $p_{\mathrm{beam}}=1.642$ GeV/c. These observables are estimated using the efficiency independent method. Top-left: $C^{\bar{Y}Y}_{xx}$, top-right: $C^{\bar{Y}Y}_{yy}$ and bottom-left: $C^{\bar{Y}Y}_{zz}$ of the $\bar{\Lambda}\Lambda$ pair. Bottom-right: The average $(C^{\bar{Y}Y}_{xz} + C^{\bar{Y}Y}_{zx})/2$. The vertical error bars represent statistical uncertainties, the horizontal bars the bin widths and the solid curve the input distributions. For $C^{\bar{Y}Y}_{xx}$, $C^{\bar{Y}Y}_{xx}$ and $C^{\bar{Y}Y}_{xx}$ correlations, we do not expect agreement with the input model (solid curve) due to large high order terms. In the case of $C^{\bar{Y}Y}_{yy}$ correlations, higher order terms were found to be negligible.}
	\label{fig:obsll}
\end{figure*}

The diagonal spin correlations, \textit{i.e.} $C^{\bar{Y}Y}_{xx}$, $C^{\bar{Y}Y}_{yy}$ and $C^{\bar{Y}Y}_{zz}$, are shown in Figure \ref{fig:llbarspincorrii}. In the top left and right panels, as well the bottom left, the correlations are extracted using efficiency corrections. The bottom right panel display the average correlation $(C^{\bar{Y}Y}_{xz}+C^{\bar{Y}Y}_{zx})/2$ extracted with the efficiency independent method. In most cases, the reconstructed distributions agree fairly well with the input distributions. However, significant deviations are observed when applying the efficiency independent parameter estimation method, as seen in Fig. \ref{fig:obsll}. This is expected since we concluded in Sect. \ref{sec:effind} that higher order terms could not be neglected in this case. With the efficiency dependent method, all deviations are small and do not follow any obvious trend. Furthermore, it is clear that the statistical precision will be greatly improved compared to the PS185 measurements \cite{PS185164}.

\subsubsection{The $\bar{p}p \to \bar{\Xi}^+\Xi^-$ reaction}
Two studies have been performed at beam momenta of $p_{\mathrm{beam}}=4.6$ GeV/c and $p_{\mathrm{beam}}=7.0$ GeV/c, using $5.86\cdot10^5$ and $4.52\cdot10^5$ reconstructed $\bar{p}p \to \bar{\Xi}^+\Xi^-$ events, respectively. The sample at $p_{\mathrm{beam}}=4.6$ GeV/c can be collected in 21 days while the sample at $p_{\mathrm{beam}}=7.0$ GeV/c requires 55 days of data taking, in line with the planned 80 days campaign at an energy around the $X(3872)$ mass. Here, we assume a luminosity of $10^{31}$cm$^{-2}$s$^{-1}$, which will be achievable at these energies during the first phase of data taking with PANDA. 

\begin{figure*}[ht]
    \centering
    \begin{minipage}{.5\textwidth}
        \centering
            \includegraphics[width=1.\linewidth]{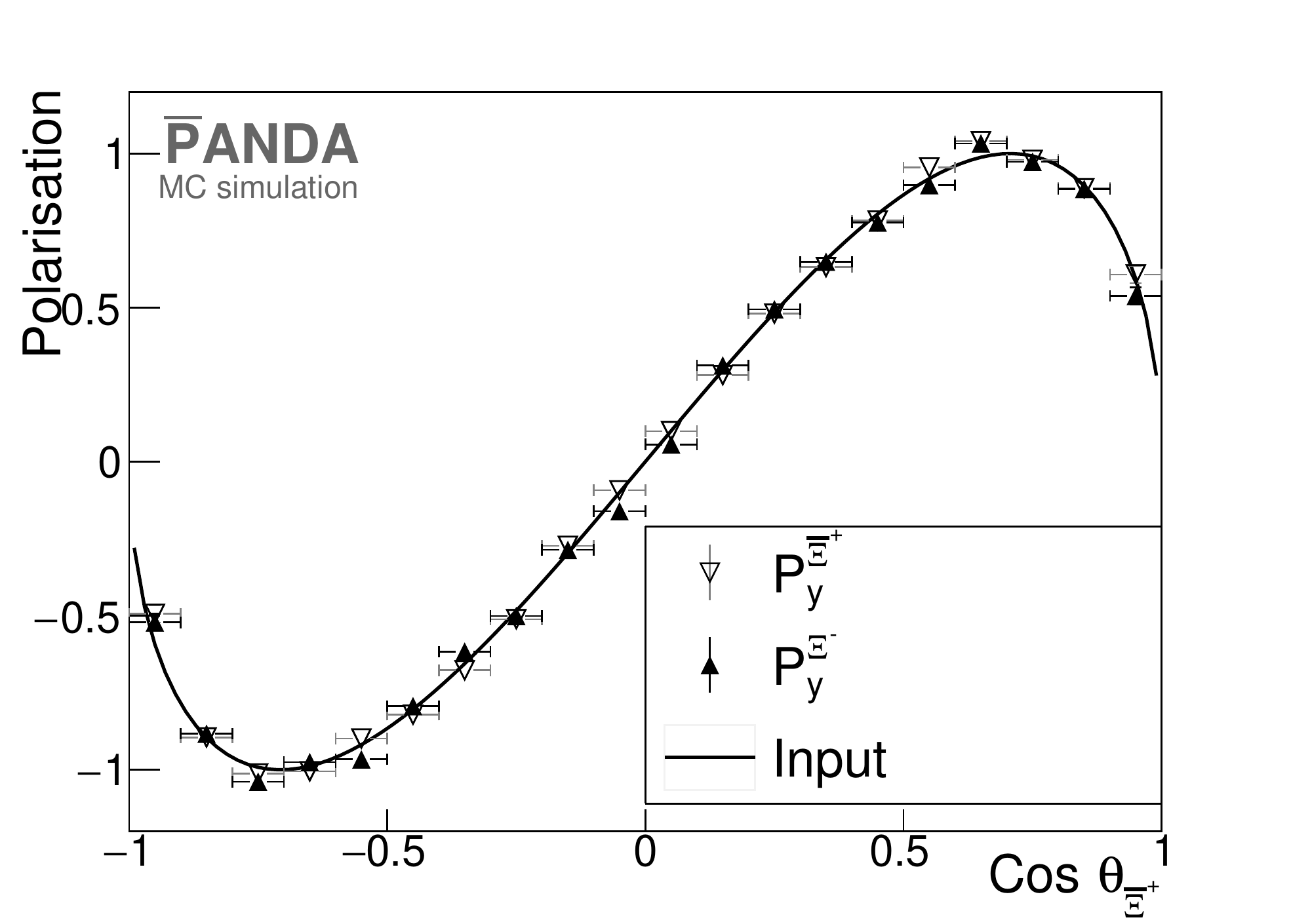}\\
            \includegraphics[width=1.\linewidth]{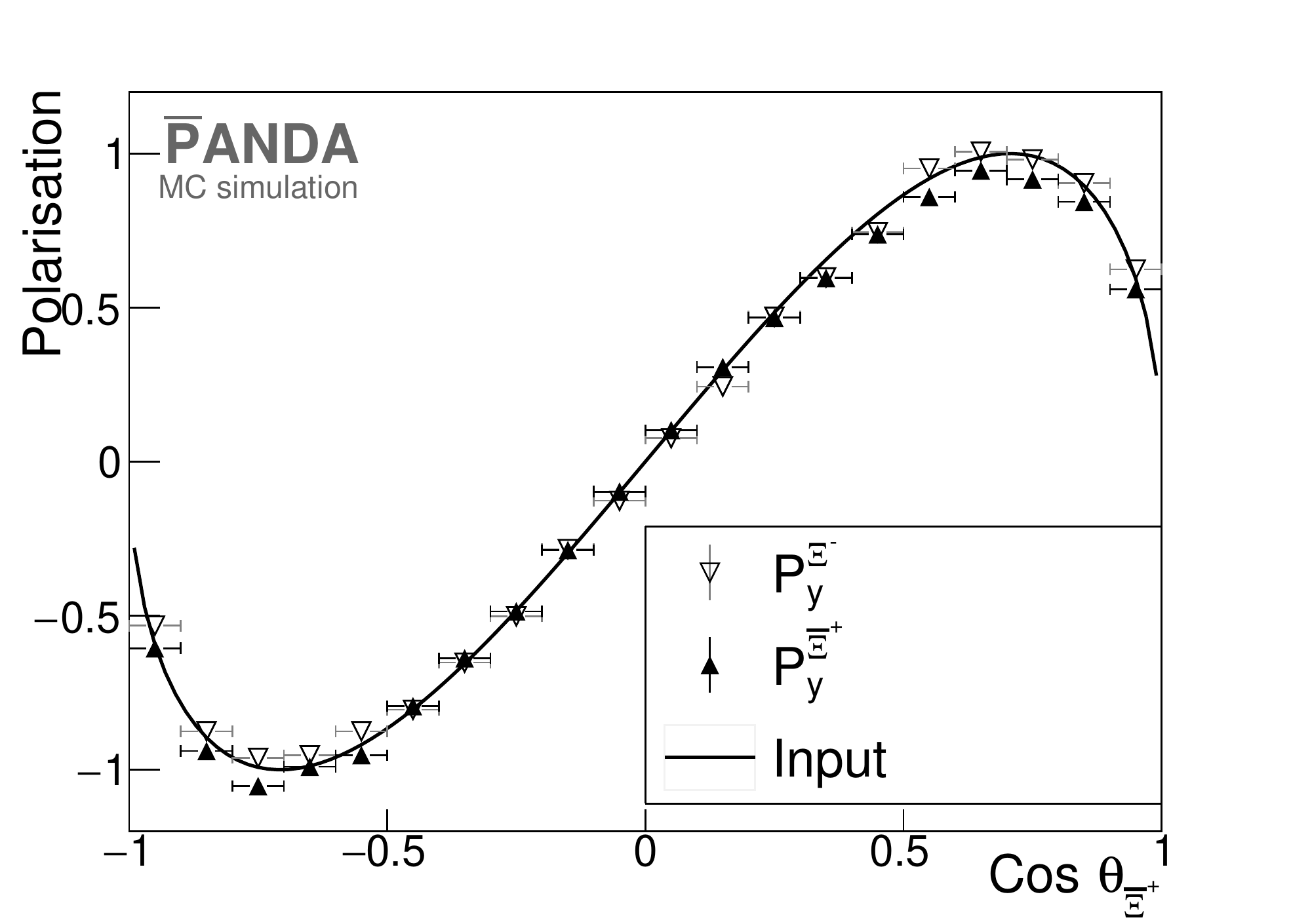}

    \end{minipage}%
    \begin{minipage}{0.5\textwidth}
        \centering
            \includegraphics[width=1.\linewidth]{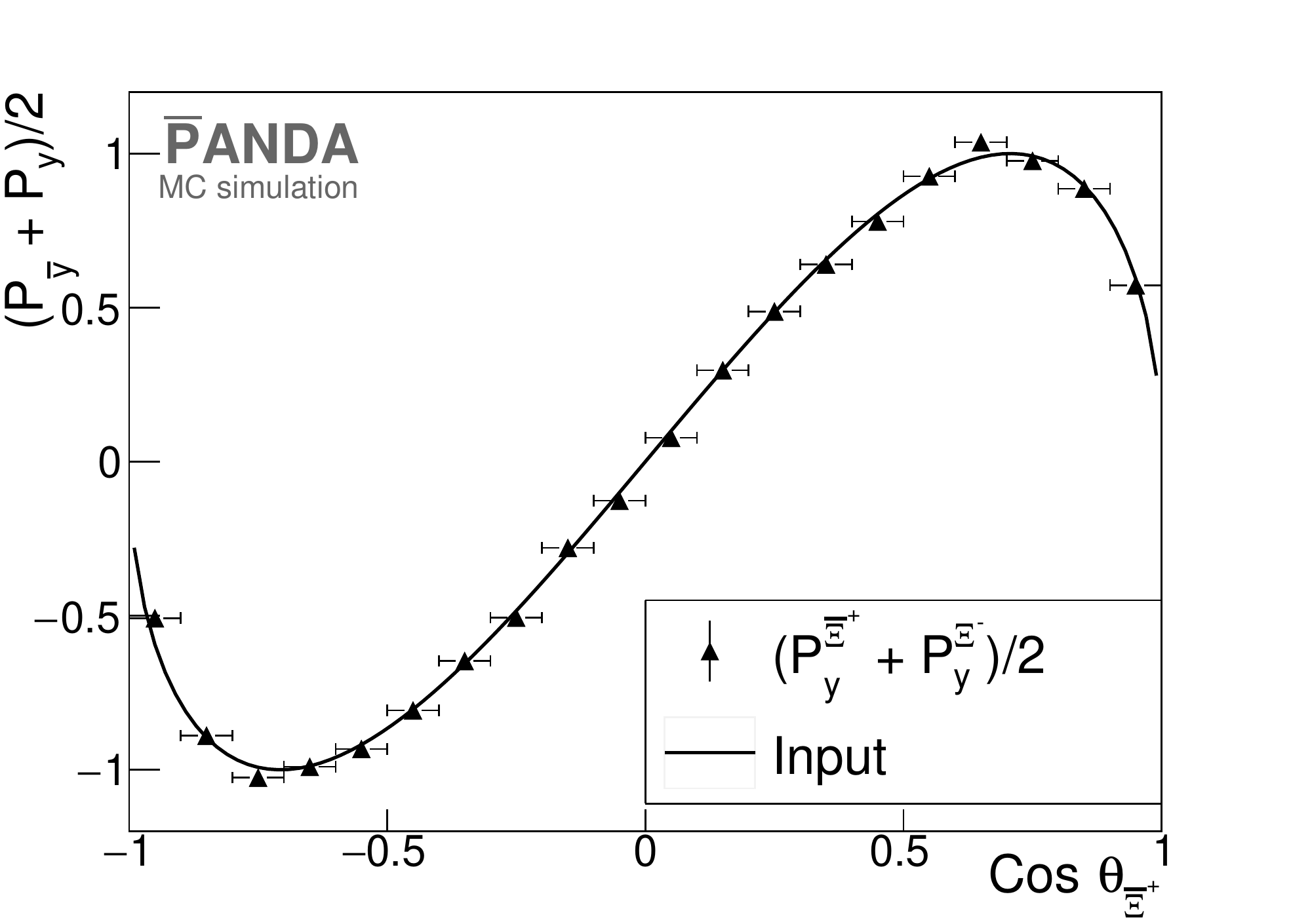}\\
            \includegraphics[width=1.\linewidth]{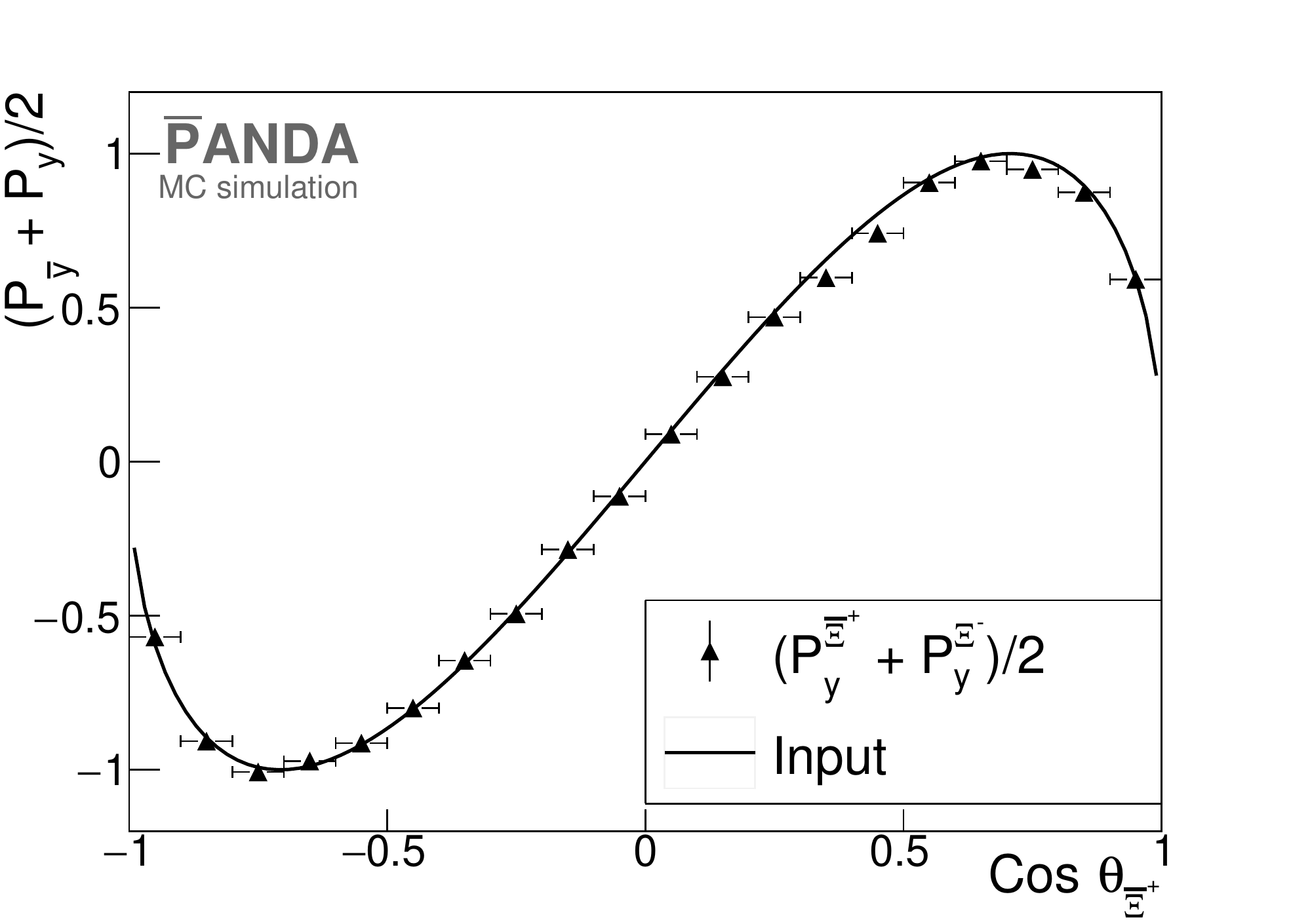}
    \end{minipage}
	\caption{Top-left: Reconstructed polarisation of the $\bar{\Xi}^+$ (black) and the $\Xi^-$ (open) at $p_{\mathrm{beam}}=4.6$ GeV/c using the efficiency dependent method with 2D efficiency matrices. Top-right: Average values of the two reconstructed polarisations. Bottom-left: Polarisations reconstructed with the efficiency independent method. Bottom-right: The average of the polarisations reconstructed with the efficiency independent method. The vertical error bars represent statistical uncertainties, the horizontal bars the bin widths and the solid curves the input model.}
	\label{fig:xixibarpol46}
\end{figure*}

In Fig. \ref{fig:xixibarpol46}, the polarisation at 4.6 GeV/$c$ of the $\Xi^-$ and the $\bar{\Xi}^+$ are shown individually (top-left panel) and averaged (top-right panel) for efficiency corrected data. The agreement between $\Xi^-$ and $\bar{\Xi}^+$ as well as between the input distributions and the reconstructed ones, is excellent and the statistical uncertainties are small. Also when using the efficiency independent method, there is good agreement between reconstructed data and the input model (bottom-left and bottom-right panels). This is expected since the simulations showed that all criteria are fulfilled for this reaction at this beam momentum.

\begin{figure*}[ht]
    \centering
    \begin{minipage}{.5\textwidth}
        \centering
        \includegraphics[width=1.\linewidth]{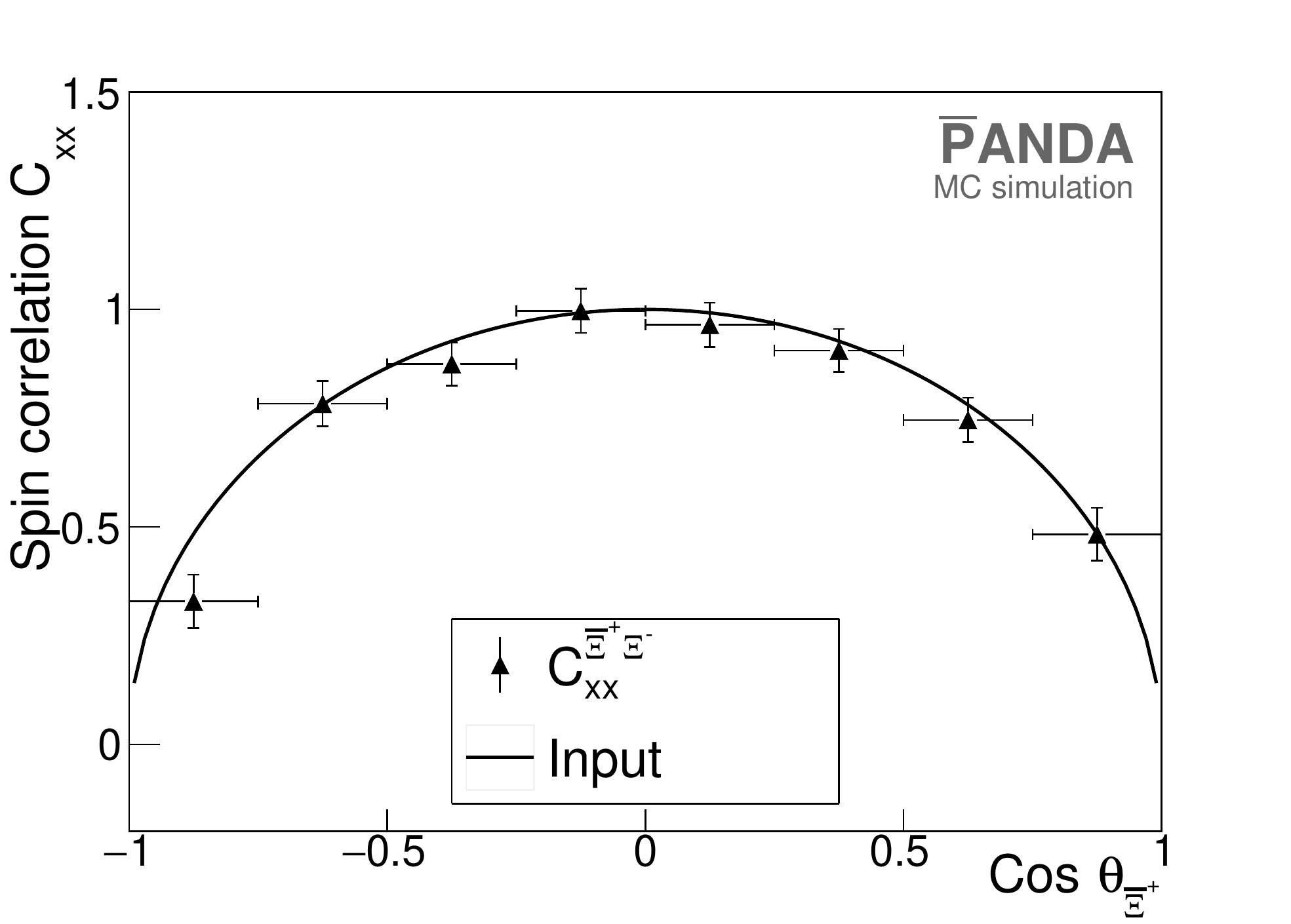}\\
        \includegraphics[width=1.\linewidth]{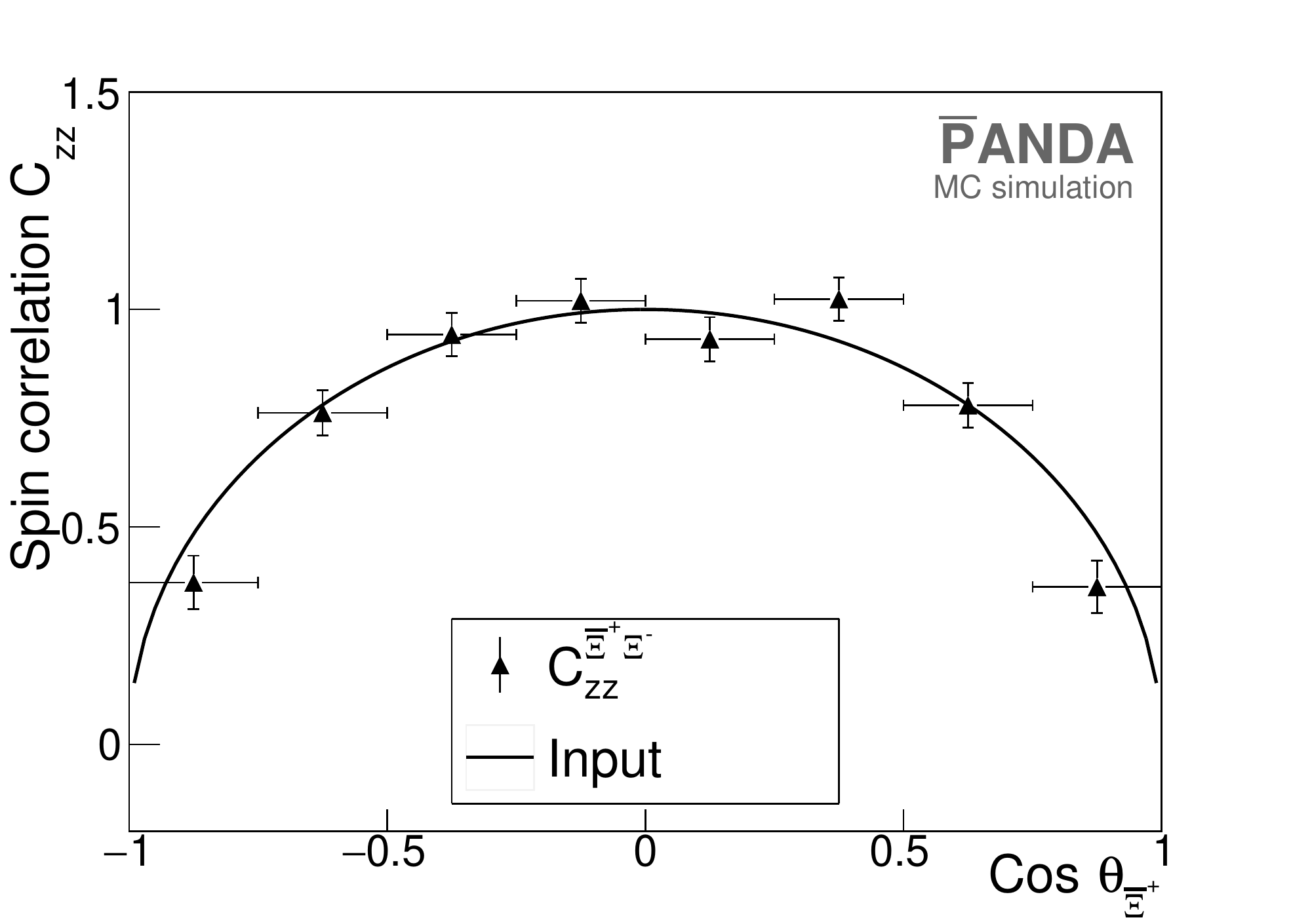}
    \end{minipage}%
    \begin{minipage}{0.5\textwidth}
        \centering
        \includegraphics[width=1.\linewidth]{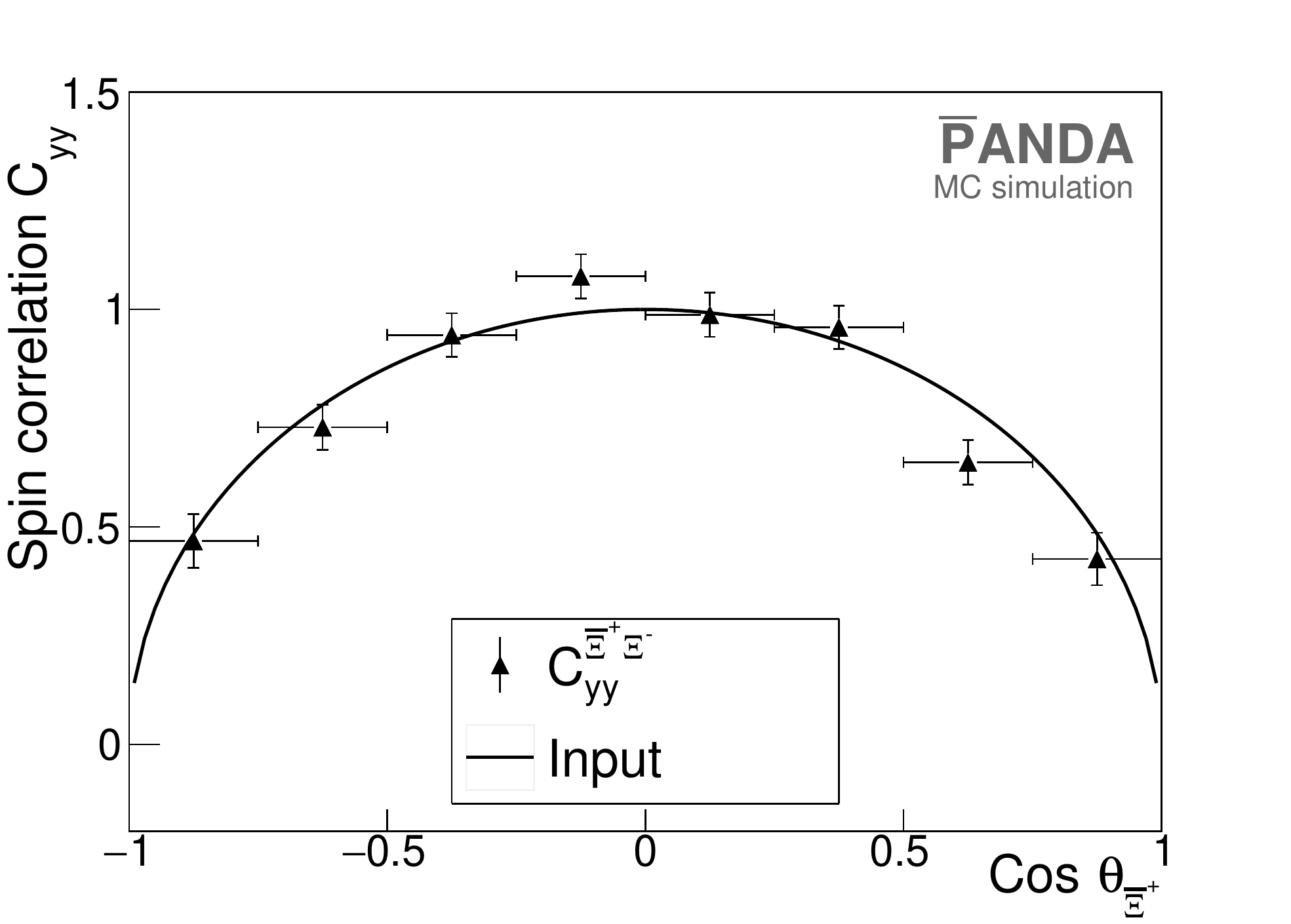}\\
        \includegraphics[width=1.\linewidth]{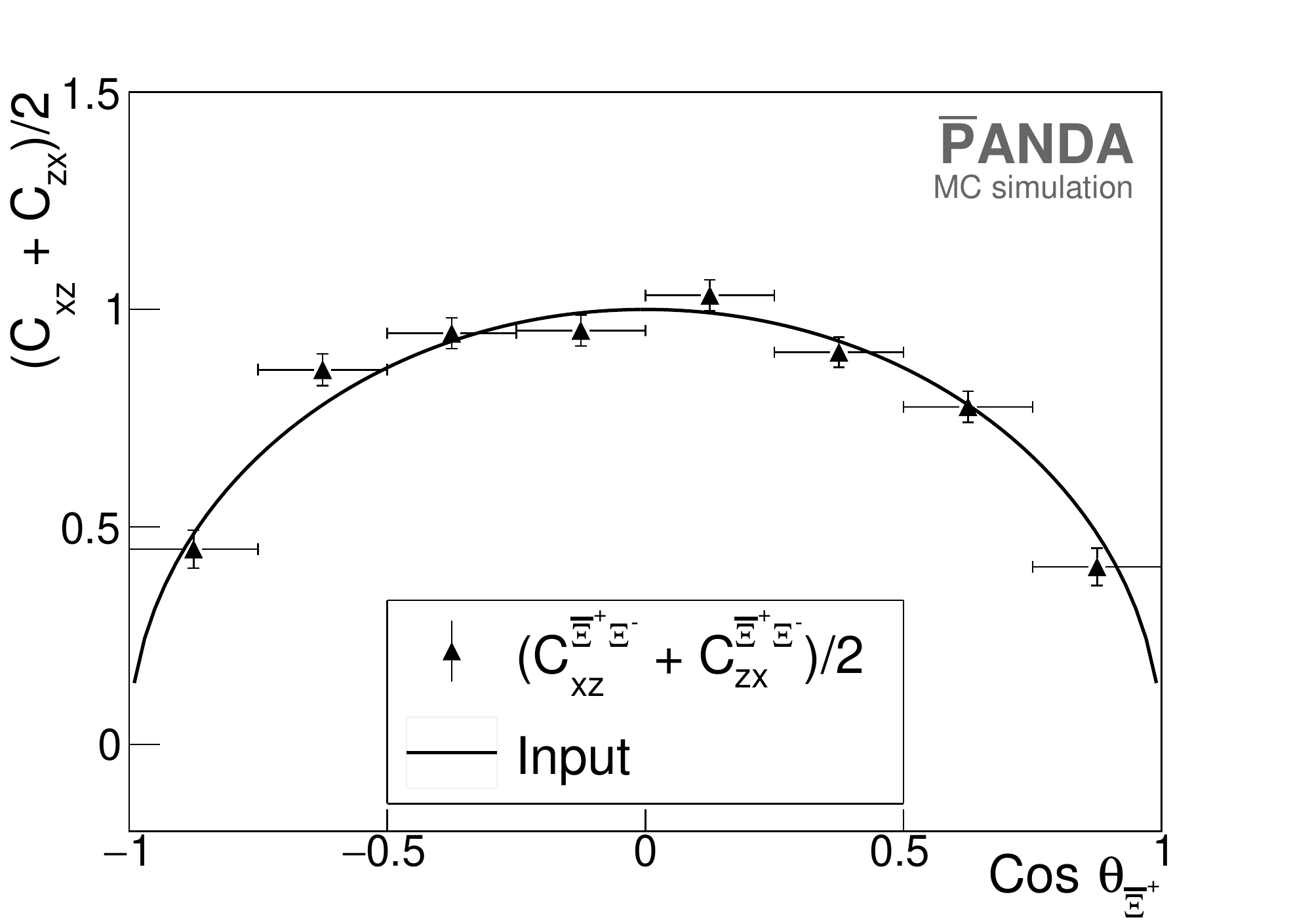}
    \end{minipage}
	\caption{Reconstructed spin correlations of the $\bar{\Xi}^+\Xi^-$ pair at $p_{\mathrm{beam}}=4.6$ GeV/c. Top-left: the $C^{\bar{Y}Y}_{xx}$ correlation. Top-right: $C^{\bar{Y}Y}_{yy}$. Bottom-left: $C^{\bar{Y}Y}_{zz}$ . Bottom-right: the reconstructed average $(C^{\bar{Y}Y}_{xz} + C^{\bar{Y}Y}_{zx})/2$. The spin correlations are reconstructed using the efficiency independent method. The vertical error bars represent statistical uncertainties, the horizontal bars the bin widths and the solid curve the input model.}
	\label{fig:xixibar46spincorr}
\end{figure*}

\begin{figure*}[ht]
    \centering
    \begin{minipage}{.5\textwidth}
        \centering
        \includegraphics[width=1.\linewidth]{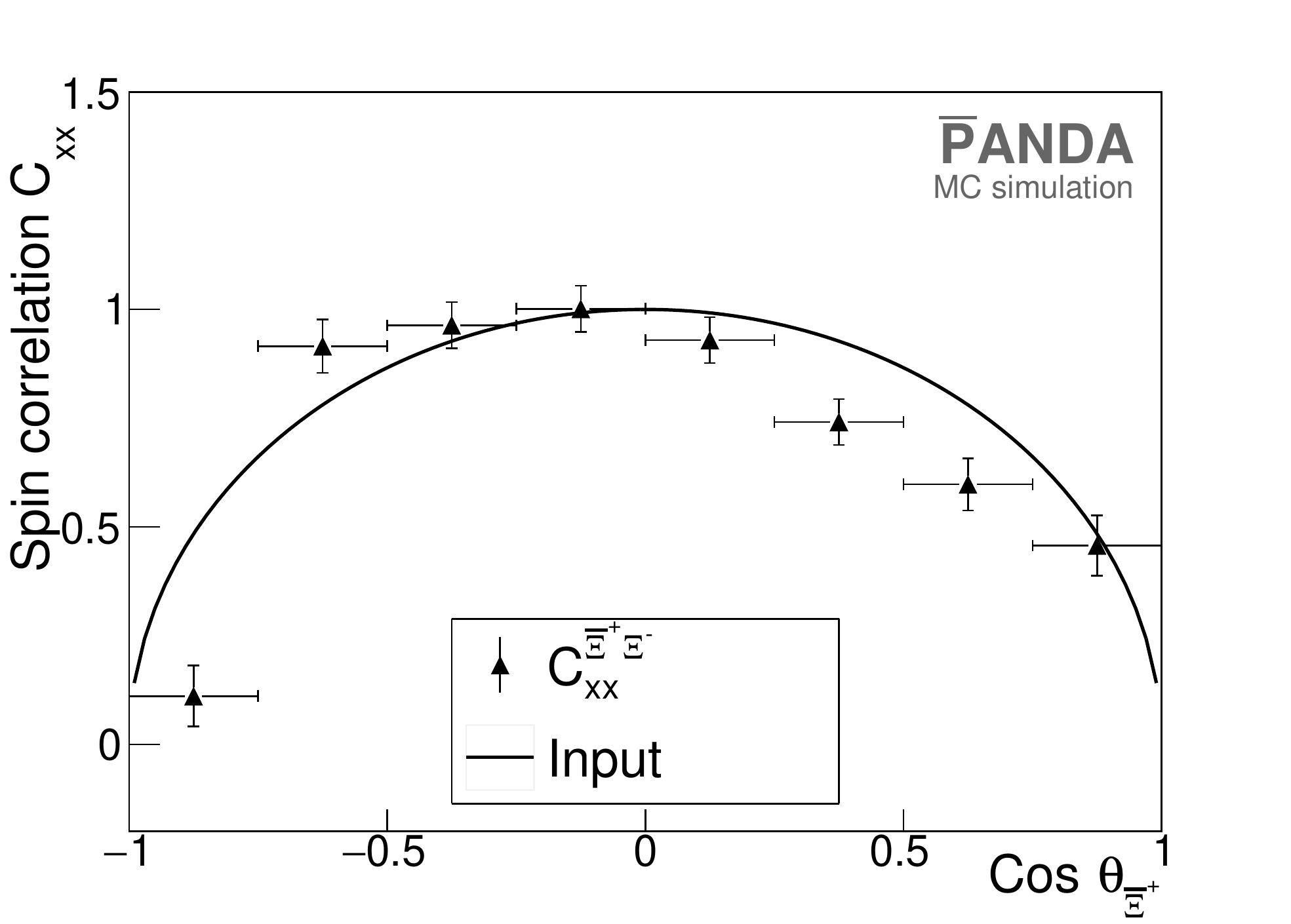}
    \end{minipage}%
    \begin{minipage}{0.5\textwidth}
        \centering
        \includegraphics[width=1.\linewidth]{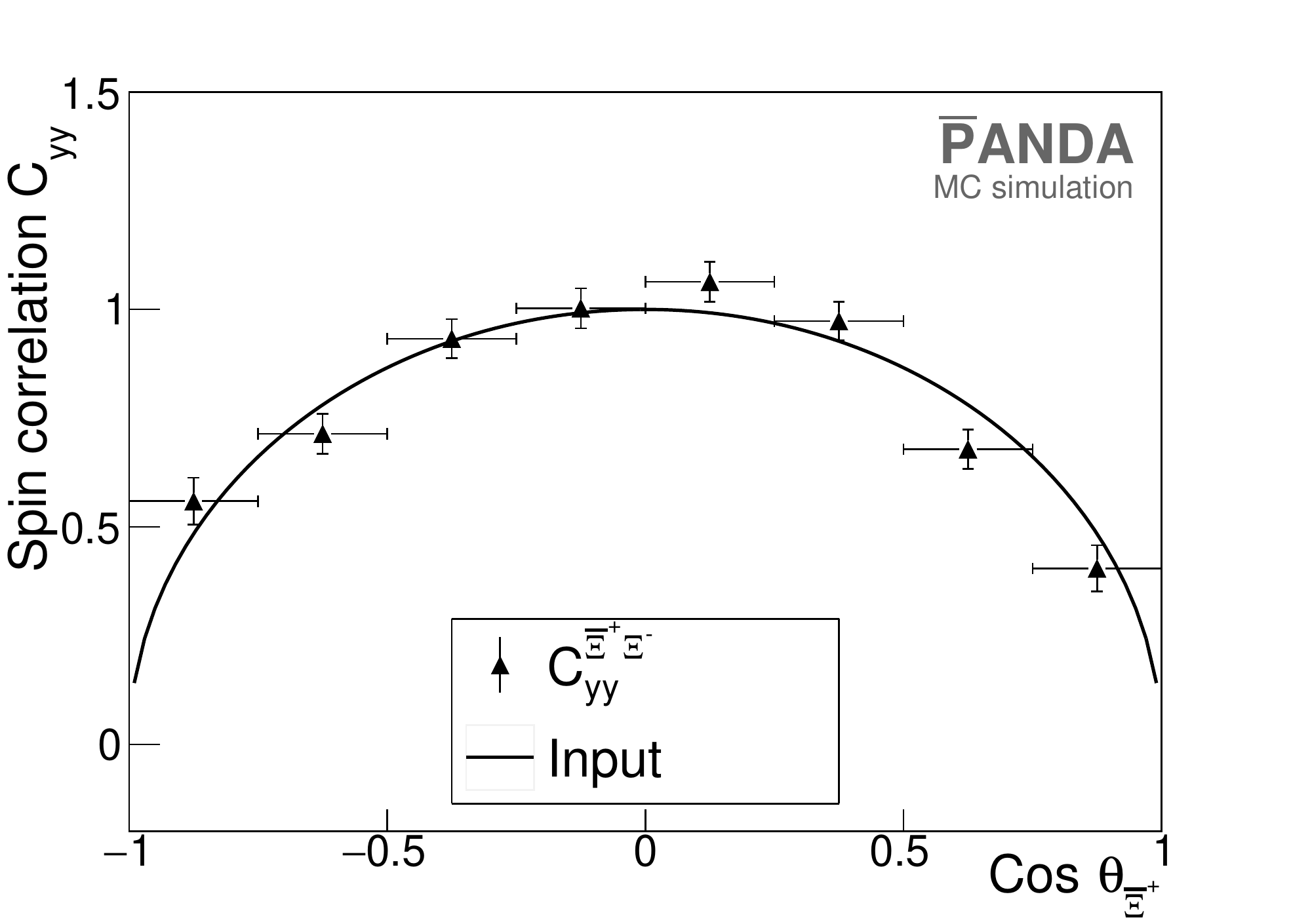}
    \end{minipage}
	\caption{Spin correlations of the $\bar{\Xi}^+\Xi^-$ pair at $p_{\mathrm{beam}}=4.6$ GeV/c, reconstructed with the efficiency independent method. Left: the $C^{\bar{Y}Y}_{xx}$ correlation. Right: the $C^{\bar{Y}Y}_{yy}$ correlation. The vertical error bars represent statistical uncertainties, the horisontal the bin widths and the red curves the input model.}
	\label{fig:obs46}
\end{figure*}

The spin correlations $C^{\bar{Y}Y}_{xx}$, $C^{\bar{Y}Y}_{yy}$, $C^{\bar{Y}Y}_{zz}$ and the average $(C^{\bar{Y}Y}_{xz}+C^{\bar{Y}Y}_{zx})/2$ are shown in Fig. \ref{fig:xixibar46spincorr} for the same beam momentum. The agreement between the input distributions and the reconstructed distributions is good. Fig. \ref{fig:obs46} displays two examples of spin correlations reconstructed with the efficiency independent method. The $C^{\bar{Y}Y}_{yy}$ correlation agrees well with the input model whereas some deviations are seen in the case of $C^{\bar{Y}Y}_{xx}$, despite the fact that the criteria outlined in Sect. \ref{sec:effind} are fulfilled. This shows that this observable is more sensitive to the efficiency than the $C^{\bar{Y}Y}_{yy}$ and that the efficiency independent method has to be used with caution. 

\begin{figure*}[ht]
    \centering
    \begin{minipage}{.5\textwidth}
        \centering
            \includegraphics[width=1.\linewidth]{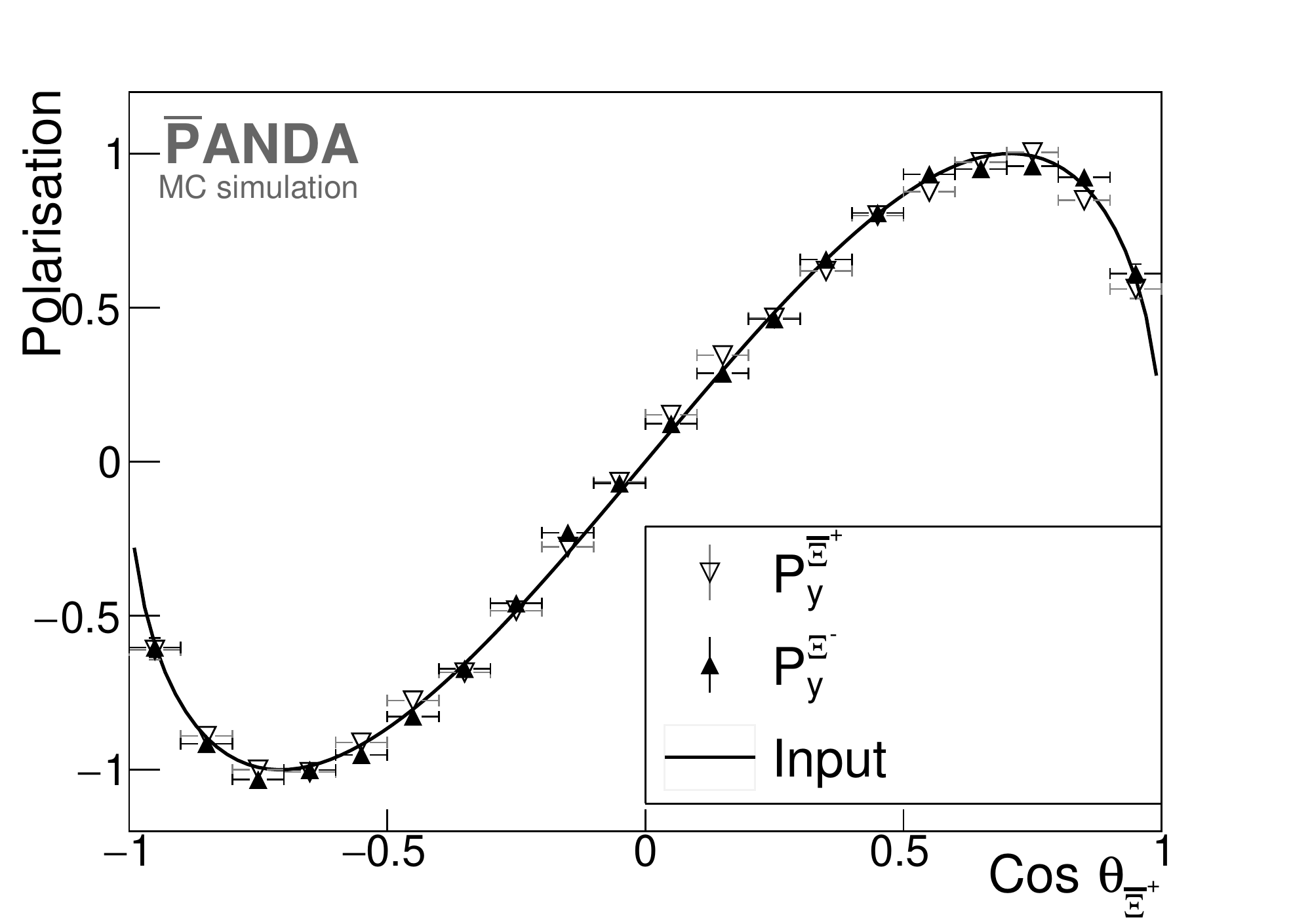}
    \end{minipage}%
    \begin{minipage}{0.5\textwidth}
        \centering
            \includegraphics[width=1.\linewidth]{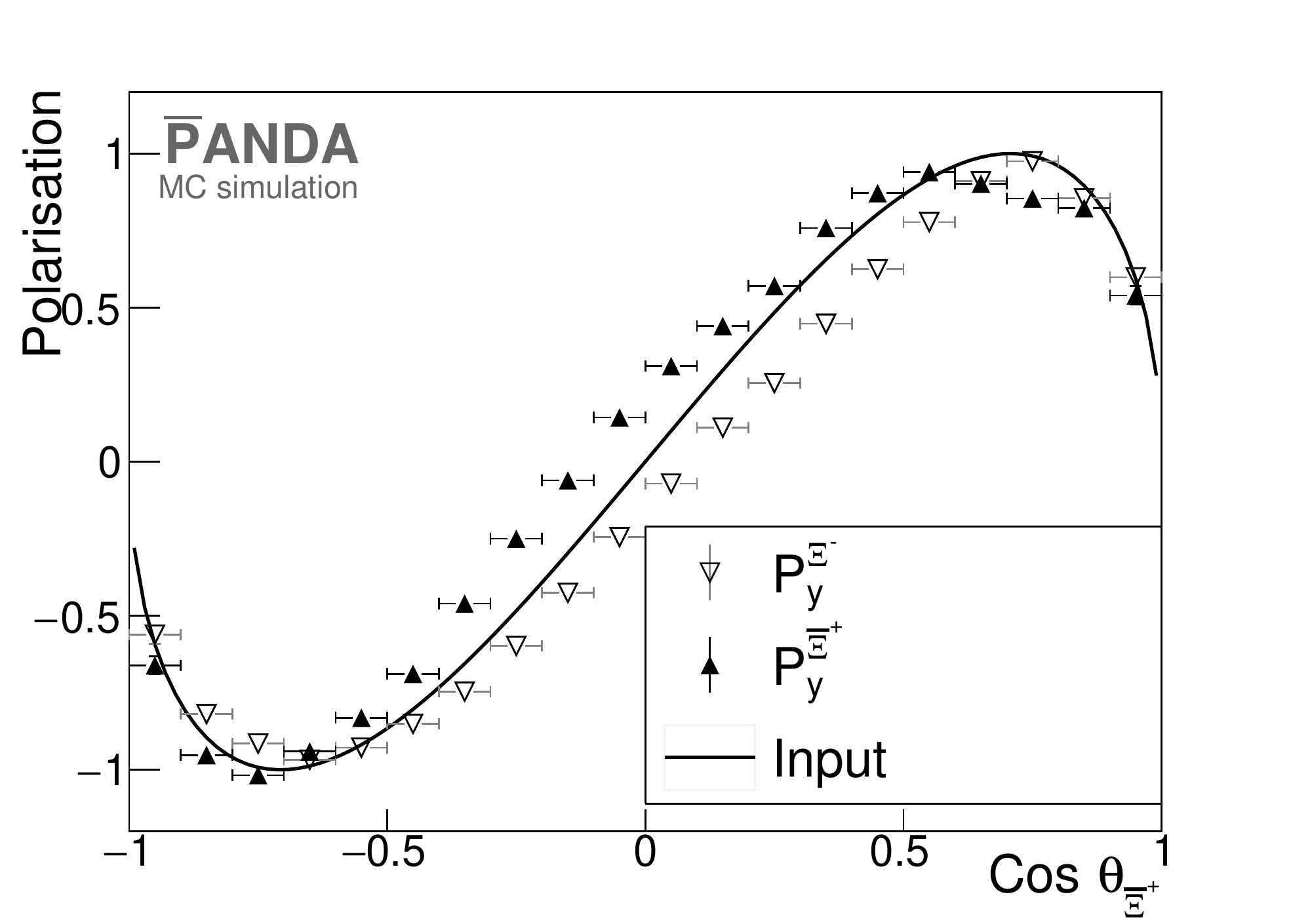}
    \end{minipage}
	\caption{The polarisation of the $\bar{\Xi}^+$ (black) and the $\Xi^-$ (open) hyperons at $p_{\mathrm{beam}}=7.0$ GeV/c. Left: data reconstructed using the efficiency dependent method. Right: data reconstructed using the efficiency independent method. The vertical error bars represent  statistical uncertainties, the horizontal bars the bin widths and the solid curve the input model.}
	\label{fig:xixibarpol7}
\end{figure*}

In Fig. \ref{fig:xixibarpol7}, the polarisations of the $\bar{\Xi}^+$ and $\Xi^-$ at 7.0 GeV/$c$ are shown. In the left panel, where the efficiency dependent method has been used, we see that the reconstructed polarisations agree well with the input model. In the right panel, the efficiency independent method is used. Here, some disagreement is observed with respect to the input model, as expected since one of the criteria in Sect. \ref{sec:effind} is not fulfilled. Furthermore, we observe that the $\bar{\Xi}^+$ polarisation disagrees with the $\Xi^-$ polarisation. This shows that a comparison between hyperon and antihyperon observables serve as a consistency check.

\begin{figure*}[ht]
    \centering
    \begin{minipage}{.5\textwidth}
        \centering
        \includegraphics[width=1.\linewidth]{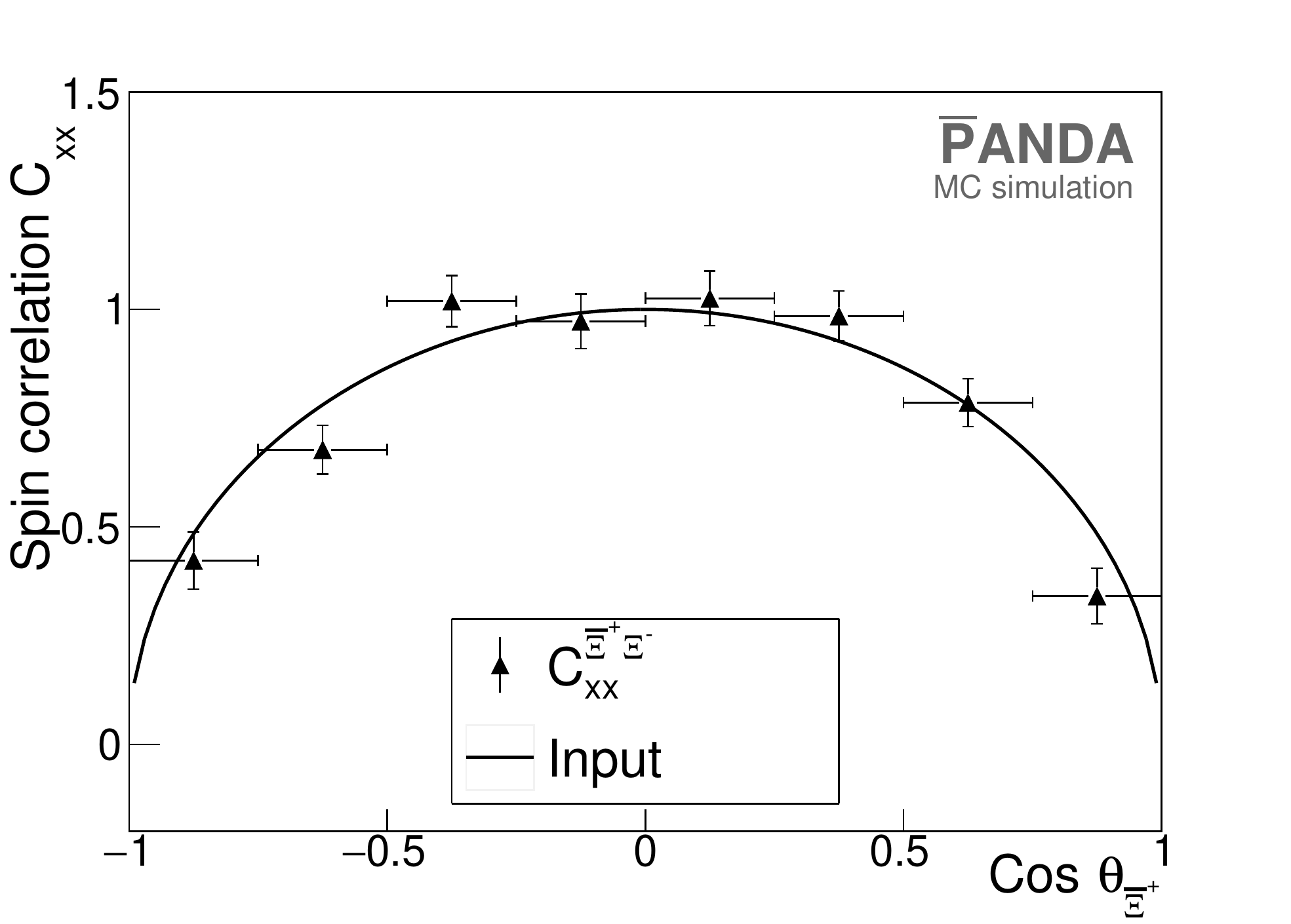}\\
        \includegraphics[width=1.\linewidth]{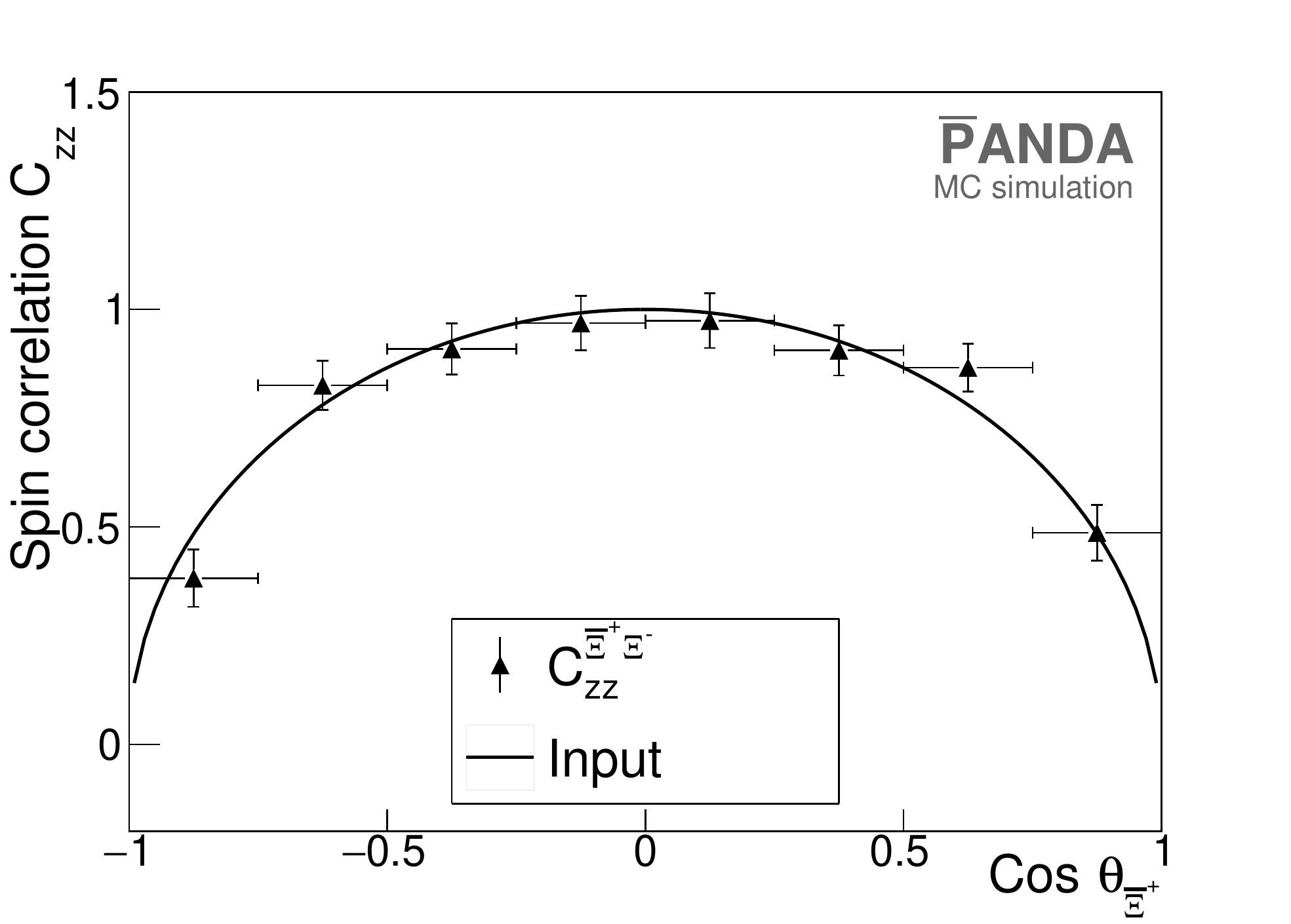}
    \end{minipage}%
    \begin{minipage}{0.5\textwidth}
        \centering
        \includegraphics[width=1.\linewidth]{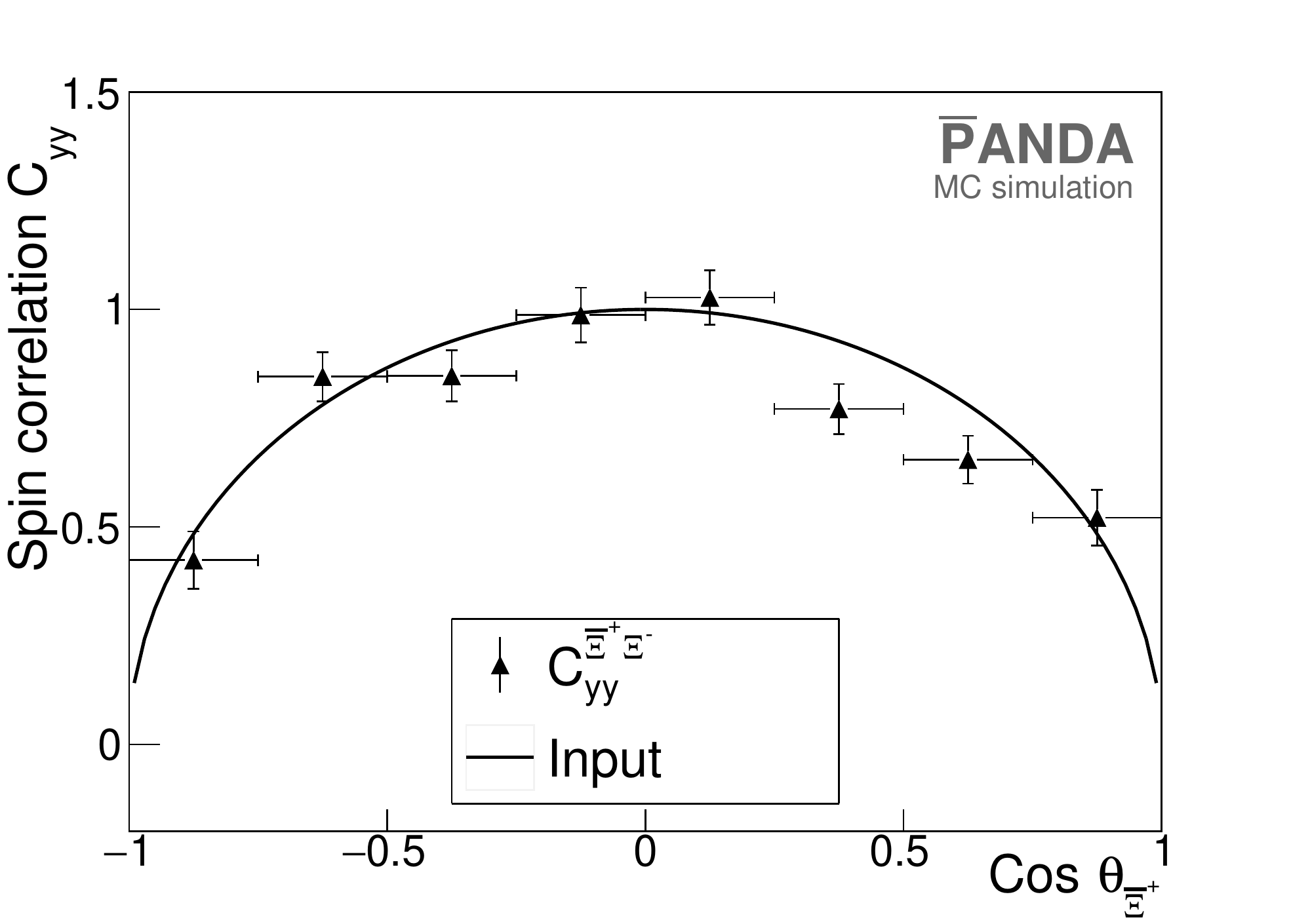}\\
        \includegraphics[width=1.\linewidth]{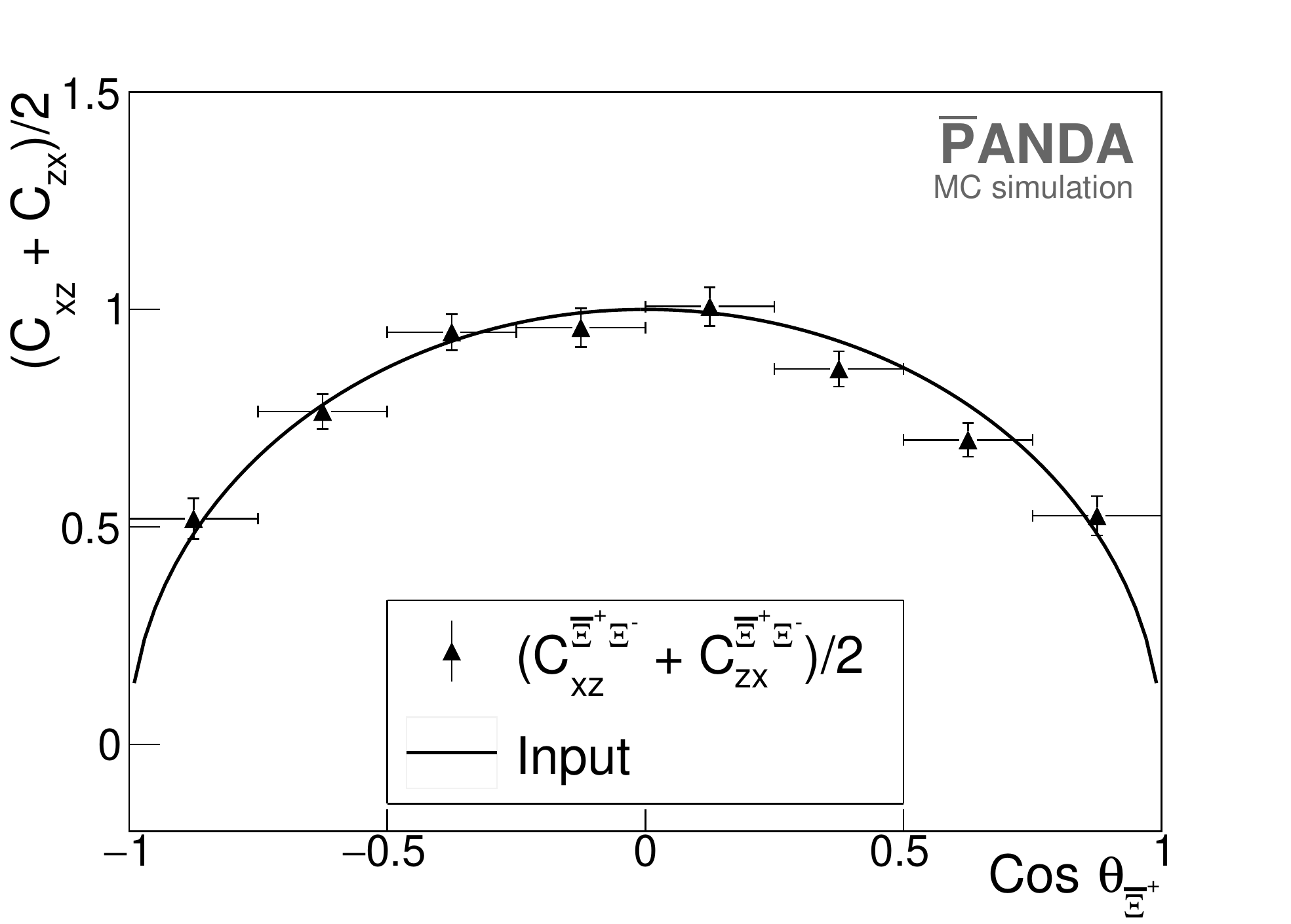}
    \end{minipage}
	\caption{Reconstructed spin correlations (top-left) $C^{\bar{Y}Y}_{xx}$, (top-right) $C^{\bar{Y}Y}_{yy}$ and (bottom-left) $C^{\bar{Y}Y}_{zz}$ of the $\bar{\Xi}^+\Xi^-$ pair at $p_{\mathrm{beam}}=7.0$ GeV/c. (bottom-right) Reconstructed average $(C^{\bar{Y}Y}_{xz} + C^{\bar{Y}Y}_{zx})/2$. The spin correlations are reconstructed at $p_{\mathrm{beam}}=7.0$ GeV/c using acceptance corrections. The vertical error bars represent statistical uncertainties, the horizontal bars the bin widths and the solid curves the input model.}
	\label{fig:xixibar7spincorr}
\end{figure*}

\begin{figure*}[ht]
    \centering
    \begin{minipage}{.5\textwidth}
        \centering
        \includegraphics[width=1.\linewidth]{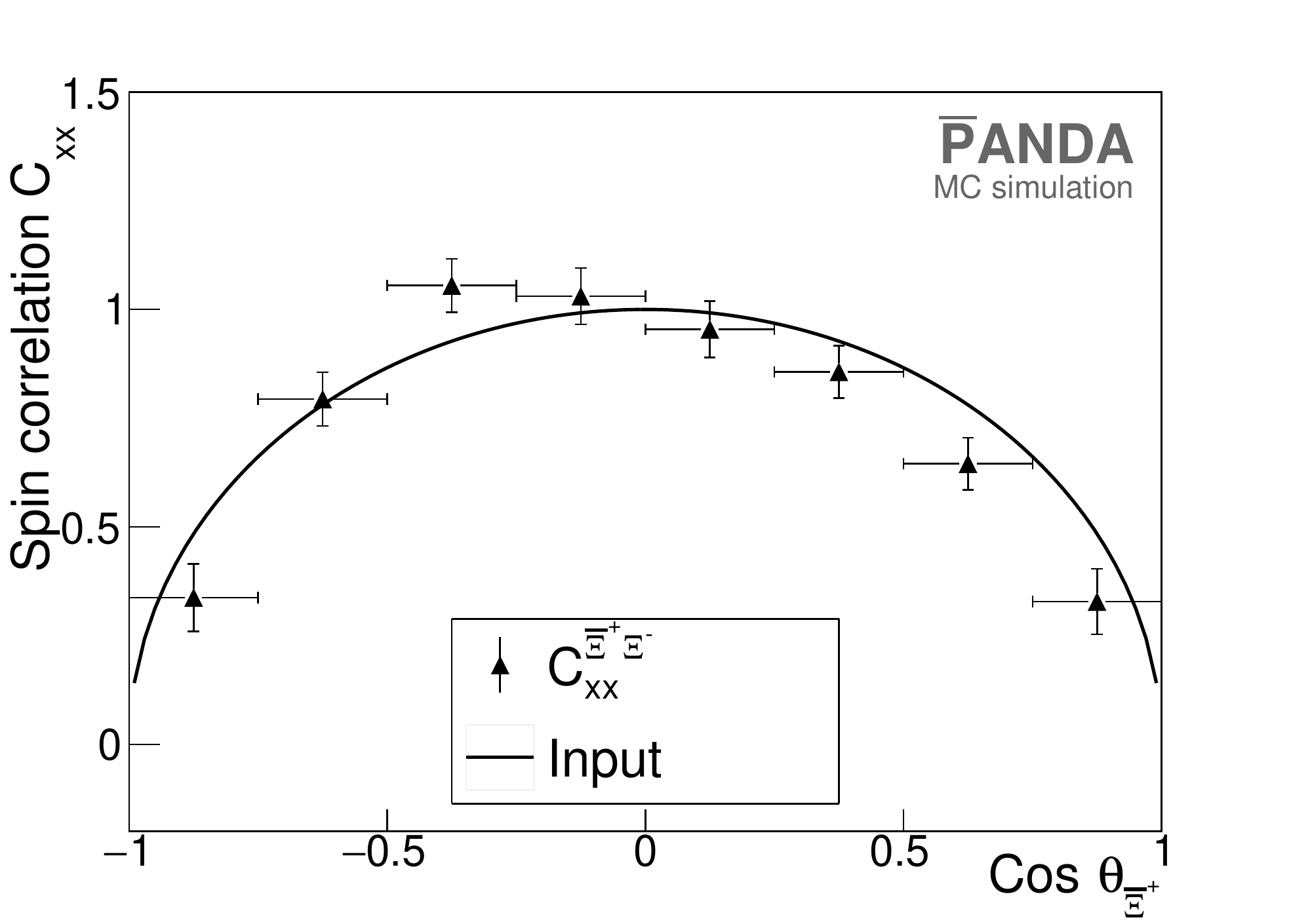}
    \end{minipage}%
    \begin{minipage}{0.5\textwidth}
        \centering
        \includegraphics[width=1.\linewidth]{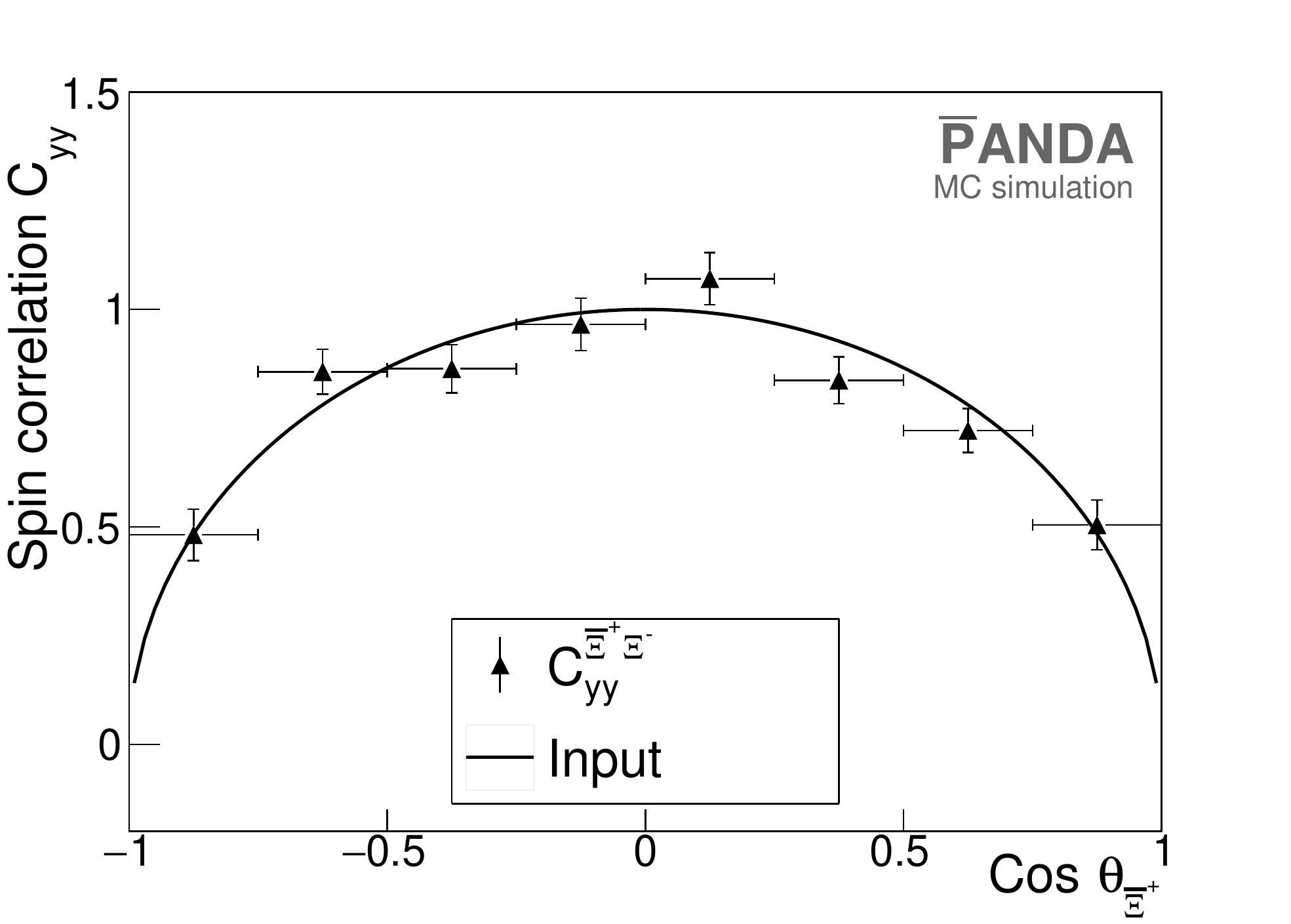}
    \end{minipage}
	\caption{Reconstructed spin correlations of the $\bar{\Xi}^+\Xi^-$ pair at $p_{\mathrm{beam}}=7.0$ GeV/$c$, reconstructed with the efficiency independent method. Left:The $C^{\bar{Y}Y}_{xx}$ correlation. Right: The $C^{\bar{Y}Y}_{yy}$ correlation. The solid curves represent the input model.}
	\label{fig:obs7}
\end{figure*}

In Fig. \ref{fig:xixibar7spincorr}, the spin correlations of the $\bar{\Xi}^+\Xi^-$ pair are shown at 7.0 GeV/$c$, reconstructed with the efficiency dependent method. The reconstructed distributions agree with the input ones, indicating that the reconstruction and analysis procedure do not impose any bias. In Fig. \ref{fig:obs7}, the $C^{\bar{Y}Y}_{xx}$ and $C^{\bar{Y}Y}_{yy}$ spin correlations are shown, reconstructed with the efficiency independent method. Even in this case, the reconstructed distributions agree well with the input models.

\begin{figure*}[ht]
    \centering
    \begin{minipage}{.5\textwidth}
        \centering
            \includegraphics[width=1.\linewidth]{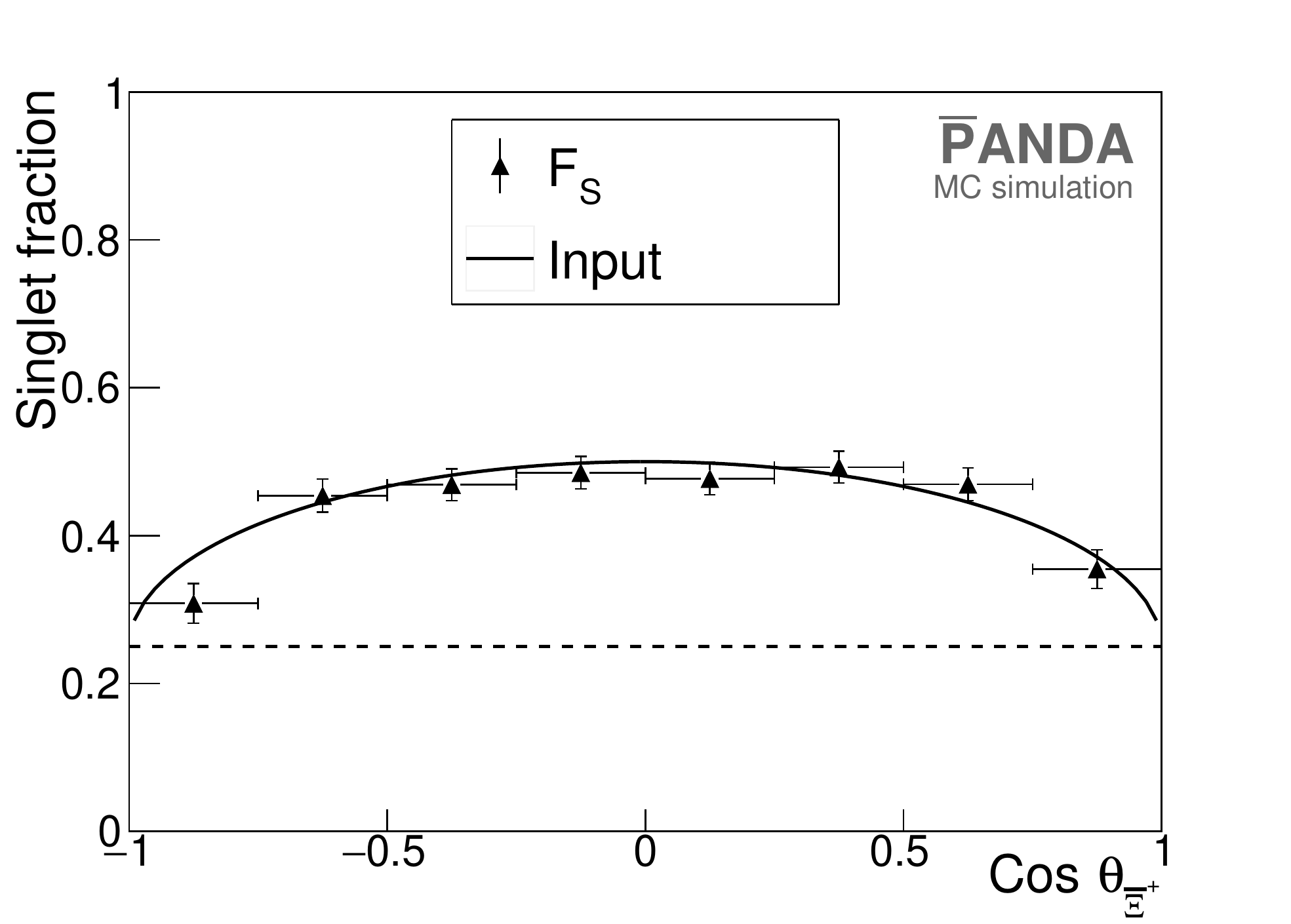}

    \end{minipage}%
    \begin{minipage}{0.5\textwidth}
        \centering
            \includegraphics[width=1.\linewidth]{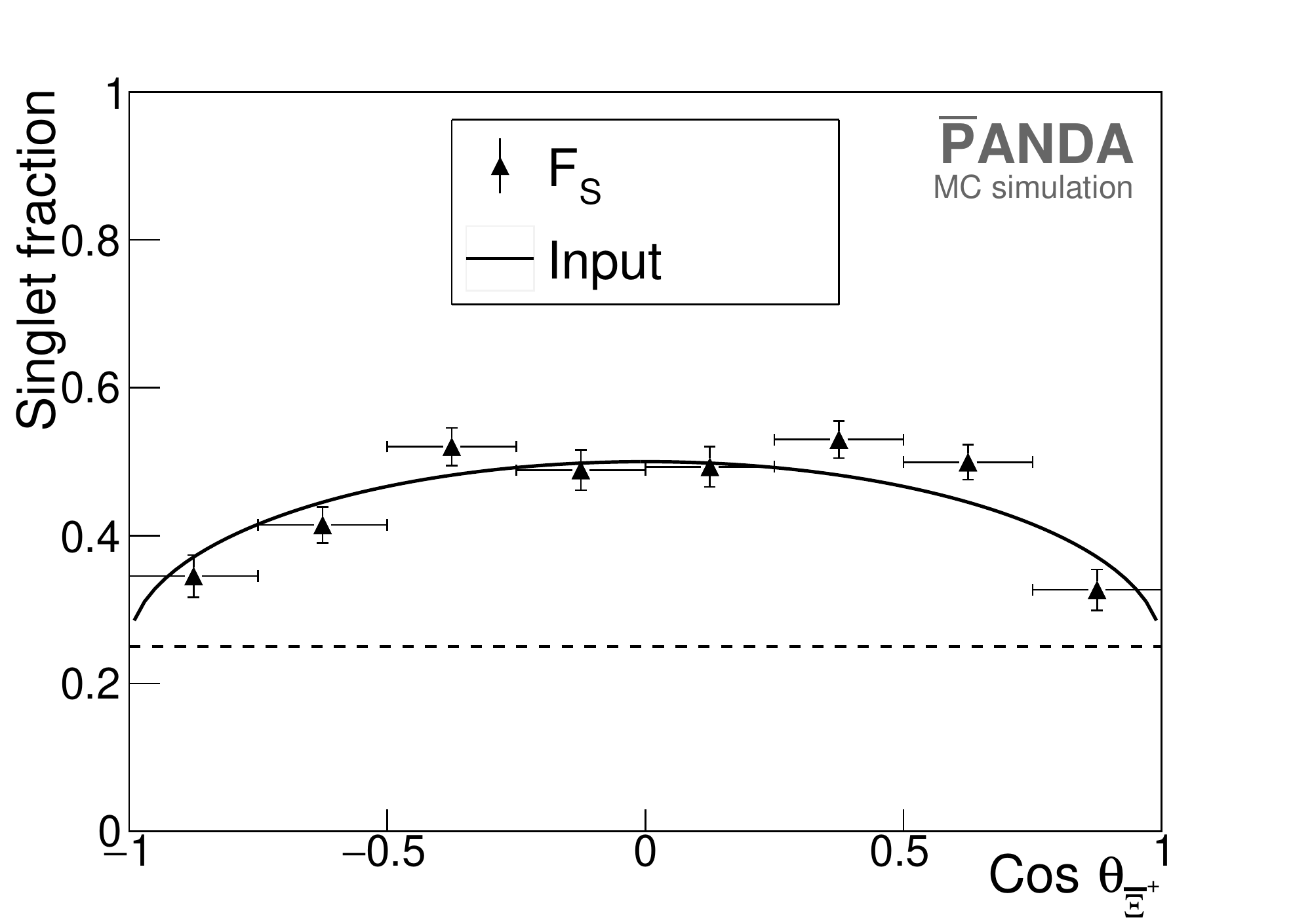}
    \end{minipage}
	\caption{Reconstructed singlet fractions at $p_{\mathrm{beam}}=4.6$ GeV/c (left) and $p_{\mathrm{beam}}=7.0$ GeV/c (right). The vertical errorbars are statistical uncertainties only. The horizontal bars are the bin widths.}
	\label{fig:xixibarsf}
\end{figure*}

The singlet fractions of the $\bar{\Xi}^+\Xi^-$ pair, calculated from the spin correlations according to Eq. (\ref{eq:spinfrac}), are shown in Fig. \ref{fig:xixibarsf} as a function of the $\bar{\Xi}^+$ scattering angle. The results show that the prospects of measuring the singlet fraction, and thereby establish in which spin state the produced $\bar{\Xi}^+\Xi^-$ is, are very good. It will also be possible to test the predictions from Ref. \cite{XiMEX}.

\subsection{Systematic uncertainties}
\label{sec:systematic}

It is hard to evaluate systematic uncertainties before the experiment is taken into operation, since effects such as trigger efficiencies or imperfections in tracking or in the Monte Carlo implementation of the detector are difficult to estimate without real data. 

In the feasibility study of electromagnetic form factors in PANDA \cite{EMFFPanda} as well as in the simulation of the foreseen energy scan around the $X(3872)$ \cite{xscan}, uncertainties in the estimated luminosity and background constitute the most important sources of systematics. While being very important in cross section measurements, effects from the uncertainty in the luminosity are expected to be negligible in measurements of differential distributions. This is because such uncertainties should be uniformly distributed over the angles of the final state particles. Regarding the background, the displaced decay vertices of hyperons result in a very distinct event topology that allows for a very strong suppression of background. Furthermore, the cross section of the hyperon channels studied in this work are several orders of magnitude larger than in Refs. \cite{EMFFPanda} and \cite{xscan}. 

Non-negligible systematic effects can arise from model-dependencies in the efficiency correction. The method of moments introduces an uncertainty for each measured variable that is integrated out when calculating each moment. In multi-dimensional problems like the ones presented here, this needs a thorough investigation. Therefore, we have carried out three comparative studies: i) between generated distributions on one hand and reconstructed and efficiency corrected distributions on the other ii) between extracted hyperon and antihyperon parameters iii) between two different parameter estimation techniques. Significant differences only appear for the efficiency independent method and are well understood since in these cases, the necessary criteria for using the efficiency independent method are not fulfilled. However, for the high-precision studies enabled by the design luminosity, it will likely be necessary to use a model-independent method for extracting the spin observables, \textit{e.g.} a Maximum Likelihood-based method similar to the one in Refs. \cite{bes3prl,bes3nature}. For $\bar{p}p$ reactions, a dedicated formalism and analysis framework will be needed for this purpose.

\section{Summary and discussion}
\label{sec:summary}

The feasibility of exclusive reconstruction of two antihyperon-hyperon reactions in the foreseen antiproton experiment PANDA at FAIR has been investigated: $\bar{p}p \to \bar{\Lambda}\Lambda$ and $\bar{p}p \to \bar{\Xi}^+\Xi^-$. The former has been studied with the PS185 experiment and will be used for quality assurance and fine-tuning of detectors, data acquisition, reconstruction and analysis. However, even at the modest luminosity during the start-up phase of PANDA, a world-record sample can be collected in a few days. Furthermore, the background can be suppressed to a very low level. This will allow PANDA to push forward the state of the art in the measurement of spin observables. The double-strange $\Xi^-$ has barely been studied with antiproton probes before and the studies proposed here will therefore be pioneering. The foreseen high data rates and the low background level will enable a complete spin decomposition of the reaction already during the first year of data taking. This demonstrates PANDAs potential as a strangeness factory.

The method of moments applied in this work is suitable for sample sizes of the first phase of PANDA. Two different approaches were applied: a standard efficiency dependent one, and a more unusual efficiency independent method. The applicability of the latter however relies on approximations whose validity need to be evaluated on a case-by-case basis. After only a few years at the initial luminosity, and even more, when the design luminosity is available, the hyperon spin studies will reach high statistical precision. For this, a multi-dimensional and model-independent analysis framework needs to be developed in order to match accuracy and precision. This could open up for large-scale searches for CP violation in hyperon decays and its feasibility will be investigated in the future. 

\section*{Acknowledgement}
We acknowledge financial support from 
the Bhabha Atomic Research Centre (BARC) and the Indian Institute of Technology Bombay, India; 
the Bundesministerium f\"ur Bildung und Forschung (BMBF), Germany; 
the Carl-Zeiss-Stiftung 21-0563-2.8/122/1 and 21-0563-2.8/131/1, Mainz, Germany; 
the Center for Advanced Radiation Technology (KVI-CART), Groningen, Netherlands; 
the CNRS/IN2P3 and the Universit\'{e} Paris-Sud, France; 
the Czech Ministry (MEYS) grants LM2015049, CZ.02.1.01/0.0/0.0/16 and 013/0001677, 
the Deutsche Forschungsgemeinschaft (DFG), Germany; 
the Deutscher Akademischer Austauschdienst (DAAD), Germany; 
the European Union's Horizon 2020 research and innovation programme under grant agreement No 824093.
the Forschungszentrum J\"ulich, Germany; 
the Gesellschaft f\"ur Schwerionenforschung GmbH (GSI), Darmstadt, Germany; 
the Helmholtz-Gemeinschaft Deutscher Forschungszentren (HGF), Germany; 
the INTAS, European Commission funding; 
the Institute of High Energy Physics (IHEP) and the Chinese Academy of Sciences, Beijing, China; 
the Istituto Nazionale di Fisica Nucleare (INFN), Italy; 
the Ministerio de Educacion y Ciencia (MEC) under grant FPA2006-12120-C03-02; 
the Polish Ministry of Science and Higher Education (MNiSW) grant No. 2593/7, PR UE/2012/2, and the National Science Centre (NCN) DEC-2013/09/N/ST2/02180, Poland; 
the State Atomic Energy Corporation Rosatom, National Research Center Kurchatov Institute, Russia; 
the Schweizerischer Nationalfonds zur F\"orderung der Wissenschaftlichen Forschung (SNF), Swiss; 
the Science and Technology Facilities Council (STFC), British funding agency, Great Britain; 
the Scientific and Technological Research Council of Turkey (TUBITAK) under the Grant No. 119F094
the Stefan Meyer Institut f\"ur Subatomare Physik and the \"Osterreichische Akademie der Wissenschaften, Wien, Austria; 
the Swedish Research Council and the Knut and Alice Wallenberg Foundation, Sweden.


\begin{thebibliography}{99}

\bibitem{pdg} M.~Tanabashi \textit{et al.} (Particle Data Group), Phys. Rev. D \textbf{98}, 030001 (2018).


\bibitem{pspin}C.~A.~Aidala \textit{et al.,} Rev. Mod. Phys. \textbf{85}, 655 (2013).

\bibitem{pspin2} C.~Alexandrou \textit{et al.,} Phys. Rev. Lett. \textbf{119}, 142002 (2017).

\bibitem{pradius1}R.~Pohl \textit{et al.,} Nature \textbf{466} (2010) 213; C.~E.~Carlson, Prog. Part. Nucl. Phys. \textbf{82}, 59 (2015).

\bibitem{pradius2}N.~Bezginov \textit{et al.,} Science \textbf{365}, Issue 6457, 1007-1012 (2019).

\bibitem{pradius3} H.-W.~Hammer and U-G.~Meissner, Sci.Bull. \textit{65}, 257-258 (2020).

\bibitem{pstructure}G.~A.~Miller, Phys. Rev. Lett. \textbf{99}, 112001 (2007).

\bibitem{pstructure2} M.~Ablikim \textit{et al.,} Phys. Rev. Lett. \textbf{124}, 042001 (2020).

\bibitem{pasym}A.~D.~Sakharov, Pisma Zh. Eksp. Teor. Phys. Fiz. \textbf{5}, 32 (1967).

\bibitem{granados}C.~Granados,~S.~Leupold and E.~Perotti, Eur. Phys. J. \textbf{A 53}, 17 (2017).

\bibitem{fair} M.~Durante {\it et al.}, Physica Scripta \textbf{94}, 033001 (2019).

\bibitem{panda} W.~Erni \textit{et al.} (PANDA collaboration), \textit{Strong Interaction Studies with PANDA}, Physics Performance Report (2009) arXiv[hep-ex]:0903.3905.

\bibitem{hesr} R.~Maier {\it et al.}, HESR Technical Design Report V. 3.1.2 (2008).

\bibitem{target} A.~Khoukaz {\it et al.}, \textit{The PANDA Internal Targets}, Technical Design Report (2012).

\bibitem{pandadet} The PANDA Collaboration, PANDA Technical Progress Report, FAIR-ESAC (2005).

\bibitem{luminosity} M.~Fritsch {\it et al.}, \textit{The PANDA Luminosity Detector}, Technical Design Report (2018).

\bibitem{PANDAROOT} S.~Spataro {\it et al.}, \textit{The PandaRoot framework for simulation,reconstruction and analysis}, J. Phys.: Conf. Ser. \textbf{331}, 032031 (2011).

\bibitem{FAIRROOT} M.~Al-Turany {\it et al.}, \textit{The FairRoot Framework}, J. Phys.: Conf. Ser. \textbf{396}, 022001 (2012).

\bibitem{ROOT} R.~Brun and F.~Rademakers,
\textit{ROOT - An Object Oriented Data Analysis Framework}, Nucl. Inst. \& Meth. in Phys. Res. A \textbf{389}, 81-86 (1997)

\bibitem{bigibook} I.\ I.\ Bigi, A.\ I.\ Sanda, \textit{CP violation} Cambridge Monographs on Particle Physics, Nuclear Physics and Cosmology, {\bf vol.\ 9}, 
Cambridge University Press (2000).

\bibitem{jennypthesis} J.~Puetz, \textit{Study of Excited $\Xi$ Baryons in Anti-Proton Proton Collisions with the PANDA Detector}, PhD Thesis, Rheinischen Friedrich-Wilhelms-Universität, Bonn (2020).

\bibitem{leeyang}T.~D.~Lee and C.~N.~Yang, Phys. Rev. \textbf{108}, 1645 (1957).

\bibitem{tord} T.~Johansson, Proceedings of the \textit{8th International Conference on Low-Energy Antiproton Physics (LEAP 2005)} 95 (2005).

\bibitem{quarkgluon} M.~Kohno and W.~Weise, Phys. Lett. B \textbf{179} 15 (1986); H.~R.~Rubinstein and H.~Snellman Phys. Lett. B \textbf{165} 187 (1985); S.~Furui and A.~Faessler, Nucl. Phys. A \textbf{486} 669 (1987); M.~Burkardt and M.~Dillig Phys. Rev. C \textbf{37} 1362 (1988); M.~A.~Alberg \textit{et al.} Z. Phys. A \textbf{331} 207 (1988).

\bibitem{kaonexchange} F.~Tabakin and R.~A.~Eisenstein Phys. Rev. C \textbf{31} 1857 (1985); M.~Kohno and W.~Weise, Phys. Lett. B. \textbf{179} 15 (1985); P.~La France \textit{et al.} Phys. Lett. B \textbf{214} 317 (1988); R.~G.~E.~Timmermans \textit{et al.} 1992  Phys. Rev. D \textbf{45} 2288 (1992); J.~Haidenbauer \textit{et al.} Phys. Rev. C \textbf{46} 2516 (1992).

\bibitem{quarkgluonhadron} P.~G.~Ortega \textit{et al.} Phys. Lett. B. \textbf{696} 352 (2011).

\bibitem{XiQG}P.~Kroll and W.~Schweiger, Nucl. Phys. A \textbf{474}, 608 (1987);P.~Kroll, B.~Quadder and W.~Schweiger, Nucl. Phys. B \textbf{316}, 373 (1989);H.~Genz, M.~Nowakowski and D.~Woitschitzsky, Phys. Lett. B \textbf{260}, 179 (1991).

\bibitem{kaidalov} A.~B.~Kaidalov and P.~E.~Z.~Volkovitsky, Phys. C \textbf{63} 51 (1994).

\bibitem{XiMEX}J.~Haidenbauer, K.~Holinde and J.~Speth, Phys. Rev. C \textbf{47}, 2982 (1993).

\bibitem{alberg}M.~Alberg, Nucl. Phys. A \textbf{655}, p. c179-184 (1999).

\bibitem{haidenbauerFSI} J.~Haidenbauer and U.-G.~Meissner, Phys. Lett. B \textbf{761}, 456-461 (2016).

\bibitem{PS185} P.~D.~Barnes \textit{et al.}
Nucl. Phys. A \textbf{526}, 575 (1991); E.~Klempt,~F.~Bradamante, A.~Martin, J.-M.~Richard, Phys. Rep. \textbf{368}, p 119 (2002);K.~D.~Paschke \textit{et al.},
Phys. Rev. C \textbf{74}, 015206 (2006); P.~D.~Barnes \textit{et al.}, Phys. Rev. C \textbf{62}, 055203 (2000).

\bibitem{haidenbauerPot} J. Haidenbauer \emph{et al.}, Phys. Rev. C {\bf 45}, 931 (1992).

\bibitem{bes3prl}~M.~Ablikim \textit{et al.,} Phys. Rev. Lett. \textbf{123}, 122003 (2019).

\bibitem{goran} G.~F\"aldt, Eur. Phys. J. A \textbf{52}, 141 (2016).

\bibitem{goranand} G.~F\"aldt and A.~Kupsc, Phys. Lett. B \textbf{772}, 16 (2017).




\bibitem{bes3nature} M.~Ablikim \textit{et al.} (BESIII Collaboration), Nature Phys. \textbf{15}, p. 631–634 (2019).

\bibitem{PS185164} K.~D.~Paschke \textit{et al.,} Phys. Rev. C \textbf{74}, 015206 (2006).

\bibitem{PS185164_1} P.~D.~Barnes \textit{et al.,} Phys. Rev. C \textbf{54}, 1877 (1996).

\bibitem{Musgrave1965}B.~Musgrave and G.~Petmezas, Nuovo Cim. \textbf{35}, 735 (1965);C.~Baltay \textit{et al.,} Phys. Rev. B \textbf{140}, 1027(1965).

\bibitem{LHCbcharm} 
  R.~Aaij {\it et al.} (LHCb collaboration), Phys.\ Rev.\ Lett.\  {\bf 122}, 211803 (2019).

\bibitem{cabibbo} N.~Cabibbo, Phys.\ Rev.\ Lett.\ \textbf{10}, 531 (1963).
\bibitem{kobayashi} M.~Kobayashi, T.~Maskawa, Prog.\ Th.\ Phys.\ \textbf{49}, 652 (1973).

\bibitem{werner} W.~Bernreuther, Lect. Notes Phys. \textbf{591}, 237 (2002); L.~Canetti \textit{et al.,} New J. Phys. \textbf{14}, 095012 (2012).


\bibitem{LHCb} R.~Aaij {\it et al.} (LHCb Collaboration), Nature Phys. \textbf{13}, 391 (2017).

\bibitem{LHCb2} R.~Aaij {\it et al.} (LHCb Collaboration), Phys. Rev. D \textbf{102},051101(R) (2020).

\bibitem{hyperCP} HyperCP Collaboration, Phys. Rev. Lett. \textbf{93}, 262001 (2004).

\bibitem{cronin} J.~W.~Cronin and O.~E.~Overseth, Phys. Rev. \textbf{129}, 1795 (1963).

\bibitem{claslambda} D.~G.~Ireland \textit{et al.,} Phys. Rev. Lett. \textbf{123}, 182301 (2019).

\bibitem{erikthesis} E.~Thomé, \textit{Multi-Strange and Charmed Antihyperon-Hyperon Physics for PANDA}, PhD thesis, Uppsala University (2012).

\bibitem{paschke} K.~D.~Paschke and B.~Quinn, Phys. Lett. B \textbf{495}, p49 (2000).

\bibitem{jennynstar} J.~Puetz \textit{et al.,}, Eur. Phys. J Web Conf. \textbf{241}, 03004 (2020).

\bibitem{xscan}G.~Barucca \textit{et al.,} Eur. Phys. J. A \textbf{55}, 42 (2019).

\bibitem{durand} L.~Durand III and J.~Sandweiss, Phys. Rev. \textbf{135}, B540 (1964).

\bibitem{walter}W.~Ikegami Andersson, \textit{Exploring the Merits and Challenges of Hyperon Physics with PANDA at FAIR}, PhD thesis, Uppsala University (2020).

\bibitem{michael}M.~Papenbrock \textit{et al.,} Eur. Phys. J. Conf. \textbf{214}, 02026 (2019).

\bibitem{jenny} J.~Regina \textit{et al.,} Proc-CTD19-094 (2019).

\bibitem{phaseone} PANDA Collaboration, \textit{PANDA Phase One}, in preparation (2020).

\bibitem{evtgen} A.~Ryd \textit{et al.}, \textit{EvtGen: A Monte Carlo Generator for B-Physics}, EVTGEN-V00-11-07 (2005).

\bibitem{sigma6} H.~Becker \textit{et al.,} Nucl. Phys. B \textbf{141}, 48 (1978).


\bibitem{pbarx} A.~Capella \textit{et al.,} Phys. Rep. \textbf{236}, p225-329.

\bibitem{CERNHERA} V.~Flaminio \textit{et al.}, Report CERN–HERA–84–01 (1984).

\bibitem{EASTMAN197329} P.~S.~Eastman \textit{et al.,} Nucl. Phys. B \textbf{51}, p29-56 (1973).

\bibitem{LYS1973610} J.~Lys \textit{et al.}, Phys. Rev. D \textbf{7}, p610-621 (1973).

\bibitem{dpm7} G.~D.~Patel \textit{et al.,} Z. Phys. C \textbf{12}, p189-202 (1982).

\bibitem{dpm46} D.~Everett \textit{et al.}, Nucl. Phys. B \textbf{73}, p449-464 (1974).

\bibitem{llbarpippim46} H.~W.~Atherton \textit{et al.,} Nucl. Phys. B \textbf{29}, No. 2, p477-503 (1971).

\bibitem{llbarpippim7} C.~Y.~Chien \textit{et al.,} Phys. Rev. \textbf{152}, 1171-1188 (1966).

\bibitem{nonres7} G.~Alexander \textit{et al.,} Nucl. Phys. B \textbf{35}, No. 1, p45-60 (1971).

\bibitem{nonres46} H.~Braun \textit{et al.,} Nuovo Cim. A \textbf{4}, No. 4, p703-714 (1971).

\bibitem{nonres462} K.~B\"ockman \textit{et al.,} Nuovo Cim. A \textbf{42}, No. 4, p954-996 (1966).

\bibitem{HULSBERGEN2005566} W.~D.~Hulsbergen, Nucl. Instr. Meth. A \textbf{552}, No. 3, p566 (2005).

\bibitem{EMFFPanda} B.~Singh \textit{et al.,}     Eur. Phys. J. A \textbf{52}, No. 10, p. 325 (2016).

\bibitem{tayloethesis} R.~L.~Tayloe, \textit{A measurement of the $\bar{p}p \to \bar{\Lambda}\Lambda$ and $\bar{p}p \to \bar{\Sigma}^0\Lambda + c.c$ reactions at 1.726 GeV/c}, Ph. D. thesis, University of Illinois, USA (1995).


\end{thebibliography}
\end{document}